\documentclass[12pt]{article}
\usepackage{indentfirst}
\usepackage{mathrsfs}
\usepackage{amsbsy}
\usepackage{amsmath,bm}
\usepackage{amssymb}
\usepackage{multicol}
\usepackage{times,epsfig,verbatim,lscape}
\usepackage{amsthm}
\usepackage{graphicx}
\usepackage{float}
\usepackage{subfigure}
\usepackage{color}
\usepackage{multirow}
\usepackage[textwidth=14.5cm]{geometry}
\usepackage{blindtext}
\usepackage{blkarray}
\usepackage{algorithm}
\usepackage{algorithmicx}
\usepackage{booktabs}
\usepackage[titletoc]{appendix}
\usepackage[round]{natbib}
\usepackage{epstopdf}
\usepackage[tiny]{titlesec}%
\usepackage{titlesec}
       \titleformat{\section}
             {\normalfont\large\bfseries}{\thesection}{11pt}{\large}
\usepackage{titlesec}
      \titleformat{\subsection}
            {\normalfont\bfseries}{\thesubsection}{11pt}
\allowdisplaybreaks
\renewcommand{\baselinestretch} {1.2}
\makeatletter 
\def\singlespace{\def\baselinestretch{1}\@normalsize}

\makeatletter

\newtheorem{condition}{Condition}

\newcommand{\lv}[2]{l_{#1}^{#2}}

\def\non{\nonumber}

\def\bse{\begin{eqnarray*}}
\def\ese{\end{eqnarray*}}
\def\be{\begin{eqnarray}}
\def\ee{\end{eqnarray}}
\def\bsq{\begin{equation*}}
\def\esq{\end{equation*}}
\def\bq{\begin{equation}}

\def\th{^{\mbox{\tiny\rm th}}}

\def\AIC{\mbox{AIC}}
\def\BIC{\mbox{BIC}}

\def\bb{{\boldsymbol\beta}}
\def\bg{{\boldsymbol\gamma}}
\def\bd{{\boldsymbol\delta}}
\def\btheta{{\boldsymbol\theta}}
\def\bTheta{{\boldsymbol\Theta}}
\def\bbeta{{\boldsymbol\eta}}
\def\bxi{{\boldsymbol\xi}}
\def\bkappa{{\boldsymbol\kappa}}

\def\bOme{{\boldsymbol \Omega}}

\def\bq{\begin{equation}}
\def\eq{\end{equation}}

\def\overp{\stackrel{p}\rightarrow}
\def\overas{\stackrel{a.s.}\longrightarrow}

\def\log{\hbox{log}}

\def\squarebox#1{\hbox to #1{\hfill\vbox to #1{\vfill}}}
\def\btheta{{\boldsymbol \theta}}

\def\bse{\begin{eqnarray*}}
\def\ese{\end{eqnarray*}}
\def\be{\begin{eqnarray}}
\def\ee{\end{eqnarray}}
\def\bsq{\begin{equation*}}
\def\esq{\end{equation*}}
\def\bq{\begin{equation}}
\def\eq{\end{equation}}

\def\log{\hbox{log}}

\def\th{{\mbox{\tiny\rm th}}}

\newtheorem{thm}{\underline{\bf Theorem}}

\newtheorem{lemma}{\underline{\bf Lemma}}

\newtheorem{remark}{\underline{\bf Remark}}

\allowdisplaybreaks

\def\bse{\begin{eqnarray*}}
\def\ese{\end{eqnarray*}}
\def\be{\begin{eqnarray}}
\def\ee{\end{eqnarray}}
\def\bsq{\begin{equation*}}
\def\esq{\end{equation*}}
\def\bq{\begin{equation}}
\def\eq{\end{equation}}

\def\th{^{\mbox{\tiny\rm th}}}

\def\AIC{\mbox{AIC}}
\def\BIC{\mbox{BIC}}

\def\bb{{\boldsymbol\beta}}
\def\bg{{\boldsymbol\gamma}}

\def\0{{\bf 0}}

\def\x{{\bf x}}

\def\bOme{{\bf \Omega}}

\def\non{\nonumber}

\def\overp{\stackrel{p}\rightarrow}
\def\overd{\stackrel{d}\rightarrow}
\def\overas{\stackrel{a.s.}\rightarrow}

\def\bq{\begin{equation}}
\def\eq{\end{equation}}

\def\log{\hbox{log}}

\def\squarebox#1{\hbox to #1{\hfill\vbox to #1{\vfill}}}

\def\bse{\begin{eqnarray*}}
\def\ese{\end{eqnarray*}}
\def\be{\begin{eqnarray}}
\def\ee{\end{eqnarray}}
\def\bsq{\begin{equation*}}
\def\esq{\end{equation*}}
\def\bq{\begin{equation}}
\def\eq{\end{equation}}

\def\log{\hbox{log}}

\def\bqe{\begin{eqnarray}}
\def\bse{\begin{eqnarray*}}
\def\eqe{\end{eqnarray}}
\def\ese{\end{eqnarray*}}

\def\th{^{\mbox{\tiny\rm th}}}

\def\xic{{\mbox{\tiny\rm xIC}}}
\def\xIC{{\mbox{\rm xIC}}}

  \def\calO{{\cal O}}

  \def\calU{{\cal U}}

\def\sn{\sqrt{n}}
\def\AIC{{\mbox{\rm AIC}}}
\def\BIC{{\mbox{\rm BIC}}}
\def\aic{{\mbox{\tiny\rm AIC}}}
\def\bic{{\mbox{\tiny\rm BIC}}}

\def\tit.arg{\textbf{On the asymptotic distribution of model averaging based on information criterion}}
\def\shorttitle.arg{
On the asymptotic distribution of model averaging using AIC and BIC
}

\def\key.arg{Asymptotic distribution; General fixed parameter setup; Model averaging; Smoothed AIC; Smoothed BIC.}

\def\abst.arg{Smoothed AIC (S-AIC) and Smoothed BIC (S-BIC) are very widely used in model averaging and are very easily to implement.
Especially, the optimal model averaging method MMA and JMA have only been well developed
in linear models. Only by modifying, they can be applied to other models.
But S-AIC and S-BIC can be used in all situations where AIC and BIC
can be calculated. In this paper, we study the asymptotic behavior of two commonly used model averaging estimators, the S-AIC and S-BIC estimators, under the standard asymptotic with general fixed parameter setup.
In addition, the resulting coverage probability in \cite{S1997Model} is not studied accurately, but it is claimed that it will be close to the intended.
Our derivation make it possible to study accurately.
Besides, we also prove that the confidence interval construction method in \cite{hjort2003averaging} still works in linear regression with normal distribution error. Both the simulation and applied example support our theory conclusion.
}

\def\author.arg{Miaomiao Wang$^{1,2,3,}$\footnote{Email: wangmm@amss.ac.cn}, Xinyu Zhang$^{2}$, Alan T.K. Wan$^{4}$, Guohua Zou$^{5}$
\vskip 0.2cm
{\it \small $^{1}$School of Chinese Pharmacy, Beijing University of Chinese Medicine, Beijing, China. \\
$^{2}$Academy of Mathematics and Systems Science, Chinese Academy of Sciences, Beijing, China. \\
$^{3}$University of the Chinese Academy of Sciences, Beijing, China. \\
$^{4}$Department of Management Sciences, City University of Hong Kong, Kowloon, Hong Kong.\\
$^{5}$School of Mathematical Sciences, Capital Normal University, Beijing, China.
}\\
\medskip
}

\allowdisplaybreaks
\begin{document}
\thispagestyle{empty}

\begin{center}
{\large \tit.arg}

\vskip 3mm

\author.arg
\end{center}

\vskip 3mm

\centerline{\small SUMMARY} \abst.arg
\\


\clearpage\pagebreak\newpage
\pagenumbering{arabic}
\setcounter{page}{1}

\section{Introduction}\label{chap:introduction}
\setcounter{equation}{0}
\baselineskip=24pt
Since Model Averaging was advocated, it has become one of the most important tools for estimation and inference available to statisticians. Unlike model selection, which picks
a single model among the candidate models, model averaging incorporates all available information by averaging over all potential models. Model averaging is more robust than model selection because the averaging estimator considers the uncertainty across different
models as well as the model bias from each candidate model. The central questions of concern are how to optimally assign the weights for candidate models and how to make inferences based on the averaging estimator.

Bayesian model averaging (BMA) and frequentist model averaging (FMA) are two branches of Model averaging, and here we restrict our attention
to FMA. Some FMA strategies have been developed and studied, in
which, weighting strategy based on the Akaike information criterion (AIC) or the Bayesian information criterion (BIC) scores proposed
by \cite{S1997Model} is probably the most widely used method
because of convenience. See, for example, \cite{burnham2004multimodel} and \cite{layton2006embracing}. However, in \cite{S1997Model}, there is a variance component due to model selection uncertainty that should be incorporated into estimates of precision. That is, one needs estimates that are ¡°unconditional¡± on the selected model. They propose
a simple estimator of the unconditional variance for the maximum likelihood estimator under the assumption that the parameters are perfect relevant. This inference can only be applied to parameters that are in common to all contending model.

However, a large number of authors have examined the inference of model averaging
estimators.
\cite{hjort2003averaging} and \cite{Claeskens2008Model} studied the asymptotic properties of the FMA maximum likelihood estimator.
\cite{Liu2015Distribution} derived the limiting distributions of least squares averaging estimators for linear regression models. Other works on the asymptotic property of averaging estimators include \cite{leung2006information}, \cite{Benedikt2006The}, \cite{Hansen2009Averaging} and \cite{Hansen2010Averaging}. All of them used local misspecification assumption. Local misspecification assumption means that some parameters are of order $1/\sqrt{n}$, where $n$ is sample size. Although this assumption provides a suitable framework for studying the asymptotic theories of FMA estimators, it also draws comments from \cite{Raftery2003Discussion} because of its realism.
Researchers aiming at estimating effects or parameters are more likely to use larger sample sizes. For example, the asymptotic properties in above papers are under large sample. However, as the sample size increases, the parameter space becomes smaller. So parameters under local misspecification assumption will be closer to the true value. In other words, all candidate models are closer to true model. This assumption is unreal under large sample. Furthermore, the local asymptotic framework induces the local parameters in the asymptotics, which generally cannot be estimated consistently. It is not just a technical regularity condition, but a key assumption about the way the world works.
Recently, there have some studies about inference of model averaging estimators under general fixed parameter settings. For example, based on the Mallows criterion and the jackknife, \cite{zhang2018INFERENCE} derived the asymptotic distributions of the nested least squares averaging estimators. Without the local misspecification assumption, \cite{Mitra2019A} established the limiting distributions of model averaging estimators but they must consider possible modeling biases.

In the current paper, under general fixed parameter
setup, which do not rely on the assumption that all uncertain parameters
are small, we investigate limiting distributions of model averaging estimators by weighting strategy based on the smoothed AIC (S-AIC) or smoothed BIC (S-BIC) scores.
Our motivation is that S-AIC and S-BIC are very widely used model averaging and are very easily to implement. Especially, the optimal model averaging method  Mallows model averaging (MMA) (\cite{hansen2007least}) and jackknife model averaging (JMA) (\cite{hansen2012jackknife}) have only been well developed in linear models, but S-AIC and S-BIC can be used in all situations where AIC and BIC can be calculated. Besides, \cite{S1997Model} did't present a asymptotic distributions theory of model averaging estimators. They
advocated a simple estimator of variance for the maximum likelihood estimator under the perfect correlation assumption. It will result in the overestimation and wider confidence interval.

Recently, \cite{Charkhi2018Asymptotic} studied the asymptotic distribution of a special case of model averaging estimators, called Post-AIC-selection estimator, in which the weight for the model with the smallest AIC
value is equal to one and the remaining weights are equal to zero, with the weights summing to one and being random. Let $\{\AIC_{1}, \AIC_{2},\cdots,\AIC_{M}\}$ be the AIC scores for all candidate models, more generally, \cite{S1997Model} suggested estimators of the model average form using weights proportional to $\exp(-\AIC_{m}/2)$, not just by means of indicator functions. These are two different weighting methods.
S-AIC weights can approximate to the special form $1$ for the selected model and $0$ for all other models when the quantity $\exp(-\AIC_{m}/2)$ of one model is large enough, that of other models are relatively small compared to it.
The performance of their inference depends not only on the selected model,
but also on candidate model set. They concerned two cases, including the classical setting in which a true model exists and is included in the candidate
set of models and all candidate models are misspecified.
But, we cannot construct confidence interval by the asymptotic distribution accurately if relaxing the assumption about the existence of a true model. More details refer to \cite{Charkhi2018Asymptotic}.

Therefore, under general fixed parameter setup, we study asymptotic distributions of model averaging estimators by weighting strategy based on the S-AIC or S-BIC scores and statistical inference, constructing the confidence intervals based on these asymptotic distributions.
\cite{S1997Model} and \cite{Burnham2002Model} provided an intuitive interval construction method. The resulting coverage probability of their confidence interval is not studied accurately, but it is claimed that it will be close to the intended value.
We will confirm the validity of their method.
Besides, we also prove that the confidence interval construction method for the linear function of the parameters in \cite{hjort2003averaging} still works in linear regression with the normally distributed error. We will use the simulation study and real data analysis to support our methods.

The reminder of the article is organized as follows. We present the candidate model set and the model averaging estimators in Section \ref{sec:MA Estimator}. In Section \ref{sec:Inference}, the theoretical properties of the proposed model averaging method are exploited. Section \ref{sec:CI} construct the confidence intervals based on our asymptotic distributions and briefly discuss each estimator and how to construct the confidence intervals for each
estimator in the existing literature.
The derivation for the methods in \cite{S1997Model} and \cite{hjort2003averaging} are provided in Section \ref{D:S1997} and Section \ref{D:H2003}, respectively.
Simulations and real data analysis results are reported in Sections \ref{sec:sim} and \ref{sec:real}, to assess the performance of the proposed procedure compared to several existing model selection and model averaging methods. All proofs of technical details are provided in \ref{sec:A}.

\section{Model Average Estimators}
\label{sec:MA Estimator}
Suppose independent random variables $y_{1},\ldots,y_{n}$ come from density $f$. Inference is sought for a certain parameter $\mu=\mu(f)$. In \cite{hjort2003averaging},
they concern with a model of the type
\begin{eqnarray}
\label{setup}
f(y,\btheta),
\end{eqnarray}
where $\btheta \subset \bTheta \subset \mathbb{R}^{q}$, $\btheta=\left(\theta_1,\ldots,\theta_q\right)^{T}=\left(\bb^{T},\bg^{T}\right)^{T}$ with a $q_{1}\times1$ vector of parameters $\boldsymbol{\bb}$ and an additional $q_{2}\times1$ vector of $\bg$ parameters, and $\bg=\bg_{0}$ is fixed and known, which corresponds to the narrow model in the sense that $f(y,\btheta)=f\left(y,\left(\bb^{T},\bg_{0}^{T}\right)^{T}\right)$. They investigate the asymptotic theory in a local  misspecification framework
$$f_{true}(y)=f_{n}(y)=f\left(y,\left(\bb_{0}^{T},\left(\bg_{0}+\bd/\sqrt{n}\right)^{T}\right)^{T}\right).$$
The $\bd=\left(\delta_{1},\ldots,\delta_{q_{2}}\right)^{T}$ parameters signify the degrees of the model departures in directions $1,...,q_{2}$, with due influence on the estimand $\mu_{true}=\mu\left(\left(\bb_{0}^{T},\left(\bg_{0}+\bd/\sqrt{n}\right)^{T}\right)^{T}\right)$.

In our paper, we study the asymptotic distribution of model averaging using AIC and BIC under general fixed parameter setup \eqref{setup}. Without loss of generality, we assume $\bg_{0}=\boldsymbol{0}$. Then we present the candidate models setting first.
Let $\btheta_{0}$ denote the true value of the parameters. Suppose that we combine $M$ sub-models of \eqref{setup} and the $m\th$ sub-model has parameter $\bb$ and some of the parameter $\gamma_{j}$, where $\gamma_{j}$ is the $j\th$ element of $\bg$. Denote $k_{m}$ as the number of parameters including in the $m\th$ sub-model.
Typically, $M=2^{q_2}$ if all possible models are considered and $M=q_2+1$ when combining nested models, one for each subset $S$ of $\{1,\ldots,q_{2}\}.$
For the $m\th$ sub-model, the subset $S_{m}=\left\{i_{1},\cdots,i_{k_{m}-q_{1}}\}\subset\{1,\ldots,q_{2}\right\}$
and $S_{m}^{c}=\left\{i_{k_{m}-q_{1}+1},\cdots,i_{q_{2}}\right\}$ are the complement of $S_{m}$ on set $\{1,\ldots,q_{2}\}$. To facilitate the presentation, we write $\bg_{S_{m}}=\left(\gamma_{i_{1}},\cdots,\gamma_{i_{k_{m}-q_{1}}}\right)^{T}$ and
$\bg_{S_{m}^{c}}=\left(\gamma_{i_{k_{m}-q_{1}+1}},\cdots,\gamma_{i_{q_{2}}}\right)^{T}$, then the parameter in the $m\th$ model is $\btheta_{m}=\left(\bb^{T},\bg_{m}^{T}\right)^{T}$ where $\bg_{m}=\left(\gamma_{1},\gamma_{2},\cdots,\gamma_{q_{2}}\right)^{T}$ with $\bg_{S_{m}^{c}}=\boldsymbol{0}.$ 
In addition,
through a series of permutations for $\btheta_{m}$, it can be written under the form $\btheta_{m}^{\prime}=\left(\bb^{T},\bg_{S_{m}}^{T},\boldsymbol{0}^{T}\right)^{T}=\left(\btheta_{S_{m}}^{T},\boldsymbol{0}^{T}\right)^{T}$, where $\btheta_{S_{m}}=\left(\bb^{T},\bg_{S_{m}}^{T}\right)^{T}.$ We give a permutation matrix $\Pi_{m}$ so that $\btheta_{m}^{\prime}=\Pi_{m}\btheta_{m}$, where $\Pi_{m}$ can be obtained by row permutation for unit matrix $\boldsymbol{I}_{q}=\left(\boldsymbol{e}_{1}^{T},\boldsymbol{e}_{2}^{T},\cdots,\boldsymbol{e}_{q}^{T}\right)^{T}$ and $\boldsymbol{e}_{j}$ is a unit vector with the $j\th$ element being $1$, i.e. $$\Pi_{m}=\left(\boldsymbol{e}_{1}^{T},\cdots,\boldsymbol{e}_{q_{1}}^{T},
\boldsymbol{e}_{q_{1}+i_{1}}^{T},\cdots,\boldsymbol{e}_{q_{1}+i_{k_{m}-q_{1}}}^{T},
\boldsymbol{e}_{q_{1}+i_{k_{m}-q_{1}+1}}^{T},\cdots,\boldsymbol{e}_{q_{1}+i_{q_{2}}}^{T}\right)^{T}.$$

Denote $\widehat{\btheta}_{S_{m}}$ as the maximum-likelihood estimator of $\btheta_{S_{m}}$ under the $m\th$ sub-model over parameter space $\bTheta_{m}\subset \mathbb{R}^{k_{m}}$, which is the solution of the $\log$ likelihood equation
$\sum_{t=1}^{n}\partial\log f(y_{t},\btheta_{m})/\partial\btheta_{S_{m}}=0.$ Define $\widehat\btheta_{m}^{\prime}=(\widehat{\btheta}_{S_{m}}^{T},\boldsymbol{0}^{T})^{T}.$ \cite{Akaike1973Information} has noted that since $l_{n}(\btheta_{m})$ is a natural alternative for $\mathrm{E}\{\log f(y_{t},\btheta_{m})\},$ $\widehat\btheta_{S_{m}}$ is a natural estimator for $\btheta_{S_{m}}^{\ast},$ the parameter vector which minimizes the Kullback-Leibler Information Criterion (KLIC) defined in \cite{kullback1951information},
\be
I\left\{f(y,\btheta_{0}):f(y,\btheta_{m})\right\}
   =\mathrm{E}\left[\log\left\{\frac{f(y,\btheta_{0})}{f(y,\btheta_{m})}\right\}\right].
\ee
Here, and in what follows, expectations are taken with respect to the true distribution. Hence,
\be
I\left\{f(y,\btheta_{0}):f(y,\btheta_{m})\right\}
   =\int f(y,\btheta_{0})\log f(y,\btheta_{0})\mathrm{d}y-\int f(y,\btheta_{0})\log f(y,\btheta_{m})\mathrm{d}y.
\ee
Adopting inverse permutations $\Pi_{m}^{T}$ for $(\btheta_{S_{m}}^{\ast T},\boldsymbol{0}^{T})^{T}$, we can obtain $\btheta_{m}^{\ast}$, which has the same order of parameter $\btheta_{0}$.
Define $\widehat\btheta_{m}$ as the estimator of $\btheta_{m}$ under the $m\th$ sub-model over $\bTheta_{m}\subset \mathbb{R}^{k_{m}},$ so the model averaging estimator of $\btheta$ is
$
\widehat\btheta(\boldsymbol{w})=\sum_{m=1}^{M} w_m\widehat\btheta_m, \label{hatthetaw}
$
where $w_{m}$ is the weight corresponding to the $m\th$ sub-model and $\boldsymbol{w}=(w_1,\ldots,w_M)^{T}$, belonging to weight set ${\cal{W}} = \{\boldsymbol{w} \in \left[0,1\right]^M:\sum\nolimits_{m = 1}^{M}w_m = 1\}$.

Next, we introduce some frequentist model averaging strategies.
In the $m\th$ candidate model, AIC and BIC scores are
$\AIC_{m}=-2\sum_{t=1}^{n}\log f(y_{t},\widehat{\btheta}_{m})+2k_{m}$ and
$\BIC_{m}=-2\sum_{t=1}^{n}\log f(y_{t},\widehat{\btheta}_{m})+k_{m}\log n$, respectively.
Define the following weights:
\be
\widehat w_{\xic,m}=\exp(-\x IC_m/2)/\sum_{m=1}^{M}\exp(-\x IC_m/2),\quad m=1,\ldots, M,\label{hatwxic}
\ee
where $\xIC_m$ is the AIC or BIC value from the $m\th$ sub-model.
The corresponding average estimators are commonly called S-AIC or S-BIC estimators:
\be
\label{12}
\widehat\btheta(\widehat{\boldsymbol{w}}_{\aic})=\sum_{m=1}^{M}\widehat{w}_{\aic,m}\widehat\btheta_m,\\
\widehat\btheta(\widehat{\boldsymbol{w}}_{\bic})=\sum_{m=1}^{M}\widehat{w}_{\bic,m}\widehat\btheta_m.
\ee
\section{Inference after weighting strategy based on the AIC or BIC scores} \label{sec:Inference}
Before presenting asymptotic results, we give some notations and list the regularity conditions, where all limiting processes here and throughout the text are with respect to $n\to\infty$ and notations ¡°$\overd$¡±, ¡°$\overas$¡± and ¡°$\overp$¡± denote convergence in distribution, almost surely and in probability, respectively.

The likelihood function is defined as
$$L_{n}(\btheta)=\prod_{t=1}^{n}f(y_{t},\btheta),$$ and the $\log$ likelihood function is denoted by $l_{n}(\btheta)=\log L_{n}(\btheta).$ We assume that the first and second partial derivatives of $f(y,\btheta)$ with respect to $\btheta$ exist. Let
\be
\Psi(y,\btheta)&=&\partial\log f(y,\btheta)/\partial\btheta,\ \ \text{a}\ q\times1\ \text{vector},\non\\
\dot{\Psi}(y,\btheta)&=&\partial^{2}\log f(y,\btheta)/\partial\btheta\partial\btheta^{T},\ \ \text{a}\ q\times q\ \text{matrix}\non
\ee
and then
\be
A_{n}(\btheta)&=&n^{-1}\sum_{t=1}^{n}\dot{\Psi}(y_{t},\btheta).
\ee
Then the first and second partial derivatives of $l_{n}(\btheta)$ with respect to $\btheta$ are $\dot{l}_{n}(\btheta)=\sum_{t=1}^{n}\Psi(y_{t},\btheta)$ and $\ddot{l}_{n}(\btheta)=\sum_{t=1}^{n}\dot{\Psi}(y_{t},\btheta),$ respectively.
Fisher Information is defined as
$$\mathcal{F}(\btheta)=E\left\{\Psi(y,\btheta)\Psi(y,\btheta)^{T}\right\},\ \ \text{a}\ q\times q\ \text{matrix}.$$

Correspondingly, for the $m\th$ sub-model, define  $\btheta_{0,m}=\Pi_{m}\btheta_{0}=(\btheta_{0,S_{m}}^{T},\btheta_{0,S_{m}^{c}}^{T})^{T}$, where $\btheta_{0,S_{m}}$ is the first $k_{m}$ elements of $\btheta_{0,m}$ and $\btheta_{0,S_{m}^{c}}$ is the remaining $q-k_{m}$ elements. Denote
$$\mathcal{F}_{m}(\btheta_{0,m})=E\left\{\Psi(y,\btheta_{0,m})\Psi(y,\btheta_{0,m})^{T}\right\}
=\Pi_{m}\mathcal{F}(\btheta_{0})\Pi_{m}^{T},\ \ \text{a}\ q\times q\ \text{matrix}.$$
When the partial derivatives exist, we define the vectors and matrices
\be
\Psi_{m}(y,\btheta_{S_{m}})&=&\partial\log f(y,\btheta_{m})/\partial\btheta_{S_{m}},\ \ \text{a}\ k_{m}\times1\ \text{vector},\non\\
\dot{\Psi}_{m}(y,\btheta_{S_{m}})&=&\partial^{2}\log f(y,\btheta_{m})/\partial\btheta_{S_{m}}\partial\btheta_{S_{m}}^{T},\ \ \text{a}\ k_{m}\times k_{m}\ \text{matrix},\non\\
A_{m,n}(\btheta_{S_{m}})&=&n^{-1}\sum_{t=1}^{n}\dot{\Psi}_{m}(y_{t},\btheta_{S_{m}}),\non\\
B_{m,n}(\btheta_{S_{m}})&=&n^{-1}\sum_{t=1}^{n}\Psi_{m}(y_{t},\btheta_{S_{m}})\Psi_{m}(y_{t},\btheta_{S_{m}})^{T},\non\\
\dot{l}_{m,n}(\btheta_{S_{m}})&=&\sum_{t=1}^{n}\Psi_{m}(y_{t},\btheta_{S_{m}}),\non\\ \text{and}~\ddot{l}_{m,n}(\btheta_{S_{m}})&=&\sum_{t=1}^{n}\dot{\Psi}_{m}(y_{t},\btheta_{S_{m}}).\non
\ee
If expectations also exist, we define the matrices
\be
A_{m}(\btheta_{S_{m}})&=&\mathrm{E}\left\{\dot{\Psi}_{m}(y,\btheta_{S_{m}})\right\},\non\\
B_{m}(\btheta_{S_{m}})&=&\mathrm{E}\left\{\Psi_{m}(y,\btheta_{S_{m}})\Psi_{m}(y,\btheta_{S_{m}})^{T}\right\}\non.\ee
We define
\be C_{m,n}(\btheta_{S_{m}})&=&A_{m,n}(\btheta_{S_{m}})^{-1}B_{m,n}(\btheta_{S_{m}})A_{m,n}(\btheta_{S_{m}})^{-1},\non\\
C_{m}(\btheta_{S_{m}})&=&A_{m}(\btheta_{S_{m}})^{-1}B_{m}(\btheta_{S_{m}})A_{m}(\btheta_{S_{m}})^{-1},\non\ee
where the inverse matrixes are assumed to exist.
The Fisher Information matrix of $\btheta_{0,S_{m}}$ is denoted by
$$\mathcal{F}_{m}(\btheta_{0,S_{m}})=E\left\{\Psi_{m}(y,\btheta_{0,S_{m}})\Psi_{m}(y,\btheta_{0,S_{m}})^{T}\right\}
,\ \ \text{a}\ k_{m}\times k_{m}\ \text{matrix}.$$

Assume that the $m_{o}\th$ model is the true model, which means that
$\btheta_{m_{o}}=\btheta_{0}$, and $\widehat{\btheta}_{0}$ and $\widehat{\btheta}_{0,m}$ are the maximum-likelihood estimator of $\btheta_{0}$ and $\btheta_{0,m}$ under the true model over $\bTheta$, respectively. Any sub-model including
all regressors of the true model is called overfitted model, contained by set $\calO$. The remaining sub-models are underfitted models, contained by set $\calU$.
The notations $m\in\calO$ and $m\in\calU$ imply the $m^{th}$ model is overfitted model and underfitted model, respectively.

We now give some regular conditions required for our theorems.
\begin{condition}
\label{c1}
The density $f(y,\btheta)$ is measurable in $y$ for every $\btheta$ in $\bTheta,$ a compact subset of $\mathbb{R}^{q},$ continuous in $\btheta$ for every $y$ in $\bOme,$ a measurable Euclidean space, and the true parameter point $\btheta_{0}$ is identifiable.
\end{condition}
\begin{condition}
\label{c2}
\emph{(a)} $\mathrm{E}\{\log f\left(y_{t},\btheta_{m}\right)\}$ exists and $\left|\log f\left(y,\btheta_{m}\right)\right|\leq K_{m}(y)$ for all $\btheta_{m}$ in $\bTheta,$ where $K_{m}(y)$ is integrable;

\emph{(b)} $I\left\{f(y,\btheta_{0,m}):f(y,\btheta_{m})\right\}$ has a unique minimum at $\btheta_{m}^{\ast}$ with $\btheta_{S_{m}}^{\ast}$ being in $\bTheta_{m}$.
\end{condition}
\begin{condition}
\label{c3}
\emph{(a)} $\partial\log f(y,\btheta_{m})/\partial\theta_{i}, i=1,\ldots,k_{m},$ are bounded in absolute value by a function integrable uniformly in some neighborhood of $\btheta_{0}.$

\emph{(b)} Second partial derivative of $f(y,\btheta_{m})$ with respect to $\btheta_{m}$ exists and is continuous for all $y$, and may be passed under the integral sign in $\int f(y,\btheta_{m})\mathrm{d}y.$
\end{condition}
\begin{condition}
\label{c4}
$\left|\partial^{2}\log f(y,\btheta_{m})/\partial\theta_{i}\partial\theta_{j}\right|$ and $\left|\partial\log f(y,\btheta_{m})/\partial\theta_{i}\cdot\partial\log f(y,\btheta_{m})/\partial\theta_{j}\right|,\ i,j=1,\ldots,k_{m}$ are dominated by functions integrable for all $y$ in $\bOme$ and $\btheta_{m}$ in $\bTheta$.
\end{condition}
\begin{condition}
\label{c5}
\emph{(a)} $\btheta_{S_{m}}^{\ast}$ is interior to $\bTheta_{m};$ (b) $B(\btheta_{S_{m}}^{\ast})$ is nonsingular; (c) $\btheta_{S_{m}}^{\ast}$ is a regular point of $A_{m}(\btheta_{S_{m}})$, which means $\btheta_{S_{m}}$ is a value such that $A_{m}(\btheta_{S_{m}})$ has constant rank in some open neighborhood of $\btheta_{S_{m}}^{\ast}.$
\end{condition}
\begin{condition}
\label{c6}
Fisher Information $\mathcal{F}(\btheta_{0})$ is a positive definite matrix.
\end{condition}
\begin{condition}
\label{c7}
$\left|\partial\left\{\partial f(y,\btheta)/\theta_{i}\cdot f(y,\btheta)\right\}/\partial\theta_{j}\right|,\ i,j=1,\cdots,q,$ are dominated
by functions integrable for all $\btheta$ in $\bTheta$, and the minimal support of $f(y,\btheta)$ does not depend on $\btheta.$
\end{condition}
\begin{remark}
Condition \emph{\ref{c2}} ensures that the KLIC is well-defined. Condition \emph{\ref{c3}(a)} allows us to apply a Uniform Law of Large Numbers. Condition \emph{\ref{c3}(b)} ensures that the first two derivatives with respect to $\btheta_{S_{m}}$. Condition \emph{\ref{c4}} ensures that the derivatives are appropriately dominated by functions integrable, which ensures that $A_{m}(\btheta_{S_{m}})$ and $B_{m}(\btheta_{S_{m}})$ are continuous in $\btheta_{S_{m}}$ and that we can apply a Uniform Law of Large Numbers to $A_{m,n}(\btheta_{S_{m}})$ and $B_{m,n}(\btheta_{S_{m}}).$ We can find the same assumptions in \emph{\cite{white1982maximum}} and \emph{\cite{Ferguson1996A}}.
\end{remark}

Under the $m\th$ model, the likelihood ratio test statistic is
$$\lambda_{m,n}=\frac{sup_{\btheta\in\bTheta_{m}}\prod_{t=1}^{n}f(y_{t},\btheta)}
{sup_{\btheta\in\bTheta}\prod_{t=1}^{n}f(y_{t},\btheta)}=\frac{\prod_{t=1}^{n}f(y_{t},\widehat{\btheta}_{m})}{\prod_{t=1}^{n}f(y_{t},\widehat{\btheta}_{o})}.$$
Let $\boldsymbol{I}_{q}$ be an identity matrix. We can draw the following conclusion.
\begin{lemma}
\label{lem1}
If Conditions \emph{\ref{c1}}, \emph{\ref{c3}} and \emph{\ref{c6}} are satisfied, then when model $m\in\cal O,$
\be
-2\log\lambda_{m,n}&=&(\Pi_{m}\bxi_{n})^{T}\left[\left\{\Pi_{m}\mathcal{F}(\btheta_{0})\Pi_{m}^{T}\right\}^{-1}-H_{m}\right]\Pi_{m}\bxi_{n}+o_{a.s.}(1)\non\\
&\overd& \bkappa^{T}P_{m}\bkappa\sim \chi^{2}(q-k_{m}),\non
\ee where $\bxi_{n}=\frac{1}{\sqrt{n}}\dot{l}_{n}(\btheta_{0})$, $\bkappa\sim\mathcal{N}(0,\boldsymbol{I}_{q})$ and
$$P_{m}=\left\{\Pi_{m}\mathcal{F}(\btheta_{0})\Pi_{m}^{T}\right\}^{\frac{1}{2}}\left[\left\{\Pi_{m}\mathcal{F}(\btheta_{0})\Pi_{m}^{T}\right\}^{-1}-H_{m}\right]\left\{\Pi_{m}\mathcal{F}(\btheta_{0})\Pi_{m}^{T}\right\}^{\frac{1}{2}}.$$
\begin{proof}
See \ref{sec:lem1}.
\end{proof}
\end{lemma}
\begin{lemma}
\label{lem2}
For any underfitted model $m\in\cal U,$ under Conditions \emph{\ref{c1},\ref{c2}} and Conditions \emph{\ref{c3}(b)-\ref{c5}}, we have
\be
n^{-1}l_{n}\left(\widehat{\btheta}_{m}\right)
=\mathrm{E}\left\{\log f(y_{t},\btheta_{m}^{\ast})\right\}+o_{p}(1).
\ee
\begin{proof}
See \ref{sec:lem2}.
\end{proof}
\end{lemma}
Let
$$G_{m}=\exp\left\{(\Pi_{m}\mathcal{F}(\btheta_{0})\bbeta)^{T}\left\{H_{m}-(\Pi_{m}\mathcal{F}(\btheta_{0})\Pi_{m}^{T})^{-1}\right\}\Pi_{m}\mathcal{F}(\btheta_{0})\bbeta/2-k_{m}\right\}$$
with $\bbeta\sim \mathcal{N}\left\{0,\mathcal{F}^{-1}(\btheta_{0})\right\}$.
Define  $\boldsymbol{w}_{m_o}^o$ to be a
vector with the $m\th$ element taking on the value of unity and other
elements zeros, and
$\boldsymbol{w}_{\aic}$ to be an $M\times1$ vector with the $m^{th}$ element
$$w_{\aic,m}=\frac{\boldsymbol{1}\{m\in \calO\}G_m}{G},$$
where $\boldsymbol{1}\{\cdot\}$ is an indicator function and $G=\sum_{m\in\calO}G_{m}$. Then we can derive the asymptotic distributions of the S-AIC and S-BIC weights as follows.
\begin{lemma} \label{lem3}
Let $\widehat{\boldsymbol{w}}_{\aic}=(\widehat{w}_{\aic,1},\ldots,\widehat{w}_{\aic,M})^{T}$
and $\widehat{\boldsymbol{w}}_{\bic}=(\widehat{w}_{\bic,1},\ldots,\widehat{w}_{\bic,M})^{T}$.
If Conditions \emph{\ref{c1}-\ref{c6}} are satisfied, then
\be
\widehat{\boldsymbol{w}}_{\aic}\overd \boldsymbol{w}_{\aic}
\ee
and
\be
\boldsymbol{\widehat{w}}_{\bic}\overp \boldsymbol{w}_{m_{o}}^{o}.
\ee
\begin{proof}
See \ref{sec:lem3}.
\end{proof}
\end{lemma}
Let $\Delta_{m}$ be a diagonal selection matrix so that
$$\Delta_{m}=
\begin{pmatrix}
\boldsymbol{I}_{q_{1}}&&&&\\
&\delta_{1}^{m}&&&\\
&&\delta_{2}^{m}&&\\
&&&\ddots&\\
&&&&\delta_{q_{2}}^{m}\\
\end{pmatrix},$$
where
\[\delta_{j}^{m}=\left\{\begin{array}{ll}
1,&j\in S_{m},\\
0,&j\not\in S_{m}.
\end{array}\right.\]
We denote $\widehat\theta_{j}(\widehat{\boldsymbol{w}}_{\aic})$, $\widehat\theta_{j}(\widehat{\boldsymbol{w}}_{\bic})$ and $\theta_{j,0}$ as the $j\th$ components of $\widehat\btheta(\widehat{\boldsymbol{w}}_{\aic})$, $\widehat\btheta(\widehat{\boldsymbol{w}}_{\bic})$ and $\btheta_{0}$, respectively.
Then we can establish the asymptotic distributions for the S-AIC and S-BIC estimators.
\begin{lemma}
\label{lem4}
If Conditions \emph{\ref{c1}-\ref{c7}} are satisfied, then
\be\sn\left\{\widehat\btheta( \widehat{\boldsymbol{w}}_{\aic})-\btheta_{0}\right\}\overd \sum_{m\in\calO}(G_{m}/G)\Pi_{m}^{T}H_{S_{m}}\Pi_{m}\Delta_{m}\mathcal{F}(\btheta_{0})\bbeta\ee
and $\widehat{\theta}_{j}(\boldsymbol{\widehat{w}}_{\bic})\overd 0$ for $j\not\in S_{m_{o}}$ and $\widehat{\theta}_{j}(\boldsymbol{\widehat{w}}_{\bic})\overd Z_{j,m_{o}}$ for $j\in S_{m_{o}}$,
where
$Z_{j,m_{o}}\sim \mathcal{N}(0,\sigma_{j,m_{o}}^{2})$ and $\sigma_{j,m_{o}}^{2}$ is the $j\th$ element on the diagonal of matrix $$\Sigma_{m_{o}}=\Pi_{m_{o}}^{T}H_{S_{m_{o}}}\Pi_{m_{o}}\Delta_{m_{o}}\mathcal{F}(\btheta_{0})(\Pi_{m_{o}}^{T}H_{S_{m_{o}}}\Pi_{m_{o}}\Delta_{m_{o}})^{T}.$$
\begin{proof}
See \ref{sec:lem4}.
\end{proof}
\end{lemma}

However, S-AIC and S-BIC estimators are not very appropriate to estimate $\btheta_{0}$, because some elements of $\widehat{\btheta}_{m}$ are directly set to $0$ so that the sum of weight associated with estimators is not $1$. Therefore, as in \cite{S1997Model}, we use the scaled S-AIC and scaled S-BIC estimators, seeking weights that can be associated with estimators drawn under each of the contending models including parameter $\theta_{j}$ and denoted by $l_{v}^{j}$, $v=1,\cdots,M_{j}$. Denote $\lv{v}{j}$
$$\boldsymbol{\widehat{w}}_{\aic_{s}}^{j}
=\left(\frac{\widehat{w}_{\aic,\lv{1}{j}}}{\sum_{v=1}^{M_{j}}\widehat{w}_{\aic,\lv{v}{j}}}, \cdots,\frac{\widehat{w}_{\aic,\lv{M_{j}}{j}}}{\sum_{v=1}^{M_{j}}\widehat{w}_{\aic,\lv{v}{j}}}\right)^{T},$$
$$\boldsymbol{\widehat{w}}_{\bic_{s}}^{j}
=\left(\frac{\widehat{w}_{\bic,\lv{1}{j}}}{\sum_{v=1}^{M_{j}}\widehat{w}_{\bic,\lv{v}{j}}}, \cdots,\frac{\widehat{w}_{\bic,\lv{M_{j}}{j}}}{\sum_{v=1}^{M_{j}}\widehat{w}_{\bic,\lv{v}{j}}}\right)^{T}$$ and
$\widehat{\btheta}_{m}=(\widehat{\theta}_{1,m},\cdots,\widehat{\theta}_{q,m})^{T}$, where $\widehat{\theta}_{j,m}$ is the $j$th component of $\widehat{\btheta}_{m}$. Therefore the scaled S-AIC and scaled S-BIC estimators for $\theta_{j}$ are \be \label{13}
\widehat{\theta}_{j}\left(\boldsymbol{\widehat{w}}_{\aic_{s}}^{j}\right)=\sum_{v=1}^{M_{j}}\frac{\widehat{w}_{\aic,\lv{v}{j}}}{\sum_{v=1}^{M_{j}}\widehat{w}_{\aic,\lv{v}{j}}}\widehat{\theta}_{j,\lv{v}{j}}
\ \textrm{and}\  \widehat{\theta}_{j}\left(\boldsymbol{\widehat{w}}_{\bic_{s}}^{j}\right)=\sum_{v=1}^{M_{j}}\frac{\widehat{w}_{\bic,\lv{v}{j}}}{\sum_{v=1}^{M_{j}}\widehat{w}_{\bic,\lv{v}{j}}}\widehat{\theta}_{j,\lv{v}{j}},
\ee respectively. Denote $$\widehat{\btheta}(\boldsymbol{\widehat{w}}_{\aic_{s}})=\left(\widehat{\theta}_{1}\left(\boldsymbol{\widehat{w}}_{\aic_{s}}^{j}\right),\cdots,\widehat{\theta}_{q}\left(\boldsymbol{\widehat{w}}_{\aic_{s}}^{j}\right)\right)^{T}$$
and
$$\widehat{\btheta}(\boldsymbol{\widehat{w}}_{\bic_{s}})=\left(\widehat{\theta}_{1}\left(\boldsymbol{\widehat{w}}_{\bic_{s}}^{j}\right),\cdots,\widehat{\theta}_{q}\left(\boldsymbol{\widehat{w}}_{\bic_{s}}^{j}\right)\right)^{T}.$$
Next, we establish the asymptotic distributions of $\widehat{\btheta}(\boldsymbol{\widehat{w}}_{\aic_{s}})$ and $\widehat{\btheta}(\boldsymbol{\widehat{w}}_{\bic_{s}})$.
\begin{thm}
\label{th1}
If Conditions \emph{\ref{c1}-\ref{c7}} are satisfied, then
\be\sn\left\{\widehat{\theta}_{j}\left(\boldsymbol{\widehat{w}}_{\aic_{s}}^{j}\right)-\theta_{j,0}\right\}\overd \sum_{v=1}^{M_{j}}\boldsymbol{1}\{\lv{v}{j}\in\calO\}\frac{G_{\lv{v}{j}}}{\sum_{v=1}^{M_{j}}\boldsymbol{1}\{\lv{v}{j}\in\calO\}G_{\lv{v}{j}}}
\widetilde{\eta}_{j,\lv{v}{j}}\non\ee
and $\widehat{\theta}_{j}\left(\boldsymbol{\widehat{w}}_{\bic_{s}}\right)\overd 0$
for $j\not\in S_{m_{o}}$ and $\widehat{\theta}_{j}\left(\boldsymbol{\widehat{w}}_{\bic_{s}}\right)\overd Z_{j,m_{o}}$ for $j\in S_{m_{o}}$,
where $\widetilde{\eta}_{j,\lv{v}{j}}$ is the $j^{th}$ element of the vector $\Pi_{\lv{v}{j}}^{T}H_{S_{\lv{v}{j}}}\Pi_{\lv{v}{j}}\Delta_{\lv{v}{j}}\mathcal{F}(\btheta_{0})\bbeta$ and other notations are the same as those in Lemma \emph{\ref{lem4}}.
\begin{proof}
We can establish it easily from the proof of Lemma \ref{lem4}.
\end{proof}
\end{thm}

Further, for the interested parameter $\mu=\mu\left(\btheta_{0}\right)$, which is
a smooth real-valued function. Let $\widehat{\mu}_{m}=\mu\left(\widehat{\btheta}_{m}\right)$
denote the submodel estimators. Then the averaging estimators of $\mu$ based on S-AIC weight and S-BIC weight are \be \label{14}
\widehat{\mu}(\widehat{\boldsymbol{w}}_{\aic})=\sum_{m=1}^{M}\widehat{w}_{\aic,m}\widehat{\mu}_{m}\ \  \textrm{and}\ \   \widehat{\mu}(\widehat{\boldsymbol{w}}_{\bic})=\sum_{m=1}^{M}\widehat{w}_{\bic,m}\widehat{\mu}_{m},\ee
respectively.
Then by Lemma \ref{lem4} and Delta Method, we can obtain the following conclusion.
\begin{thm}\label{th2}
Let $\dot{\mu}(\btheta)=\partial\mu/\partial\btheta$ be continuous in a neighborhood of $\btheta_{0}$. Then under the same assumptions of Lemma \emph{\ref{lem4}}, we have
\be\sn\left\{\widehat{\mu}\left(\widehat{\boldsymbol{w}}_{\aic}\right)-\mu\left(\btheta_{0}\right)\right\}\overd \sum_{m\in\calO}\left(G_{m}/G\right)\dot{\mu}(\btheta_{0})^{T}\Pi_{m}^{T}H_{S_{m}}\Pi_{m}\Delta_{m}\mathcal{F}(\btheta_{0})\bbeta\ee
and
\be \sn\left\{\widehat{\mu}\left(\widehat{\boldsymbol{w}}_{\bic}\right)-\mu(\btheta_{0})\right\}\overd \mathcal{N}\left(0,\dot{\mu}(\btheta_{0})^{T}\Sigma_{m_{o}}\dot{\mu}(\btheta_{0})\right)
\ee
where the notations are the same as that in Lemma \emph{\ref{lem4}}.
\begin{proof}
See \ref{sec:lem4}.
\end{proof}
\end{thm}

\section{Construction of confidence intervals based on model averaging}\label{sec:CI}
Under general fixed parameter setup, in order to compare our method with the existing methods in the literature, in this section, we construct the confidence intervals based on our asymptotic distributions and briefly discuss each estimator and how to construct the confidence intervals for each estimator in the existing literature, including \cite{S1997Model}, \cite{hjort2003averaging} and \cite{Charkhi2018Asymptotic}. We also present the confidence intervals based on the largest model.

\subsection{Confirmation of the validity of the method in \cite{S1997Model}.}\label{D:S1997}
\cite{S1997Model} presented a simple estimator of variance for the maximum likelihood estimator under the perfect correlation assumption and then advocated simulated inference by bootstrap.
Based on the variance estimators of model averaging estimators in \cite{S1997Model}, \cite[p.164]{Burnham2002Model} proposed ad-hoc confidence intervals and found it is more stable than the bootstrap method.
Further, \cite{Burnham2002Model} gave many examples to illustrate that such intervals have excellent achieved coverage probability. But there is no theoretical explanation.
Therefore, we will study the resulting coverage probability of the confidence interval in \cite{Burnham2002Model}.

Let $z_{1-\alpha/2}$ be the $1-\alpha/2$ normal quantile. As in \cite{S1997Model} and \cite{Burnham2002Model}, the $1-\alpha$ confidence interval of $\theta_{j}$  based on SAIC weight (labeled by SAIC$_{97}$) is constructed as \be \label{15}
CI_{\theta_{j},n}\left(\text{SAIC}_{97}\right)&=&\left[\widehat{\theta}_{j}\left(\boldsymbol{\widehat{w}}_{\aic_{s}}\right)-z_{1-\alpha/2}\sqrt{\widehat{\mathrm{var}}\left\{\widehat\theta_{j}\left(\widehat{\boldsymbol{w}}_{\aic_{s}}\right)\right\}}
,\right.\non\\
&&~~\left.\widehat{\theta}_{j}\left(\boldsymbol{\widehat{w}}_{\aic_{s}}\right)+z_{1-\alpha/2}\sqrt{\widehat{\mathrm{var}}\left\{\widehat\theta_{j}\left(\widehat{\boldsymbol{w}}_{\aic_{s}}\right)\right\}}\right],\ee
where $\widehat{\mathrm{var}}\left\{\widehat\theta_{j}\left(\widehat{\boldsymbol{w}}_{\aic_{s}}\right)\right\}$ is the unconditional variance estimator proposed by \cite{S1997Model}, which is given by
$$ \widehat{\mathrm{var}}\left\{\widehat\theta_{j}\left(\widehat{\boldsymbol{w}}_{\aic_{s}}\right)\right\}=\left[\sum_{v=1}^{M_{j}}\frac{\widehat{w}_{\aic_{s},\lv{v}{j}}}{\sum_{v=1}^{M_{j}}\widehat{w}_{\aic_{s},\lv{v}{j}}}
\sqrt{\widehat{\mathrm{var}}\left(\widehat{\theta}_{j,\lv{v}{j}}|\lv{v}{j}\right)+\left\{\widehat{\theta}_{j,\lv{v}{j}}-\widehat\theta_{j}\left(\widehat{\boldsymbol{w}}_{\aic_{s}}\right)\right\}^{2}}\right]^{2}$$
with $\widehat{\mathrm{var}}\left(\widehat{\theta}_{j,\lv{v}{j}}|\lv{v}{j}\right)$ being found by the normal inference methods based on model $\lv{v}{j}$.
Similarly,
the $1-\alpha$ confidence interval of $\mu(\btheta_{0})$ based on SAIC$_{97}$ method is constructed as \be \label{mu}
CI_{\mu,n}\left(\text{SAIC}_{97}\right)&=&\left[\widehat{\mu}\left(\boldsymbol{\widehat{w}}_{\aic}\right)-z_{1-\alpha/2}\sqrt{\widehat{\mathrm{var}}\left\{\widehat{\mu}(\boldsymbol{\widehat{w}}_{\aic})\right\}}
,\right.\non\\
&&~~\left.\widehat{\mu}(\boldsymbol{\widehat{w}}_{\aic})+z_{1-\alpha/2}\sqrt{\widehat{\mathrm{var}}\left\{\widehat{\mu}\left(\boldsymbol{\widehat{w}}_{\aic}\right)\right\}}\right],\ee
where
$$ \widehat{\mathrm{var}}\left\{\widehat{\mu}\left(\boldsymbol{\widehat{w}}_{\aic}\right)\right\}=\left[\sum_{m=1}^{M}\widehat{w}_{\aic,m}
\sqrt{\widehat{\mathrm{var}}\left(\widehat{\mu}_{m}|m\right)+\left\{\widehat{\mu}_{m}-\widehat{\mu}\left(\widehat{\boldsymbol{w}}_{\aic}\right)\right\}^{2}}\right]^{2},$$
with $\widehat{\mathrm{var}}\left(\widehat{\mu}_{m}|m\right)$ being found by normal inference methods under model $m$.
For the method in \cite{S1997Model} based on SBIC weight (SBIC$_{97}$), we only need to replace $\widehat{\boldsymbol{w}}_{\aic_{s}}$ by $\widehat{\boldsymbol{w}}_{\bic_{s}}$ and $\widehat{\boldsymbol{w}}_{\aic}$ by $\widehat{\boldsymbol{w}}_{\bic}$ in the above confidence interval formula.

In the following, we confirm the validity of these intervals.
Let $\sigma_{j,\lv{v}{j}}^{2}$ denote the $j^{th}$ diagonal element of the matrix $$\Sigma_{\lv{v}{j}}=\Pi_{\lv{v}{j}}^{T}H_{S_{\lv{v}{j}}}\Pi_{\lv{v}{j}}\Delta_{\lv{v}{j}}\mathcal{F}(\btheta_{0})\left(\Pi_{\lv{v}{j}}^{T}H_{S_{\lv{v}{j}}}\Pi_{\lv{v}{j}}\Delta_{\lv{v}{j}}\right)^{T}.$$
From Lemma \ref{lem3}, we see that when $\lv{v}{j}\in\calU$, $\widehat{w}_{\aic,\lv{v}{j}}\overp 0$. From the proof of Lemma \ref{lem4}, we can conclude that when $\lv{v}{j}\in\calO$,
$\widehat{\theta}_{j,\lv{v}{j}}\overas\theta_{j,0}$ and $\sqrt{n}\left(\widehat{\theta}_{j,\lv{v}{j}}-\theta_{j,0}\right)\overd\widetilde{\eta}_{j,\lv{v}{j}}$,
where $\widetilde{\eta}_{j,\lv{v}{j}}\sim\mathcal{N}\left(0,\sigma_{j,\lv{v}{j}}^{2}\right)$.
Because $\widehat{\mathrm{var}}\left(\widehat{\theta}_{j,\lv{v}{j}}|\lv{v}{j}\right)$ is found by normal inference methods under model $\lv{v}{j}$,
Then $n\widehat{\mathrm{var}}\left(\widehat{\theta}_{j,\lv{v}{j}}|\lv{v}{j}\right)\overp\sigma_{j,\lv{v}{j}}^{2}$.
Since $\sqrt{n}\left\{\widehat{\theta}_{j}\left(\boldsymbol{\widehat{w}}_{\aic_{s}}\right)-\theta_{j,0}\right\}$
and $\sqrt{n\widehat{\mathrm{var}}\left\{\widehat\theta_{j}\left(\widehat{\boldsymbol{w}}_{\aic_{s}}\right)\right\}}$
can be expressed the function of the same random vector in probability, by
$$\sqrt{n}\left\{\widehat{\theta}_{j}\left(\boldsymbol{\widehat{w}}_{\aic_{s}}\right)-\theta_{j,0}\right\}
\overd\sum_{v=1}^{M_{j}}\boldsymbol{1}\{\lv{v}{j}\in\calO\}\frac{G_{\lv{v}{j}}}{\sum_{v=1}^{M_{j}}\boldsymbol{1}\{\lv{v}{j}\in\calO\}G_{\lv{v}{j}}}\widetilde{\eta}_{j,\lv{v}{j}}$$
and
\be
&&\sqrt{n\widehat{\mathrm{var}}\left\{\widehat\theta_{j}\left(\widehat{\boldsymbol{w}}_{\aic_{s}}\right)\right\}}\non\\
&=&\sum_{v=1}^{M_{j}}\frac{\widehat{w}_{\aic_{s},\lv{v}{j}}}{\sum_{k=1}^{M_{j}}\widehat{w}_{\aic_{s},\lv{v}{j}}}
\sqrt{n\widehat{\mathrm{var}}(\widehat{\theta}_{j,\lv{v}{j}}|\lv{v}{j})+\left[\sqrt{n}\left\{\widehat{\theta}_{j,\lv{v}{j}}-\widehat\theta_{j}(\widehat{\boldsymbol{w}}_{\aic_{s}})\right\}\right]^{2}}\non\\
&\overd&\sum_{v=1}^{M_{j}}\frac{\boldsymbol{1}\{\lv{v}{j}\in\calO\}G_{\lv{v}{j}}}{\sum_{v=1}^{M_{j}}\boldsymbol{1}\{\lv{v}{j}\in\calO\}G_{\lv{v}{j}}}
\sqrt{\sigma_{j,\lv{v}{j}}^{2}+\left[\widetilde{\eta}_{j,\lv{v}{j}}-\sum_{v=1}^{M_{j}}\frac{\boldsymbol{1}\{\lv{v}{j}\in\calO\}G_{\lv{v}{j}}}{\sum_{v=1}^{M_{j}}\boldsymbol{1}\{\lv{v}{j}\in\calO\}G_{\lv{v}{j}}}\widetilde{\eta}_{j,\lv{v}{j}}\right]^{2}}\non,
\ee
we have the following conclusion
\be
&&\frac{\widehat{\theta}_{j}\left(\boldsymbol{\widehat{w}}_{\aic_{s}}\right)-\theta_{j,0}}
{\sqrt{\widehat{\mathrm{var}}\left\{\widehat\theta_{j}\left(\widehat{\boldsymbol{w}}_{\aic_{s}}\right)\right\}}}
=\frac{\sqrt{n}\left(\widehat{\theta}_{j}(\boldsymbol{\widehat{w}}_{\aic_{s}})-\theta_{j,0}\right)}
{\sqrt{n\widehat{\mathrm{var}}\left\{\widehat\theta_{j}\left(\widehat{\boldsymbol{w}}_{\aic_{s}}\right)\right\}}}\non\\
&\overd&\Lambda_{j}=\frac{\sum_{v=1}^{M_{j}}\frac{\boldsymbol{1}\{\lv{v}{j}\in\calO\}G_{\lv{v}{j}}}{\sum_{v=1}^{M_{j}}\boldsymbol{1}\{\lv{v}{j}\in\calO\}G_{\lv{v}{j}}}\widetilde{\eta}_{j,\lv{v}{j}}}{\sum_{v=1}^{M_{j}}\frac{\boldsymbol{1}\{\lv{v}{j}\in\calO\}G_{\lv{v}{j}}}{\sum_{v=1}^{M_{j}}\boldsymbol{1}\{\lv{v}{j}\in\calO\}G_{\lv{v}{j}}}
\sqrt{\sigma_{j,\lv{v}{j}}^{2}+\left[\widetilde{\eta}_{j,\lv{v}{j}}-\sum_{v=1}^{M_{j}}\frac{\boldsymbol{1}\{\lv{v}{j}\in\calO\}G_{\lv{v}{j}}}{\sum_{v=1}^{M_{j}}\boldsymbol{1}\{\lv{v}{j}\in\calO\}G_{\lv{v}{j}}}\widetilde{\eta}_{j,\lv{v}{j}}\right]^{2}}}\non\\
&=&\frac{\sum_{v=1}^{M_{j}}\boldsymbol{1}\{\lv{v}{j}\in\calO\}G_{\lv{v}{j}}\widetilde{\eta}_{j,\lv{v}{j}}}{\sum_{v=1}^{M_{j}}\boldsymbol{1}\{\lv{v}{j}\in\calO\}G_{\lv{v}{j}}
\sqrt{\sigma_{j,\lv{v}{j}}^{2}+\left[\widetilde{\eta}_{j,\lv{v}{j}}-\sum_{v=1}^{M_{j}}\frac{\boldsymbol{1}\{\lv{v}{j}\in\calO\}G_{\lv{v}{j}}}{\sum_{v=1}^{M_{j}}\boldsymbol{1}\{\lv{v}{j}\in\calO\}G_{\lv{v}{j}}}\widetilde{\eta}_{j,\lv{v}{j}}\right]^{2}}},\non\\
\ee
where
\be
G_{\lv{v}{j}}&=&\exp\left(\left\{\Pi_{\lv{v}{j}}\mathcal{F}(\btheta_{0})\bbeta\right\}^{T}\left[H_{\lv{v}{j}}-\left\{\Pi_{\lv{v}{j}}\mathcal{F}(\btheta_{0})\Pi_{\lv{v}{j}}^{T}\right\}^{-1}\right]\Pi_{\lv{v}{j}}\mathcal{F}(\btheta_{0})\bbeta/2-k_{\lv{v}{j}}\right)\non\ee
and  $\widetilde{\eta}_{j,\lv{v}{j}}=\boldsymbol{e}_{j}^{T}\Pi_{\lv{v}{j}}^{T}H_{S_{\lv{v}{j}}}\Pi_{\lv{v}{j}}\Delta_{\lv{v}{j}}\mathcal{F}(\btheta_{0})\bbeta$ with $\bbeta\sim \mathcal{N}\left(0,\mathcal{F}^{-1}\left(\btheta_{0}\right)\right)$.
Therefore,
\be
&&\mathrm{P}\left[\widehat{\theta}_{j}(\boldsymbol{\widehat{w}}_{\aic_{s}})-z_{1-\alpha/2}\sqrt{\widehat{\mathrm{var}}\left\{\widehat\theta_{j}(\widehat{\boldsymbol{w}}_{\aic_{s}})\right\}}
\leq\theta_{j,0}\leq\widehat{\theta}_{j}(\boldsymbol{\widehat{w}}_{\aic_{s}})+z_{1-\alpha/2}\sqrt{\widehat{\mathrm{var}}\left\{\widehat\theta_{j}(\widehat{\boldsymbol{w}}_{\aic_{s}})\right\}}\right]\non\\
&&=\mathrm{P}\left[-z_{1-\alpha/2}
\leq\frac{\widehat{\theta}_{j}(\boldsymbol{\widehat{w}}_{\aic_{s}})-\theta_{j,0}}
{\sqrt{\widehat{\mathrm{var}}\left\{\widehat\theta_{j}(\widehat{\boldsymbol{w}}_{\aic_{s}})\right\}}}\leq z_{1-\alpha/2}\right]\non\\
&&\rightarrow \mathrm{P}\left(-z_{1-\alpha/2}\leq\Lambda_{j}\leq z_{1-\alpha/2}\right).\ee
The limiting coverage probability cannot be directly calculated. We first consider some special cases. For example, when the number of the overfit model is $1$, $\Lambda_{j}=\frac{\widetilde{\eta}_{j,m_{o}}}{\sigma_{j,m_{o}}}$, so the limiting coverage probability is $1-\alpha$. However, when the the number of the overfit model is more than $1$, $\Lambda$ is clearly not standard normal. But it is approximately standard normal. To illustrate this, consider a simulation-based method.
If a set of observations is approximately normally distributed, a normal quantile-quantile (Q-Q) plot of the observations will result in an approximately straight line.
The true parameter $\btheta_{0}$ in $\Lambda_{j}$ is estimated by scaled S-BIC estimator. Generate $q\times1$ normal random vector $\bbeta^{(r)}\sim \mathcal{N}\left(0,\mathcal{F}^{-1}\left(\widehat\btheta\left(\widehat w_{\bic_{s}}\right)\right)\right)$ for $r=1,\ \cdots,\ R_{0}$, where $R_{0}$ is sufficiently large. For each $r$, we compute $$\widehat{\Lambda}^{(r)}=\frac{\sum_{v=1}^{M_{j}}\boldsymbol{1}\{\lv{v}{j}\in\calO\}G_{\lv{v}{j}}\widetilde{\eta}_{j}^{(r)}}{\sum_{v=1}^{M_{j}}\boldsymbol{1}\{\lv{v}{j}\in\calO\}G_{\lv{v}{j}}
\sqrt{\sigma_{j,\lv{v}{j}}^{2}+\left(\widetilde{\eta}_{j}^{(r)}-\sum_{v=1}^{M_{j}}\frac{\boldsymbol{1}\{\lv{v}{j}\in\calO\}G_{\lv{v}{j}}}{\sum_{v=1}^{M_{j}}\boldsymbol{1}\{\lv{v}{j}\in\calO\}G_{\lv{v}{j}}}\widetilde{\eta}_{j}^{(r)}\right)^{2}}},
$$
where $\widetilde{\eta}_{j}^{(r)}=\boldsymbol{e}_{j}^{T}\Pi_{\lv{v}{j}}^{T}H_{S_{\lv{v}{j}}}\Pi_{\lv{v}{j}}\Delta_{\lv{v}{j}}\mathcal{F}(\btheta_{0})\bbeta^{(r)}$
We obtain $R_{0}$ samples from the asymptotic distributions $\Lambda$, then we can do a Q-Q plot to compared with standard normal distribution. In the next section,
We use simulated method to verify that $\Lambda_{j}$ is approximately standard normal distribution.

Similarly, from the proof of Lemma \ref{lem4}, we know $\sqrt{n}\left\{\widehat\theta_{j}(\widehat{\boldsymbol{w}}_{\bic_{s}})-\theta_{j,0}\right\}\overd \mathcal{N}(0,\sigma_{j,m_{o}}^{2})$ and when $\lv{v}{j}=m_{o}$, $\widehat{w}_{\bic,m_{o}}\overp 1$, otherwise $\widehat{w}_{\bic,\lv{v}{j}}\overp 0$. Then
\be\frac{\widehat{\theta}_{j}\left(\boldsymbol{\widehat{w}}_{\bic_{s}}\right)-\theta_{j,0}}
{\sqrt{\widehat{\mathrm{var}}\left\{\widehat\theta_{j}\left(\widehat{\boldsymbol{w}}_{\bic_{s}}\right)\right\}}}
&=&\frac{\sqrt{n}\left\{\widehat{\theta}_{j}\left(\boldsymbol{\widehat{w}}_{\bic_{s}}\right)-\theta_{j,0}\right\}}
{\sqrt{n\widehat{\mathrm{var}}\left\{\widehat\theta_{j}\left(\widehat{\boldsymbol{w}}_{\bic_{s}}\right)\right\}}}
\overd\frac{\widetilde{\eta}_{j,m_{o}}}{\sigma_{j,m_{o}}}.
\ee
Therefore,
\be
&&\mathrm{P}\left[\widehat{\theta}_{j}\left(\boldsymbol{\widehat{w}}_{\bic_{s}}\right)-z_{1-\alpha/2}\sqrt{\widehat{\mathrm{var}}\left\{\widehat\theta_{j}\left(\widehat{\boldsymbol{w}}_{\bic_{s}}\right)\right\}}
\leq\theta_{j,0}\leq\widehat{\theta}_{j}\left(\boldsymbol{\widehat{w}}_{\bic_{s}}\right)+z_{1-\alpha/2}\sqrt{\widehat{\mathrm{var}}\left\{\widehat\theta_{j}\left(\widehat{\boldsymbol{w}}_{\bic_{s}}\right)\right\}}\right]\non\\
&&=\mathrm{P}\left[-z_{1-\alpha/2}
\leq\frac{\sqrt{n}\left\{\widehat{\theta}_{j}\left(\boldsymbol{\widehat{w}}_{\bic_{s}}\right)-\theta_{j,0}\right\}}
{\sqrt{n\widehat{\mathrm{var}}\left\{\widehat\theta_{j}\left(\widehat{\boldsymbol{w}}_{\bic_{s}}\right)\right\}}}\leq z_{1-\alpha/2}\right]\non\\
&&\rightarrow \mathrm{P}\left(-z_{1-\alpha/2}\leq\frac{\widetilde{\eta}_{j,m_{o}}}{\sigma_{j,m_{o}}}\leq z_{1-\alpha/2}\right)=1-\alpha.\ee

The similar results can be concluded for the confidence interval (\ref{mu}) by the same proof technique and thus omitted.
\subsection{Confidence intervals of \cite{hjort2003averaging}}\label{D:H2003}
In this subsection, we delve into the coverage probability of the confidence intervals derived by \cite{hjort2003averaging}.
The method in \cite{hjort2003averaging} is to derive the asymptotic distribution of the model averaging estimators under local misspecification assumption and then put forward the confidence intervals of the parameters. The methods of constructing intervals are labeled by SAIC$_{L}$ and SBIC$_{L}$, whose estimators are based on S-AIC and
S-BIC weights respectively. We will further analyze the performance of these confidence intervals under the general fixed parameter settings.

In order to describe the inference in \cite{hjort2003averaging}, we first present  some notations. The Fisher Information matrix is
\begin{center}
\begin{blockarray}{ccc}
&$q_{1}\times q_{1}$     & $q_{1}\times q_{2}$\\
\begin{block}{c(cc)}
&$J_{00} $               & $J_{01}$ \\
$J_{full}=\mathcal{F}\left(\left(\bb_{0}^{T},\boldsymbol0^{T}\right)^{T}\right)=$&~&\\
 &$J_{10}$              & $J_{11}$ \\
\end{block}
&$q_{2}\times q_{1}$ & $q_{2}\times q_{2}$
\end{blockarray}~
\end{center}
with inverse matrix
\begin{center}
\begin{blockarray}{ccc}
&$q_{1}\times q_{1}$     & $q_{1}\times q_{2}$\\
\begin{block}{c(cc)}
&$J^{00}$               & $J^{01}$ \\
$J_{full}^{-1}=$&~&\\
 &$J^{10}$              & $J^{11}$ \\
\end{block}
&$q_{2}\times q_{1}$ & $q_{2}\times q_{2}$
\end{blockarray}~.
\end{center}

In \cite{hjort2003averaging}, if we focus on the parameter $\mu(\btheta_{0})$, the $1-\alpha$ confidence interval of $\mu(\btheta_{0})$ is constructed as: \be
&&\left[\widehat{\mu}\left(\boldsymbol{w}\left(D_{n}\right)\right)-\widehat{\boldsymbol{\omega}}^{T}\left\{D_{n}-\widehat{\delta}\left(D_{n}\right)\right\}/\sqrt{n}-z_{1-\alpha/2}\widehat{\varphi}/\sqrt{n}
,\right.\non\\
&&\left.~~\widehat{\mu}\left(\boldsymbol{w}\left(D_{n}\right)\right)-\widehat{\boldsymbol{\omega}}^{T}\left\{D_{n}-\widehat{\delta}\left(D_{n}\right)\right\}/\sqrt{n}+z_{1-\alpha/2}\widehat{\varphi}/\sqrt{n}\right],\ee
where $\widehat{\mu}$ is the corresponding model averaging estimator of $\mu$;
$z_{1-\alpha/2}$ is the $1-\alpha/2$ normal quantile;
$\widehat{J}_{00}^{-1}$, $\widehat{J}_{10}$, $\widehat{J}_{01}$ and $\widehat{J}_{11}$ are the sample estimators of $J_{00}^{-1}$, $J_{10}$, $J_{01}$ and $J_{11}$, respectively;  $\widehat{K}=\left(\widehat{J}_{11}-\widehat{J}_{10}\widehat{J}_{00}^{-1}\widehat{J}_{01}\right)^{-1}$
is the estimator for $K=\left(J_{11}-J_{10}J_{00}^{-1}J_{01}\right)^{-1}$;
$\widehat{\boldsymbol{\omega}}=\widehat{J}_{10}\widehat{J}_{00}^{-1}\frac{\partial\mu}{\partial\bb}-\frac{\partial\mu}{\partial\bg}$
and
$\widehat{\varphi}=\left\{\left(\frac{\partial\mu}{\partial\btheta}\right)^{T}\widehat{J}_{00}^{-1}\frac{\partial\mu}{\partial\btheta}
+\widehat{\boldsymbol{\omega}}^{T}\widehat{K}\widehat{\boldsymbol{\omega}}\right\}^{1/2}$ are consistent estimators of
$\boldsymbol{\omega}=J_{10}J_{00}^{-1}\frac{\partial\mu}{\partial\bb}-\frac{\partial\mu}{\partial\bg}$
and
$\varphi=\left\{\left(\frac{\partial\mu}{\partial\btheta}\right)^{T}J_{00}^{-1}\frac{\partial\mu}{\partial\btheta}
+\boldsymbol{\omega}^{T}K\boldsymbol{\omega}\right\}^{1/2}$, where
the partial derivatives are evaluated under the narrow model $\left(\bb_{0}^{T},\boldsymbol0^{T}\right)^{T}$;
Let $\Gamma_{m}$ be $(k_{m}-q_{1})\times q_{2}$ projection matrix mapping $\bg_{m}$ to the subvector $\bg_{S_{m}}=\Gamma_{m}\bg_{m}$ and denote
$K_{m}=\left(\Gamma_{m}K^{-1}\Gamma_{m}^{T}\right)^{-1}$;
$\widehat{\delta}(D_{n})=K^{1/2}\left\{\sum_{m=1}^{M}w_{m}\left(D_{n}\right)V_{m}\right\}K^{-1/2}D_{n}$ with
$D_{n}=\widehat{\bd}_{full}=\sqrt{n}\widehat{\bg}_{full}$ and $V_{m}=K^{-1/2}\Gamma_{m}^{T}K_{m}\Gamma_{m}K^{-1/2}$.

It is clear that
\begin{center}
\begin{blockarray}{ccc}
&$q_{1}\times q_{1}$     & $q_{1}\times (k_{m}-q_{1})$\\
\begin{block}{c(cc)}
&$J_{00} $               & $J_{01}\Gamma_{m}^{T}$ \\
$J_{S_{m}}=$&~&\\
 &$\Gamma_{m}J_{10}$              & $\Gamma_{m}J_{11}\Gamma_{m}^{T}$ \\
\end{block}
&$(k_{m}-q_{1})\times q_{1}$ & $(k_{m}-q_{1})\times (k_{m}-q_{1})$
\end{blockarray}~
\end{center}
with inverse matrix
\begin{center}
\begin{blockarray}{ccc}
&$q_{1}\times q_{1}$     & $q_{1}\times (k_{m}-q_{1})$\\
\begin{block}{c(cc)}
&$J^{00,m} $               & $J^{01,m}$ \\
$J_{S_{m}}^{-1}=$&~&\\
 &$J^{10,m}$              & $J^{11,m}$ \\
\end{block}
&$(k_{m}-q_{1})\times q_{1}$ & $(k_{m}-q_{1})\times (k_{m}-q_{1})$
\end{blockarray}~.
\end{center}

For SAIC$_{L}$ and SBIC$_{L}$, let $\widehat{\theta}_{j}(\widehat{\boldsymbol{w}}_{\aic})$ and $\widehat{\theta}_{j}(\widehat{\boldsymbol{w}}_{\bic})$ denote the $j\th$ component of $\widehat{\btheta}(\widehat{\boldsymbol{w}}_{\aic})$ and $\widehat{\btheta}(\widehat{\boldsymbol{w}}_{\aic})$, respectively.
Then the $1-\alpha$ confidence interval of $\theta_{j}$ based on SAIC$_{L}$ and SBIC$_{L}$ methods are constructed as:
\be
CI_{\theta_{j},n}\left(\text{SAIC}_{L}\right)&=&\left[\widehat{\theta}_{j}\left(\widehat{\boldsymbol{w}}_{\aic}\right)-\widehat{\boldsymbol{\omega}}^{T}\left\{D_{n}-\widehat{\delta}\left(D_{n}\right)\right\}/\sqrt{n}-z_{1-\alpha/2}\widehat{\varphi}/\sqrt{n}
,\right.\non\\
&&\left.~~\widehat{\theta}_{j}\left(\widehat{\boldsymbol{w}}_{\aic}\right)-\widehat{\boldsymbol{\omega}}^{T}\left\{D_{n}-\widehat{\delta}\left(D_{n}\right)\right\}/\sqrt{n}+z_{1-\alpha/2}\widehat{\varphi}/\sqrt{n}\right]\ee
and
\be
CI_{\theta_{j},n}\left(\text{SBIC}_{L}\right)&=&\left[\widehat{\theta}_{j}\left(\widehat{\boldsymbol{w}}_{\bic}\right)-\widehat{\boldsymbol{\omega}}^{T}\left\{D_{n}-\widehat{\delta}\left(D_{n}\right)\right\}/\sqrt{n}-z_{1-\alpha/2}\widehat{\varphi}/\sqrt{n}
,\right.\non\\
&&\left.~~\widehat{\theta}_{j}\left(\widehat{\boldsymbol{w}}_{\bic}\right)-\widehat{\boldsymbol{\omega}}^{T}\left\{D_{n}-\widehat{\delta}\left(D_{n}\right)\right\}/\sqrt{n}+z_{1-\alpha/2}\widehat{\varphi}/\sqrt{n}\right].\ee
Similarly, the $1-\alpha$ confidence interval of $\mu(\btheta_{0})$ based on SAIC$_{L}$ and SBIC$_{L}$ methods are constructed as:
\be
CI_{\mu,n}\left(\text{SAIC}_{L}\right)&=&\left[\widehat{\mu}\left(\widehat{\boldsymbol{w}}_{\aic}\right)-\widehat{\boldsymbol{\omega}}^{T}\left\{D_{n}-\widehat{\delta}\left(D_{n}\right)\right\}/\sqrt{n}-z_{1-\alpha/2}\widehat{\varphi}/\sqrt{n}
,\right.\non\\
&&\left.~~\widehat{\mu}\left(\widehat{\boldsymbol{w}}_{\aic}\right)-\widehat{\boldsymbol{\omega}}^{T}\left\{D_{n}-\widehat{\delta}\left(D_{n}\right)\right\}/\sqrt{n}+z_{1-\alpha/2}\widehat{\varphi}/\sqrt{n}\right]\ee
and
\be
CI_{\mu,n}\left(\text{SBIC}_{L}\right)&=&\left[\widehat{\mu}\left(\widehat{\boldsymbol{w}}_{\bic}\right)-\widehat{\boldsymbol{\omega}}^{T}\left\{D_{n}-\widehat{\delta}\left(D_{n}\right)\right\}/\sqrt{n}-z_{1-\alpha/2}\widehat{\varphi}/\sqrt{n}
,\right.\non\\
&&\left.~~\widehat{\mu}\left(\widehat{\boldsymbol{w}}_{\bic}\right)-\widehat{\boldsymbol{\omega}}^{T}\left\{D_{n}-\widehat{\delta}\left(D_{n}\right)\right\}/\sqrt{n}+z_{1-\alpha/2}\widehat{\varphi}/\sqrt{n}\right].\ee

Next, we explain why the results of \cite{hjort2003averaging} are not necessarily true under the general fixed parameter setup.
Let the fixed parameter setup be
$\mu(\btheta_{0})=\mu\left(y,\left(\bb_{0}^{T},\bg^{T}\right)^{T}\right)
=\mu\left(y,\left(\bb_{0}^{T},\bd'^{T}\right)^{T}\right)
=\mu\left(y,\left(\bb_{0}^{T},\bd/\sqrt{n}^{T}\right)^{T}\right)
$, then $\bd'=\bg$ is fixed and $\bd=\sqrt{n}\bd'$.
Therefore, the $1-\alpha$ confidence interval of $\mu$ becomes
\be
&&\left[\widehat{\mu}\left(\widehat{\boldsymbol{w}}_{\aic}\right)-\widehat{\boldsymbol{\omega}}^{T}\left\{D'_{n}-\widehat{\delta}\left(D'_{n}\right)\right\}
-z_{1-\alpha/2}\widehat{\varphi}/\sqrt{n}
,\right.\nonumber\\
&&~~\left.\widehat{\mu}\left(\widehat{\boldsymbol{w}}_{\aic}\right)-\widehat{\boldsymbol{\omega}}^{T}\left\{D'_{n}-\widehat{\delta}\left(D'_{n}\right)\right\}
+z_{1-\alpha/2}\widehat{\varphi}/\sqrt{n}\right],
\ee
where $D'_{n}=\widehat{\delta}'_{full}=\widehat{\bg}_{full}$.

In the following, we explain why the results of \cite{hjort2003averaging} are not necessarily true under the general fixed parameter setup.

Denoting $\widehat{\btheta}_{S_{m}}=\left(\widehat{\bb}_{S_{m}}^{T},\widehat{\bg}_{S_{m}}^{T}\right)^{T}$,
using Taylor expansion for
$$\sum_{t=1}^{n}\frac{\partial\log f\left(y_{t},\left(\widehat{\btheta}_{S_{m}}^{T},\bg_{0,S_{m}^{c}}^{T}\right)^{T}\right)}{\partial\btheta_{S_{m}}}$$
in $\left(\bb_{0}^{T},\boldsymbol0^{T}\right)^{T}$ and by $\sum_{t=1}^{n}\partial\log f\left(y_{t},\left(\widehat{\btheta}_{S_{m}}^{T},\bg_{0,S_{m}^{c}}^{T}\right)^{T}\right)\bigg/\partial\btheta_{S_{m}}=0$, we have
\be
0&=&\frac{1}{\sqrt{n}}\sum_{t=1}^{n}\frac{\partial\log f\left(y_{t},\left(\widehat{\btheta}_{S_{m}}^{T},\bg_{0,S_{m}^{c}}^{T}\right)^{T}\right)}{\partial\btheta_{S_{m}}}\non\\
&=&\frac{1}{\sqrt{n}}\sum_{t=1}^{n}\left[\frac{\partial\log f\left(y_{t},\left(\bb_{0}^{T},\boldsymbol0^{T}\right)^{T}\right)}{\partial\btheta_{S_{m}}}
+\left(\frac{\partial^{2}\log f\left(y_{t},\left(\widetilde{\bb}_{0}^{T},\widetilde{\bg}_{0}^{T}\right)^{T}\right)
}{\partial\btheta_{S_{m}}\partial\btheta_{S_{m}}^{T}},
\frac{\partial^{2}\log f\left(y_{t},\left(\widetilde{\bb}_{0}^{T},\widetilde{\bg}_{0}^{T}\right)^{T}\right)
}{\partial\btheta_{S_{m}}\partial\bg_{S_{m}^{c}}^{T}}\right)
\right.\non\\
&&\left.\left(\left(\widehat{\bb}_{S_{m}}-\bb_{0}\right)^{T},\left(\widehat{\bg}_{S_{m}}-\bg_{0,S_{m}}\right)^{T},
\boldsymbol0^{T}\right)^{T} \right]\non\\
&=&\frac{1}{\sqrt{n}}\sum_{t=1}^{n}\left[\frac{\partial\log f\left(y_{t},\left(\bb_{0}^{T},\boldsymbol0^{T}\right)^{T}\right)}{\partial\btheta_{S_{m}}}
+\frac{\partial^{2}\log f\left(y_{t},\left(\widetilde{\bb}_{0}^{T},\widetilde{\bg}_{0}^{T}\right)^{T}\right)
}{\partial\btheta_{S_{m}}\partial\btheta_{S_{m}}^{T}}\right.\non\\
&&\left.\left(\left(\widehat{\bb}_{S_{m}}-\bb_{0}\right)^{T},\left(\widehat{\bg}_{S_{m}}-\bg_{0,S_{m}}\right)^{T}\right)^{T} \right],\non
\ee
where $\left(\widetilde{\bb}_{0}^{T},\widetilde{\bg}_{0}^{T}\right)^{T}=\left(\widetilde{\bb}_{0}^{T},\widetilde{\bg}_{0,S_{m}}^{T},\bg_{0,S_{m}^{c}}^{T}\right)^{T}$ is between
$\left(\bb_{0}^{T},\bg_{0}^{T}\right)^{T}$ and $\left(\widehat{\btheta}_{S_{m}}^{T},\bg_{0,S_{m}^{c}}^{T}\right)^{T}$.
Therefore,
\be\label{nbbgg}
\left(
 \begin{matrix}
   \sqrt{n}\left(\widehat{\bb}_{S_{m}}-\bb_{0}\right)  \\
   \sqrt{n}\left(\widehat{\bg}_{S_{m}}-\bg_{0,S_{m}}\right)
  \end{matrix}
  \right)
  =\left(-\frac{1}{n}\sum_{t=1}^{n}\frac{\partial^{2}\log f\left(y_{t},\left(\widetilde{\bb}_{0}^{T},\widetilde{\bg}_{0}^{T}\right)^{T}\right)
}{\partial\btheta_{S_{m}}\partial\btheta_{S_{m}}^{T}}\right)^{-1}
  \cdot\sqrt{n}\left[\frac{1}{n}\sum_{t=1}^{n}\frac{\partial\log f\left(y_{t},\btheta_{0}\right)}{\partial\btheta_{S_{m}}}\right]
  \ee
Following the proof of Theorem 18 in \cite{Ferguson1996A}, first note that
$\mathrm{E}\left\{\Psi_{m}(y,\btheta_{S_{m}})\right\}$ is continuous in $\btheta_{S_{m}}$. For any $\varepsilon>0$, then there is a $\rho>0$ such that
$|\btheta_{S_{m}}-\btheta_{0,S_{m}}|<\rho$ implies
\be\label{eq:CJSm}
\left|\mathrm{E}\dot{\Psi}_{m}(y,\btheta_{S_{m}})+J_{S_{m}}\right|<\varepsilon.\ee
Next note from the Uniform Strong Law of Large Numbers that with probability $1$ there is an integer $N$ such that $n>N$ implies
\be\label{eq:LLNJSm}
\sup_{\btheta_{S_{m}}\in\{\btheta_{S_{m}}:|\btheta_{S_{m}}-\btheta_{0,S_{m}}|<\rho\}}
\left|\frac{1}{n}\sum_{t=1}^{n}\dot{\Psi}_{m}(y_{t},\btheta_{S_{m}})-\mathrm{E}\dot{\Psi}_{m}(y,\btheta_{S_{m}})\right|<\varepsilon.\ee
By \eqref{eq:CJSm} and \eqref{eq:LLNJSm}, when $n>N$, we have
\be
&&\left|\frac{1}{n}\sum_{t=1}^{n}\frac{\partial^{2}\log f\left(y_{t},\left(\bb_{0}^{T},\boldsymbol0^{T}\right)^{T}\right)
}{\partial\btheta_{S_{m}}\partial\btheta_{S_{m}}^{T}}
+J_{S_{m}}\right|\non\\
&\leq&\sup_{\btheta_{S_{m}}\in\{\btheta_{S_{m}}:|\btheta_{S_{m}}-\btheta_{0,S_{m}}|<\rho\}}
\left\{\left|\frac{1}{n}\sum_{t=1}^{n}\dot{\Psi}_{m}(y,\btheta_{S_{m}})-\mathrm{E}\dot{\Psi}_{m}(y,\btheta_{S_{m}})\right|
\right.\non\\
&&~~~~~~~~~~~~~~~~~~~~~~~~~~~~~~~~~~~~~~~~~~\left.+\left|\mathrm{E}\dot{\Psi}_{m}(y,\btheta_{S_{m}})+J_{S_{m}}\right|
\right\}\leq2\varepsilon
\ee
But it is no guarantee that $\left(\widehat{\btheta}_{S_{m}}^{T},\bg_{0,S_{m}^{c}}^{T}\right)^{T}\overas\left(\bb_{0}^{T},\boldsymbol0^{T}\right)^{T}$ under general fixed parameter setup. It means that there is no guarantee that
\be
\frac{1}{n}\sum_{t=1}^{n}\frac{\partial^{2}\log f\left(y_{t},\left(\widetilde{\bb}_{0}^{T},\widetilde{\bg}_{0}^{T}\right)^{T}\right)
}{\partial\btheta_{S_{m}}\partial\btheta_{S_{m}}^{T}}
\overas\frac{1}{n}\sum_{t=1}^{n}\frac{\partial^{2}\log f\left(y_{t},\left(\bb_{0}^{T},\boldsymbol0^{T}\right)^{T}\right)
}{\partial\btheta_{S_{m}}\partial\btheta_{S_{m}}^{T}}.
\ee
Note that
\be\label{Js}
&&\left|\frac{1}{n}\sum_{t=1}^{n}\frac{\partial^{2}\log f\left(y_{t},\left(\widetilde{\bb}_{0}^{T},\widetilde{\bg}_{0}^{T}\right)^{T}\right)
}{\partial\btheta_{S_{m}}\partial\btheta_{S_{m}}^{T}}
+J_{S_{m}}\right|\non\\
&=&\left|\frac{1}{n}\sum_{t=1}^{n}\frac{\partial^{2}\log f\left(y_{t},\left(\widetilde{\bb}_{0}^{T},\widetilde{\bg}_{0}^{T}\right)^{T}\right)
}{\partial\btheta_{S_{m}}\partial\btheta_{S_{m}}^{T}}
-\frac{1}{n}\sum_{t=1}^{n}\frac{\partial^{2}\log f\left(y_{t},\left(\bb_{0}^{T},\boldsymbol0^{T}\right)^{T}\right)
}{\partial\btheta_{S_{m}}\partial\btheta_{S_{m}}^{T}}\right.\non\\
&&\left.+\frac{1}{n}\sum_{t=1}^{n}\frac{\partial^{2}\log f\left(y_{t},\left(\bb_{0}^{T},\boldsymbol0^{T}\right)^{T}\right)
}{\partial\btheta_{S_{m}}\partial\btheta_{S_{m}}^{T}}
+J_{S_{m}}\right|.
\ee
Then we cannot guarantee
$$-\sum_{t=1}^{n}\partial^{2}\log f\left\{y_{t},\left(\widetilde{\bb}_{0}^{T},\widetilde{\bg}_{0}^{T}\right)^{T}\right\}
\Big/\left(n\partial\btheta_{S_{m}}\partial\btheta_{S_{m}}^{T}\right)\overas J_{S_{m}}.$$
Applying Taylor expansion for $f\left(y,\left\{\bb_{0}^{T},\bd'^{T}\right\}^{T}\right)$ in $\btheta_{0}$, we have
$$f\left(y,\left(\bb_{0}^{T},\bd'^{T}\right)^{T}\right)=f\left(y,\left(\bb_{0}^{T},\boldsymbol0^{T}\right)^{T}\right)
\left\{1+\frac{\partial\log f\left(y,\left(\bb_{0}^{T},\boldsymbol0^{T}\right)^{T}\right)}{\partial\bg}\bd'+R_{1}(y,\bd')\right\},$$
where $R_{1}(y,\bd')$ is a remainder term. Under general fixed parameter setup,
the remainder of $f(y,\left(\bb_{0}^{T},\bd'^{T}\right)^{T})$ does not necessarily tend to $0$ as $n\rightarrow\infty$.
Therefore, in \cite{hjort2003averaging}, the asymptotic
distributions of the model averaging estimators are not derived so that the confidence interval of $\mu(\btheta_{0})$ based on the method  does not achieve the nominal level.

Under the fixed parameter setup, the $1-\alpha$ confidence interval of $\mu(\btheta_{0})$ becomes
\be\label{eq:ci}
CI_{\mu,n}&=&\left[\widehat{\mu}\left(\boldsymbol{w}\left(D_{n}\right)\right)-\widehat{\boldsymbol{\omega}}^{T}\left\{D'_{n}-\widehat{\delta}\left(D'_{n}\right)\right\}
-z_{1-\alpha/2}\widehat{\varphi}/\sqrt{n}
,\right.\nonumber\\
&&~~\left.\widehat{\mu}\left(\boldsymbol{w}\left(D_{n}\right)\right)-\widehat{\boldsymbol{\omega}}^{T}\left\{D'_{n}-\widehat{\delta}\left(D'_{n}\right)\right\}
+z_{1-\alpha/2}\widehat{\varphi}/\sqrt{n}\right]
\ee
with $D'_{n}=\widehat{\bd}'_{full}=\widehat{\bg}_{full}$.

However, when $\mu$ is a linear function of the parameter $\btheta$, the result is different in linear regression model with normally distributed error.
Consider a linear regression model with a finite number of
regressors
\be\label{set:line}
y_{t}=\boldsymbol{x}_{t}^{T}\bb+\boldsymbol{z}_{t}^{T}\bg+e_{t}\ \ \ \ i=1,\ldots,n,\ee
where $y_{t}$ is a scalar dependent variable, $\boldsymbol{x}_{t}=(x_{t,1},\cdots,x_{t,q_{1}})^{T}$ and $\boldsymbol{z}_{t}=(z_{t,1},\cdots,z_{t,q_{2}})^{T}$ are vectors of regressors and independent and identically distributed, and $e_{t}$, $t=1,\ldots,n$ are taken to be independent and normally distributed $\mathcal{N}(0,\sigma^{2})$.

Note that
\be
\left(
 \begin{matrix}
   \frac{\partial\log f\left(y_{t},\left(\bb_{0}^{T},\bd'^{T}\right)^{T}\right)}{\partial\bb}   \\
   \frac{\partial\log f\left(y_{t},\left(\bb_{0}^{T},\bd'^{T}\right)^{T}\right)}{\partial\bg}
     \end{matrix}
  \right)
&=&\left(
 \begin{matrix}
   -\frac{1}{\sigma^{2}}\left(-\boldsymbol{x}_{t}y_{t}+\boldsymbol{x}_{t}\boldsymbol{x}_{t}^{T}\bb_{0}+\boldsymbol{x}_{t}\boldsymbol{z}_{t}^{T}\bd'\right)  \\
   -\frac{1}{\sigma^{2}}\left(-\boldsymbol{z}_{t}y_{t}+\boldsymbol{z}_{t}\boldsymbol{z}_{t}^{T}\bd'+\boldsymbol{z}_{t}\boldsymbol{x}_{t}^{T}\bb_{0}\right)
  \end{matrix}
  \right)\non\\
&=&\left(
 \begin{matrix}
   -\frac{1}{\sigma^{2}}\left(-\boldsymbol{x}_{t}y_{t}+\boldsymbol{x}_{t}\boldsymbol{x}_{t}^{T}\bb_{0}\right)  \\
   -\frac{1}{\sigma^{2}}\left(-\boldsymbol{z}_{t}y_{t}+\boldsymbol{z}_{t}\boldsymbol{x}_{t}^{T}\bb_{0}\right)
  \end{matrix}
  \right)
  +\left(
 \begin{matrix}
   -\frac{1}{\sigma^{2}}\boldsymbol{x}_{t}\boldsymbol{z}_{t}^{T}\bd'  \\
   -\frac{1}{\sigma^{2}}\boldsymbol{z}_{t}\boldsymbol{z}_{t}^{T}\bd'
 \end{matrix}
  \right)\non\\
&=& \left(
 \begin{matrix}
   \frac{\partial\log f\left(y_{t},\bb_{0},\bg_{0}\right)}{\partial\bb}   \\
   \frac{\partial\log f\left(y_{t},\bb_{0},\bg_{0}\right)}{\partial\bg}
     \end{matrix}
  \right)
+ \left(
 \begin{matrix}
   -\frac{1}{\sigma^{2}}\boldsymbol{x}_{t}\boldsymbol{z}_{t}^{T}  \\
   -\frac{1}{\sigma^{2}}\boldsymbol{z}_{t}\boldsymbol{z}_{t}^{T}
 \end{matrix}
  \right)\bd'
\ee
and
\be\label{eq:par2}
\frac{\partial^{2}\log f\left(y_{t},\left(\bb_{0}^{T},\bd'^{T}\right)^{T}\right)}{\partial\btheta\partial\btheta^{T}}
&=&-\frac{1}{\sigma^{2}}\left(
 \begin{matrix}
  \boldsymbol{x}_{t}\boldsymbol{x}_{t}^{T} & \boldsymbol{x}_{t}\boldsymbol{z}_{t}^{T}  \\
  \boldsymbol{z}_{t}\boldsymbol{x}_{t}^{T} & \boldsymbol{z}_{t}\boldsymbol{z}_{t}^{T}
     \end{matrix}
  \right)\non\\
&=& \frac{\partial^{2}\log f\left(y_{t},\left(\bb_{0}^{T},\boldsymbol0^{T}\right)^{T}\right)}{\partial\btheta\partial\btheta^{T}}.
\ee
Applying Taylor expansion for
$$\sum_{t=1}^{n}\frac{\partial\log f\left(y_{t},\left(\widehat{\btheta}_{S_{m}}^{T},\bg_{0,S_{m}^{c}}^{T}\right)^{T}\right)}{\partial\btheta_{S_{m}}}$$
in $\left(\bb_{0}^{T},\bd'^{T}\right)^{T}$ and by $\sum_{t=1}^{n}\partial\log f\left(y_{t},\left(\widehat{\btheta}_{S_{m}}^{T},\bg_{0,S_{m}^{c}}^{T}\right)^{T}\right)\bigg/\partial\btheta_{S_{m}}=0$, we have
\be
0&=&\frac{1}{\sqrt{n}}\sum_{t=1}^{n}\left[\frac{\partial\log f\left(y_{t},\left(\bb_{0}^{T},\bd'^{T}\right)^{T}\right)}{\partial\btheta_{S_{m}}}\right.\non\\
&&+\left(\frac{\partial^{2}\log f\left(y_{t},\left(\check{\bb}_{0}^{T},\check{\bg}_{0}^{T}\right)^{T}\right)
}{\partial\btheta_{S_{m}}\partial\btheta_{S_{m}}^{T}},
\frac{\partial^{2}\log f\left(y_{t},\left(\check{\bb}_{0}^{T},\check{\bg}_{0}^{T}\right)^{T}\right)
}{\partial\btheta_{S_{m}}\partial\bg_{S_{m}^{c}}^{T}}\right)
\non\\
&&\left.\left(\left(\widehat{\bb}_{S_{m}}-\bb_{0}\right)^{T},\left(\widehat{\bg}_{S_{m}}-\bg_{0,S_{m}}\right)^{T},
\boldsymbol0^{T}\right)^{T}
+\frac{\partial^{2}\log f\left(y_{t},\left(\check{\bb}_{0}^{T},\check{\bg}_{0}^{T}\right)^{T}\right)
}{\partial\btheta_{S_{m}}\partial\bg^{T}}\bd' \right]\non\\
&=&\frac{1}{\sqrt{n}}\sum_{t=1}^{n}\left[\frac{\partial\log f\left(y_{t},\left(\bb_{0}^{T},\bd'^{T}\right)^{T}\right)}{\partial\btheta_{S_{m}}}
+\frac{\partial^{2}\log f\left(y_{t},\left(\check{\bb}_{0}^{T},\check{\bg}_{0}^{T}\right)^{T}\right)
}{\partial\btheta_{S_{m}}\partial\btheta_{S_{m}}^{T}}\right.\non\\
&&\left.\left(\left(\widehat{\bb}_{S_{m}}-\bb_{0}\right)^{T},\left(\widehat{\bg}_{S_{m}}-\bg_{0,S_{m}}\right)^{T}\right)^{T} +\frac{\partial^{2}\log f\left(y_{t},\left(\check{\bb}_{0}^{T},\check{\bg}_{0}^{T}\right)^{T}\right)
}{\partial\btheta_{S_{m}}\partial\bg^{T}}\bd'\right],\non
\ee
where $\left(\check{\bb}_{0}^{T},\check{\bg}_{0}^{T}\right)^{T}$ is between
$\left(\bb_{0}^{T},\bd'^{T}\right)^{T}$ and $\left(\widehat{\btheta}_{S_{m}}^{T},\bg_{0,S_{m}^{c}}^{T}\right)^{T}$.
Therefore,
\be\label{nbbggline}
&&\left(
 \begin{matrix}
   \sqrt{n}\left(\widehat{\bb}_{S_{m}}-\bb_{0}\right)  \\
   \sqrt{n}\left(\widehat{\bg}_{S_{m}}-\bg_{0,S_{m}}\right)
  \end{matrix}
  \right)\non\\
  &=&\left(-\frac{1}{n}\sum_{t=1}^{n}\frac{\partial^{2}\log f\left(y_{t},\left(\check{\bb}_{0}^{T},\check{\bg}_{0}^{T}\right)^{T}\right)
}{\partial\btheta_{S_{m}}\partial\btheta_{S_{m}}^{T}}\right)^{-1}\non\\
&&\sqrt{n}\left[\frac{1}{n}\sum_{t=1}^{n}\frac{\partial\log f\left(y_{t},\left(\bb_{0}^{T},\bd'^{T}\right)^{T}\right)}{\partial\btheta_{S_{m}}}
+\frac{1}{n}\sum_{t=1}^{n}\frac{\partial^{2}\log f\left(y_{t},\left(\check{\bb}_{0}^{T},\check{\bg}_{0}^{T}\right)^{T}\right)
}{\partial\btheta_{S_{m}}\partial\bg^{T}}\bd'
\right]\non\\
&=&\left(-\frac{1}{n}\sum_{t=1}^{n}\frac{\partial^{2}\log f\left(y_{t},\left(\check{\bb}_{0}^{T},\check{\bg}_{0}^{T}\right)^{T}\right)
}{\partial\btheta_{S_{m}}\partial\btheta_{S_{m}}^{T}}\right)^{-1}\non\\
&&\sqrt{n}\left[\frac{1}{n}\sum_{t=1}^{n}\Gamma_{m}\frac{\partial\log f\left(y_{t},\left(\bb_{0}^{T},\bd'^{T}\right)^{T}\right)}{\partial\btheta}
+\frac{1}{n}\sum_{t=1}^{n}\frac{\partial^{2}\log f\left(y_{t},\left(\check{\bb}_{0}^{T},\check{\bg}_{0}^{T}\right)^{T}\right)
}{\partial\btheta_{S_{m}}\partial\bg^{T}}\bd'
\right].
\ee

Then following traditional arguments given for proving asymptotic normality of maximum likelihood estimators in fixed parametric models, see, for example, Theorem 18 in \cite{Ferguson1996A}, and by \eqref{eq:par2}, we can conclude that
random vector
\be
\sqrt{n}\left\{\frac{1}{n}\sum_{t=1}^{n}\frac{\partial\log f\left(y_{t},\left(\bb_{0}^{T},\bd'^{T}\right)^{T}\right)}{\partial\btheta}\right\}
&=&\sqrt{n}\left(
 \begin{matrix}
   \frac{1}{n}\sum_{t=1}^{n}\partial\log f\left(y_{t},\left(\bb_{0}^{T},\bd'^{T}\right)^{T}\right)/\partial\bb  \\
   \frac{1}{n}\sum_{t=1}^{n}\partial\log f\left(y_{t},\left(\bb_{0}^{T},\bd'^{T}\right)^{T}\right)/\partial\bg
 \end{matrix}
  \right)\non\\
&\doteq&\left(
 \begin{matrix}
   M_{n}  \\
   N_{n}
 \end{matrix}
  \right)
\overd\left(
 \begin{matrix}
   M  \\
   N
 \end{matrix}
  \right)\sim \mathcal{N}(0,J_{full}),\ee
\be\label{var}
-\frac{1}{n}\sum_{i=1}^{n}\frac{\partial^{2}\log f\left(y_{i},\left(\check{\bb}_{0}^{T},\check{\bg}_{0}^{T}\right)^{T}\right)}{\partial\btheta_{S_{m}}\partial\btheta_{S_{m}}^{T}}
=-\frac{1}{n}\sum_{i=1}^{n}\frac{\partial^{2}\log f\left(y_{i},\bb_{0},\bg_{0}\right)}{\partial\btheta_{S_{m}}\partial\btheta_{S_{m}}^{T}}
\overas
J_{S_{m}},\ee
\be
\frac{1}{n}\sum_{t=1}^{n}\frac{\partial^{2}\log f\left(y_{t},\left(\check{\bb}_{0}^{T},\check{\bg}_{0}^{T}\right)^{T}\right)
}{\partial\btheta_{S_{m}}\partial\bg^{T}}\bd'\overas
\label{E}
\mathrm{E}\left\{\left(
 \begin{matrix}
   -\frac{1}{\sigma^{2}}\boldsymbol{x}_{i}\boldsymbol{z}_{i}^{T}  \\
   -\frac{1}{\sigma^{2}}\Gamma_{m}\boldsymbol{z}_{i}\boldsymbol{z}_{i}^{T}
 \end{matrix}
  \right)\bd'\right\}
  =\left(
 \begin{matrix}
   J_{01}  \\
   \Gamma_{m}J_{11}
 \end{matrix}
  \right)\bd'\ee
and
\be\label{eq:pdistr}
&&\sqrt{n}\left\{\frac{1}{n}\sum_{t=1}^{n}\frac{\partial\log f\left(y_{t},\left(\bb_{0}^{T},\bd'^{T}\right)^{T}\right)}{\partial\btheta_{S_{m}}}\right\}\non\\
&&\doteq\left(
 \begin{matrix}
   M_{n}  \\
   N_{n}^{m}
 \end{matrix}
  \right)
\overd\left(
 \begin{matrix}
   M  \\
   N^{m}
 \end{matrix}
  \right)\sim\mathcal{N}\left(0,J_{S_{m}}\right).
\ee
Therefore, substituting \eqref{var}-\eqref{eq:pdistr} into \eqref{nbbggline}, we can obtain
\be\label{nbbggline}
&&\left(
 \begin{matrix}
   \sqrt{n}\left(\widehat{\bb}_{S_{m}}-\bb_{0}\right)  \\
   \sqrt{n}\left(\widehat{\bg}_{S_{m}}-\bg_{0,S_{m}}\right)
  \end{matrix}
  \right)\non\\
&&=\left\{J_{S_{m}}+o_{a.s.}(1)\right\}^{-1}
\left\{\left(
 \begin{matrix}
   J_{01}  \\
   \Gamma_{m}J_{11}
 \end{matrix}
  \right)\sqrt{n}\bd'+\left(
 \begin{matrix}
   M_{n}  \\
   N_{n}^{m}
 \end{matrix}
  \right)
\right\}
\ee
and then
\be
\left(
 \begin{matrix}
   \sqrt{n}\left(\widehat{\bb}_{S_{m}}-\bb_{0}\right)  \\
   \sqrt{n}\left(\widehat{\bg}_{S_{m}}-\bg_{0,S_{m}}\right)
  \end{matrix}
  \right)-J_{S_{m}}^{-1}\left(
 \begin{matrix}
   J_{01}  \\
   \Gamma_{m}J_{11}
  \end{matrix}
  \right)\sqrt{n}\bd'\overd \mathcal{N}\left(0,J_{S_{m}}^{-1}\right).
\ee

Let $\boldsymbol{I}_{q_{2}}$ be an identity matrix. Using the same proof skills, we have
\be
D_{n}&=&\sqrt{n}\widehat{\bg}_{full}\non\\
&=&\left(
 \begin{matrix}
 \boldsymbol{0}_{q_{2}\times q_{1}}& \boldsymbol{I}_{q_{2}}
  \end{matrix}
  \right)
  \left\{J_{full}+o_{a.s.}(1)\right\}^{-1}\sqrt{n}\left\{\left(
 \begin{matrix}
   J_{01}  \\
   J_{11}
 \end{matrix}
  \right)\bd'+\left(
 \begin{matrix}
   M_{n}  \\
   N_{n}
 \end{matrix}
  \right)
\right\},
\ee
and then
\be
D_{n}-\sqrt{n}\bd'
=J^{10}M_{n}+J^{11}N_{n}+o_{a.s.}(1)
\doteq W_{n}+o_{a.s.}(1) \overd W\sim \mathcal{N}(0,K).
\ee

Using Taylor expansion for
$\widehat{\mu}_{m}-\mu\left(\btheta_{0}\right)$, we have
\be
\widehat{\mu}_{m}-\mu\left(\btheta_{0}\right)
&=&\left\{\frac{\partial \mu\left(\btheta_{0}\right)}{\partial\bb}\right\}^{T}
\left(\widehat{\bb}_{S_{m}}-\bb_{0}\right)
+\left\{\frac{\partial \mu\left(\btheta_{0}\right)}{\partial\bg_{S_{m}}}\right\}^{T}
\left(\widehat{\bg}_{S_{m}}-\bg_{0,S_{m}}\right)
-\left\{\frac{\partial \mu\left(\btheta_{0}\right)}{\partial\bg}\right\}^{T}\bd'\non\\
&&+\left(
 \begin{matrix}
   \widehat{\bb}_{S_{m}}-\bb_{0}  \\
   \left(
 \begin{matrix}
\widehat{\bg}_{S_{m}}-\bg_{0,S_{m}}\\
\boldsymbol{0}
  \end{matrix}
  \right)
   -\bd'
  \end{matrix}
  \right)^{T}
  \left(
 \begin{matrix}
   \frac{\partial^{2}\mu(\bar{\btheta}_{0})}{\partial\bb\partial\bb^{T}}& \frac{\partial^{2}\mu(\bar{\btheta}_{0})}{\partial\bb\partial\bg^{T}} \\
   \frac{\partial^{2}\mu(\bar{\btheta}_{0})}{\partial\bg\partial\bb^{T}}&
   \frac{\partial^{2}\mu(\bar{\btheta}_{0})}{\partial\bg\partial\bg^{T}}
  \end{matrix}
  \right)
\left(
 \begin{matrix}
   \widehat{\bb}_{S_{m}}-\bb_{0}  \\
   \left(
 \begin{matrix}
\widehat{\bg}_{S_{m}}-\bg_{0,S_{m}}\\
\boldsymbol{0}
  \end{matrix}
  \right)
   -\bd'
  \end{matrix}
  \right),\non
\ee
where $\bar{\btheta}_{0}$ is between
$\btheta_{0}$ and $\left(\widehat{\btheta}_{S_{m}}^{T},\widehat{\bg}_{S_{m}}^{T},\bg_{0,S_{m}^{c}}^{T}\right)^{T}$.

When $\mu$ is a linear function of the parameter $\btheta$, then by $$\left(
 \begin{matrix}
   \frac{\partial^{2}\mu(\bar{\btheta}_{0})}{\partial\bb\partial\bb^{T}}& \frac{\partial^{2}\mu(\bar{\btheta}_{0})}{\partial\bb\partial\bg^{T}} \\
   \frac{\partial^{2}\mu(\bar{\btheta}_{0})}{\partial\bg\partial\bb^{T}}&
   \frac{\partial^{2}\mu(\bar{\btheta}_{0})}{\partial\bg\partial\bg^{T}}
  \end{matrix}
  \right)=\boldsymbol{0}$$
and the asymptotic normality of maximum likelihood estimation,
substituting \eqref{var}-\eqref{eq:pdistr} into \eqref{nbbggline}, we can obtain
\be\label{n:detla}
&&\sqrt{n}\left[\left\{\widehat{\mu}_{m}-\mu\left(\btheta_{0}\right)\right\}
+\left\{\frac{\partial \mu\left(\btheta_{0}\right)}{\partial\bg}\right\}^{T}\bd'
-\left(
 \begin{matrix}
   \frac{\partial \mu\left(\btheta_{0}\right)}{\partial\bb}  \\
   \frac{\partial \mu\left(\btheta_{0}\right)}{\partial\bg_{S_{m}}}
  \end{matrix}
  \right)^{T}J_{S_{m}}^{-1}\left(
 \begin{matrix}
   J_{01}  \\
   \Gamma_{m}J_{11}
  \end{matrix}
  \right)\bd'\right]\non\\
&=&\left(
 \begin{matrix}
   \frac{\partial \mu\left(\btheta_{0}\right)}{\partial\bb}  \\
   \frac{\partial \mu\left(\btheta_{0}\right)}{\partial\bg_{S_{m}}}
  \end{matrix}
  \right)^{T}\left[\left(
 \begin{matrix}
   \sqrt{n}\left(\widehat{\bb}_{S_{m}}-\bb_{0}\right)  \\
   \sqrt{n}\left(\widehat{\bg}_{S_{m}}-\bg_{0,S_{m}}\right)
  \end{matrix}
  \right)-J_{S_{m}}^{-1}\left(
 \begin{matrix}
   J_{01}  \\
   \Gamma_{m}J_{11}
  \end{matrix}
  \right)\sqrt{n}\bd'\right]\nonumber\\
 &=&\left(
 \begin{matrix}
   \frac{\partial \mu\left(\btheta_{0}\right)}{\partial\bb}  \\
   \frac{\partial \mu\left(\btheta_{0}\right)}{\partial\bg_{S_{m}}}
  \end{matrix}
  \right)^{T}J_{S_{m}}^{-1} \left\{\left(
 \begin{matrix}
   M_{n}  \\
   N_{n}^{m}
 \end{matrix}
  \right)+o_{a.s.}(1)\right\}\nonumber\\
  &\overd&\left(
 \begin{matrix}
   \frac{\partial \mu\left(\btheta_{0}\right)}{\partial\bb}  \\
   \frac{\partial \mu\left(\btheta_{0}\right)}{\partial\bg_{S_{m}}}
  \end{matrix}
  \right)^{T}J_{S_{m}}^{-1} \left(
 \begin{matrix}
   M \\
   N^{m}
 \end{matrix}
  \right)\doteq\Lambda_{m},
\ee

From the proof of Lemma 3.3 in \cite{hjort2003averaging}, note that
\be
&&\left(
 \begin{matrix}
   \frac{\partial \mu\left(\btheta_{0}\right)}{\partial\bb}  \\
   \frac{\partial \mu\left(\btheta_{0}\right)}{\partial\bg_{S_{m}}}
  \end{matrix}
  \right)^{T}J_{S_{m}}^{-1}
  \left(
 \begin{matrix}
   J_{01}  \\
   \Gamma_{m}J_{11}
  \end{matrix}
  \right)\bd'
  -\left\{\frac{\partial \mu\left(\btheta_{0}\right)}{\partial\bg}\right\}^{T}\bd'
\non\\
  &=&\left(
 \begin{matrix}
   \frac{\partial \mu\left(\btheta_{0}\right)}{\partial\bb}  \\
   \frac{\partial \mu\left(\btheta_{0}\right)}{\partial\bg_{S_{m}}}
  \end{matrix}
  \right)^{T}
 \left(
 \begin{matrix}
   J^{00,m} & J^{01,m} \\
   J^{10,m} & J^{11,m}
  \end{matrix}
  \right)\left(
 \begin{matrix}
   J_{01}  \\
   \Gamma_{m}J_{11}
  \end{matrix}
  \right)\bd'
  -\left\{\frac{\partial \mu\left(\btheta_{0}\right)}{\partial\bg}\right\}^{T}\bd'\non\\
  &=& \left\{\frac{\partial \mu\left(\btheta_{0}\right)}{\partial\bb}\right\}^{T}
\left(J^{00,m}J_{01}+J^{01,m}\Gamma_{m}J_{11}\right)\bd'\non\\
&&~~+\left\{\frac{\partial \mu\left(\btheta_{0}\right)}{\partial\bg_{S_{m}}}\right\}^{T}
\left(J^{10,m}J_{01}+J^{11,m}\Gamma_{m}J_{11}\right)\bd'
-\left\{\frac{\partial \mu\left(\btheta_{0}\right)}{\partial\bg}\right\}^{T}\bd'\non\\
&=&\left\{\frac{\partial \mu\left(\btheta_{0}\right)}{\partial\bb}\right\}^{T}
J_{00}^{-1}J_{01}\left(\boldsymbol I_{q_{2}}-K^{1/2}H_{m}K^{-1/2}\right)\bd'\non\\
&&~~+\left\{\frac{\partial \mu\left(\btheta_{0}\right)}{\partial\bg}\right\}^{T}
\left(\Gamma_{m}^{T}K_{m}\Gamma_{m}K^{-1}\right)\bd'
-\left\{\frac{\partial \mu\left(\btheta_{0}\right)}{\partial\bg}\right\}^{T}\bd'\non\\
&=&\boldsymbol{\omega}^{T}\left(\boldsymbol I_{q_{2}}-K^{1/2}H_{m}K^{-1/2}\right)\bd'
\ee
and
\be
\Lambda_{m}&=&\left(
 \begin{matrix}
   \frac{\partial \mu\left(\btheta_{0}\right)}{\partial\bb}  \\
   \frac{\partial \mu\left(\btheta_{0}\right)}{\partial\bg_{S_{m}}}
  \end{matrix}
  \right)^{T}J_{S_{m}}^{-1} \left(
 \begin{matrix}
   M \\
   N^{m}
 \end{matrix}
  \right)\non\\
&=&\left(
 \begin{matrix}
   \frac{\partial \mu\left(\btheta_{0}\right)}{\partial\bb}  \\
   \frac{\partial \mu\left(\btheta_{0}\right)}{\partial\bg_{S_{m}}}
  \end{matrix}
  \right)^{T}
 \left(
 \begin{matrix}
   J^{00,m} & J^{01,m} \\
   J^{10,m} & J^{11,m}
  \end{matrix}
  \right)
   \left(
 \begin{matrix}
   M \\
   N^{m}
 \end{matrix}
  \right)\non\\
&=& \left\{\frac{\partial \mu\left(\btheta_{0}\right)}{\partial\bb}\right\}^{T}
\left(J^{00,m}M+J^{01,m}N^{m}\right)+
\left\{\frac{\partial \mu\left(\btheta_{0}\right)}{\partial\bg_{S_{m}}}\right\}^{T}
\left(J^{10,m}M+J^{11,m}N^{m}\right)\non\\
&=&\left\{\frac{\partial \mu\left(\btheta_{0}\right)}{\partial\bb}\right\}^{T}
\left(J_{00}^{-1}M-J_{00}^{-1}J_{01}K^{1/2}H_{m}K^{-1/2}W\right)
+\left\{\frac{\partial \mu\left(\btheta_{0}\right)}{\partial\bg_{S_{m}}}\right\}^{T}
K_{m}\Gamma_{m}K^{-1}W\non\\
&=&\left\{\frac{\partial \mu\left(\btheta_{0}\right)}{\partial\bb}\right\}^{T}J_{00}^{-1}M
-\boldsymbol{\omega}^{T}K^{1/2}H_{m}K^{-1/2}W.
\ee
Therefore, we can get a conclusion similar to Lemma 3.3 in \cite{hjort2003averaging},
\be
&&\sqrt{n}\left[\left\{\widehat{\mu}_{m}-\mu\left(\btheta_{0}\right)\right\}
-\boldsymbol{\omega}^{T}\left(\boldsymbol I_{q_{2}}-K^{1/2}H_{m}K^{-1/2}\right)\bd'\right]\non\\
  &&\overd\left\{\frac{\partial \mu\left(\btheta_{0}\right)}{\partial\bb}\right\}^{T}J_{00}^{-1}M
-\boldsymbol{\omega}^{T}K^{1/2}H_{m}K^{-1/2}W.
\ee

Note that $W$ and $M$ are independent.
Since for all model $m$, $w_{m}\left(D_{n}\right)$
and $\widehat{\mu}_{m}$
can be expressed the function of the same random vector $\left(
   M_{n}^{T} ,
   N_{n}^{T}
  \right)^{T}$ almost surely and
$w_{m}\left(D_{n}\right)$ is continuous for $D_{n}$, by Continuous Theorem, we have
\be
&&\sqrt{n}\left(\left\{\sum_{m=1}^{M}w_{m}\left(D_{n}\right)\widehat{\mu}_{m}-\mu\left(\btheta_{0}\right)\right\}
-\boldsymbol{\omega}^{T}\left[\boldsymbol I_{q_{2}}-K^{1/2}\left\{\sum_{m=1}^{M}w_{m}\left(D_{n}\right)H_{m}\right\}K^{-1/2}\right]\bd'\right)\non\\
&=&\sqrt{n}\left[\left\{\sum_{m=1}^{M}w_{m}\left(D_{n}\right)\widehat{\mu}_{m}-\mu\left(\btheta_{0}\right)\right\}
-\boldsymbol{\omega}^{T}\left\{\boldsymbol \bd'-\widehat{\delta}\left(\bd'\right)\right\}\right]\non\\
  &\overd&\left\{\frac{\partial \mu\left(\btheta_{0}\right)}{\partial\bb}\right\}^{T}J_{00}^{-1}M
-\boldsymbol{\omega}^{T}\widehat{\bd}(W)
\doteq\Lambda,
\ee
where $\widehat{\bd}(W)=K^{1/2}\left\{\sum_{m=1}^{M}w_{m}\left(W\right)H_{m}\right\}K^{-1/2}W$, $\Lambda$ is normally distributed with mean $0$ and covariance matrix $\Sigma$ no matter the parameters are fixed or local misspecification with
\be
\Sigma=\left(\frac{\partial\mu}{\partial\btheta}\right)^{T}J_{00}^{-1}\frac{\partial\mu}{\partial\btheta}
+\boldsymbol{\omega}^{T}\mathrm{Var}\widehat{\bd}(W)\boldsymbol{\omega}.
\ee

Therefore, we can conclude that
\be
T_{n}&=&\frac{\sqrt{n}\left[\widehat{\mu}\left(\boldsymbol{w}\left(D_{n}\right)\right)-\mu\left(\btheta_{0}\right)-\widehat{\boldsymbol{\omega}}^{T}\left\{D'_{n}-\widehat{\delta}\left(D'_{n}\right)\right\}
\right]}{\widehat{\varphi}}\non\\
&=&\frac{\sqrt{n}\left[\widehat{\mu}\left(\boldsymbol{w}\left(D_{n}\right)\right)-\mu\left(\btheta_{0}\right)\right]-\widehat{\boldsymbol{\omega}}^{T}\left\{D_{n}-\widehat{\delta}\left(D_{n}\right)\right\}
}{\widehat{\varphi}}\non\\
&=&\frac{\sqrt{n}\left[\widehat{\mu}\left(\boldsymbol{w}\left(D_{n}\right)\right)-\mu\left(\btheta_{0}\right)-\boldsymbol{\omega}^{T}\left\{\boldsymbol \bd'-\widehat{\delta}\left(\bd'\right)\right\}\right]
-\widehat{\boldsymbol{\omega}}^{T}\left(D_{n}-\sqrt{n}\bd'\right)
+\widehat{\boldsymbol{\omega}}^{T}\left\{\widehat{\delta}\left(D_{n}\right)-\sqrt{n}\widehat{\delta}\left(\bd'\right)\right\}
}{\widehat{\varphi}}\non\\
&\overd&\frac{\left\{\frac{\partial \mu\left(\btheta_{0}\right)}{\partial\bb}\right\}^{T}J_{00}^{-1}M
-\boldsymbol{\omega}^{T}W}{\varphi}\sim \mathcal{N}(0,1).
\ee
We establish that the limiting coverage probability is still $1-\alpha$ when we replace $\bd$ with $\sqrt{n}\bd'$ in the $1-\alpha$ confidence interval in \cite{hjort2003averaging}.
In other words, though the result in \cite{hjort2003averaging} does not hold under general fixed parameter setup, the procedure is still valid for the linear function of the parameter $\theta$ in linear regression model with normally distributed error. The confidence interval in \cite{hjort2003averaging} may also have asymptotically correct coverage for a linear model with random errors from other distributions, which remains for future research.

In the next section, we will further verify these theoretical results through Monte Carlo simulation.
\subsection{Confidence intervals based on the asymptotic theory}
In this subsection, we provide the simulation-based confidence intervals of scaled S-AIC estimators and scaled S-BIC estimators based on the asymptotic theory, where the scaled S-AIC and scaled S-BIC estimators for $\btheta_{0}$ and $\mu(\btheta_{0})$ are defined by (\ref{13}) and (\ref{14}), respectively.
We call these two methods SAIC$_{F}$ and SBIC$_{F}$. Next, we will describe them in detail.

For SAIC$_{F}$, as shown in Theorem \ref{th1}, the asymptotic distribution of scaled S-AIC estimator is a nonstandard distribution. It cannot be directly used for inference. To address this issue, we follow \cite{Claeskens2008Model}, \cite{Lu2015A}, and \cite{Francis2016Using}, and consider a simulation-based method to construct the confidence intervals.
From Theorem \ref{th1}, we see that the asymptotic distribution of the scaled S-AIC estimator is a nonlinear function of unknown parameter vector $\btheta_{0}$, overfit model set $\mathcal{O}$ and normal random vector $\bbeta$. Suppose that $\btheta_{0}$ and $\mathcal{O}$ are known, then, by drawing the replicates from $\bbeta$, we could approximate the limiting distributions defined in Theorem \ref{th1} to arbitrary precision. This is the main idea of the simulation-based confidence intervals. In practice, we replace the unknown parameters with corresponding estimators. We then simulate the limiting distribution of the S-AIC estimator and use this simulated distribution to conduct inference. We now describe the simulation-based confidence intervals in details. The true parameter $\btheta_{0}$ in the asymptotic distribution of scaled S-AIC is estimated by scaled S-BIC estimator.
Generate a sufficiently large number of $q\times1$ normal random vector $\bbeta^{(r)}\sim \mathcal{N}\left(0,\mathcal{F}^{-1}\left(\widehat\btheta\left(\widehat w_{\bic_{s}}\right)\right)\right)$ for $r=1,\ \cdots,\ R_{0}$. For each $r$, we compute the quantities of the asymptotic distributions derived in Theorem \ref{th1}. That is, we first calculate
\be
\widehat{G}_{\lv{v}{j}}&=&\exp\left\{\left[\Pi_{\lv{v}{j}}\mathcal{F}\left(\widehat\btheta\left(\widehat w_{\bic_{s}}\right)\right)\bbeta^{(r)}\right]^{T}
\left(H_{\lv{v}{j}}-\left[\Pi_{\lv{v}{j}}\mathcal{F}\left\{\widehat\btheta\left(\widehat w_{\bic_{s}}\right)\right\}\Pi_{\lv{v}{j}}^{T}\right]^{-1}\right)\Pi_{\lv{v}{j}}\right.\nonumber\\
&&\left.\mathcal{F}\left(\widehat\btheta\left(\widehat w_{\bic_{s}}\right)\right)\bbeta^{(r)}/2-k_{\lv{v}{j}}\right\}.
\ee
Then we compute $$D_{\theta_{j}}^{(r)}=\sum_{k=1}^{M_{j}}\boldsymbol{1}\{\lv{v}{j}\in\calO\}
\frac{\widehat{G}_{\lv{v}{j}}}{\sum_{k=1}^{M_{j}}\boldsymbol{1}\{\lv{v}{j}\in\calO\}\widehat{G}_{\lv{v}{j}}}
\widetilde{\eta}_{j,\lv{v}{j}}^{(r)},$$ where $\widetilde{\eta}_{j,\lv{v}{j}}^{(r)}$ is the $j\th$ component of
$\widetilde{\bbeta}_{\lv{v}{j}}^{(r)}=\Pi_{\lv{v}{j}}^{T}H_{S_{\lv{v}{j}}}\Pi_{\lv{v}{j}}\Delta_{\lv{v}{j}}\mathcal{F}\left\{\widehat\btheta\left(\widehat w_{\bic_{s}}\right)\right\}\bbeta^{(r)}$. Let $\widehat{q}_{\aic_{s},j}(\alpha/2)$ and $\widehat{q}_{\aic_{s},j}(1-\alpha/2)$ be the $\alpha/2$ and $1-\alpha/2$ quantiles of $D^{(r)}_{\theta_{j}}$ for $r=1,\ \cdots,\ R_{0}$, respectively.
The $1-\alpha$ confidence interval of $\theta_{j}$ with SAIC$_{F}$ method is constructed as
\be
CI_{\theta_{j},n}\left(\text{SAIC}_{F}\right)=\left[\widehat\theta_{j}(\widehat{\boldsymbol{w}}_{\aic_{s}})-n^{-1/2}\widehat{q}_{\aic_{s},j}(1-\alpha/2)
,\widehat\theta_{j}(\widehat{\boldsymbol{w}}_{\aic_{s}})-n^{-1/2}\widehat{q}_{\aic_{s},j}(\alpha/2)\right].\ee
Similarly, based on the conclusion in Theorem \ref{th2}, the $1-\alpha$ confidence interval of $\mu(\btheta_{0})$ with SAIC$_{F}$ method is constructed as
\be
CI_{\mu,n}\left(\text{SAIC}_{F}\right)=\left[\widehat{\mu}(\widehat{\boldsymbol{w}}_{\aic})-n^{-1/2}\widehat{q}_{\aic}(1-\alpha/2)
,\widehat{\mu}(\widehat{\boldsymbol{w}}_{\aic})-n^{-1/2}\widehat{q}_{\aic}(\alpha/2)\right],\ee
where $\widehat{q}_{\aic}(\alpha/2)$ and $\widehat{q}_{\aic}(1-\alpha/2)$ be the $\alpha/2$ and $1-\alpha/2$ quantiles of
$$D^{(r)}_{\mu}=\sum_{m\in\widehat{\calO}}\frac{\widehat{G}_{m}}{\sum_{m\in\widehat{\calO}}\widehat{G}_{m}}
\dot{\mu}\left(\widehat\btheta\left(\widehat w_{\bic_{s}}\right)\right)^{T}\Pi_{m}^{T}H_{S_{m}}\Pi_{m}\Delta_{m}\mathcal{F}\left(\btheta_{0}\right)\bbeta^{(r)}$$ for $r=1,\ \cdots,\ R_{0}$.

For the parameter $\theta_{j}$ with $j\not\in S_{m_{o}}$, the method SBIC$_{F}$ cannot be used to calculate confidence intervals because the asymptotic distribution of the scaled S-BIC estimator is a degenerate distribution. The true parameter $\btheta_{0}$ in the asymptotic distribution of scaled S-BIC estimator is also estimated by the scaled S-BIC estimator. We first compute the $\alpha/2$ quantile of every component of random variable from the normal distribution $\mathcal{N}\left(0,\widehat{\Sigma}\right)$,
where $$\widehat{\Sigma}=\Pi_{m_{o}}^{T}H_{S_{m_{o}}}\Pi_{m_{o}}\Delta_{m_{o}}\mathcal{F}\left\{\widehat\btheta\left(\widehat w_{\bic_{s}}\right)\right\}\left(\Pi_{m_{o}}^{T}H_{S_{m_{o}}}\Pi_{m_{o}}\Delta_{m_{o}}\right)^{T}.$$
$\widehat{\sigma}_{j}$ is the $j\th$ element on the diagonal of matrix $\widehat{\Sigma}$ and the $\alpha/2$ quantile of $\mathcal{N}(0,\widehat{\sigma}_{j})$ is
denoted by $z_{bic_{s},j}(\alpha/2)$. Then the $1-\alpha$ confidence interval of $\theta_{j}$ with SBIC$_{F}$ method is constructed as
\be
CI_{\theta_{j},n}\left(\text{SBIC}_{F}\right)=\left[\widehat\theta_{j}(\widehat{\boldsymbol{w}}_{\bic_{s}})-n^{-1/2}z_{bic_{s},j}(\alpha/2)
,\widehat\theta_{j}(\widehat{\boldsymbol{w}}_{\bic_{s}})+n^{-1/2}z_{bic_{s},j}(\alpha/2)\right].\ee
Similarly, based on the conclusion in Theorem \ref{th2}, the $1-\alpha$ confidence interval of $\mu(\btheta_{0})$ with SBIC$_{F}$ method is constructed as
\be
CI_{\mu,n}\left(\text{SBIC}_{F}\right)=\left[\widehat{\mu}(\widehat{\boldsymbol{w}}_{\bic})-n^{-1/2}z_{bic}(\alpha/2)
,\widehat{\mu}(\widehat{\boldsymbol{w}}_{\bic})+n^{-1/2}z_{bic}(\alpha/2)\right],\ee
where $z_{bic}(\alpha/2)$ is the $\alpha/2$ quantile of
$\dot{\mu}\left(\widehat{\btheta}\left(\widehat{\boldsymbol{w}}_{\bic_{s}}\right)\right)^{T}\widehat{\sum}\dot{\mu}\left(\widehat{\btheta}\left(\widehat{\boldsymbol{w}}_{\bic_{s}}\right)\right).$
\subsection{Confidence intervals of \cite{Charkhi2018Asymptotic}}
Following Section 3 of \cite{Charkhi2018Asymptotic}, we construct the $1-\alpha$ confidence intervals of the parameters contained in the model selected using AIC. We call this method PAIC and more details can be found in \cite{{Charkhi2018Asymptotic}}.
Let
$\mathcal{M}=\calU \bigcup\calO$. From Definition 1 of \cite{{Charkhi2018Asymptotic}}, the selection matrix $\zeta_{\mathcal{M}}$ is a
$|\mathcal{M}|\times q$ matrix with $\{0, 1\}$ elements, constructed as
$\zeta_{\mathcal{M}} = \left(\boldsymbol{1}_{q}^{T}\pi_{1}^{T}\pi_{1},\cdots, \boldsymbol{1}_{q}^{T}\pi_{M}^{T}\pi_{M}\right)^{T}$, where $|\mathcal{M}|$ is the number of models and $\pi_{m}$ is a $k_{m}\times q$
matrix that selects those covariates  belonging to model $m$ and
$\boldsymbol{1}_{q}^{T}$ is a $q\times 1$ vector with each element being $1$.
Let
$m_{p}$ be the selected model by AIC, $t(m_{p})$ be the subvector of vector $t=(t_{1},\cdots,t_{q})^{T}$ with its subscript consistent with the parameters in the $m_{p}^{th}$ model, and
$\mathcal{F}{m_{p}}(\btheta_{0})$ be the corresponding submatrix of the Fisher information matrix  $\mathcal{F}(\btheta_{0})$.
The selection region for model $m_{p}$ is
$\mathcal{A}_{m_{p}}(\calO)
=\left\{\boldsymbol{z}\in\mathbb{R}^{q}:
\left\{\boldsymbol{1}_{(|\calO|-1)}\otimes(\boldsymbol{1}_{q_{2}}^{T}\pi_{M}^{T}\pi_{M})-\zeta_{\calO\setminus M}\right\}
\left\{(z_{1}^{2}-2),\cdots,(z_{q}^{2}-2)\right\}^{T}>0
\right\}$. Denote
$$f_{m_{p}}\left(t(m_{p})\right)
=\phi_{m_{p}}\left(t(m_{p})|\mathcal{A}_{m_{p}}(\calO);
J_{m_{p}}^{-1}(\btheta_{0})\right)$$
as the asymptotic density of $\sqrt{n}(\widehat{\btheta}_{S_{m_{p}}}-\btheta_{0,S_{m_{p}}})$,
a truncated $k_{m_{p}}$-dimensional normal density. The quantile of its $j^{th}$ component is obtained via
$$\int_{\mathcal{R}_{\alpha}}f_{m_{p}}\left(t(m_{p})\right)\mathrm{d}t(m_{p})=1-\alpha,$$
where $\mathcal{R}_{\alpha}\subset\mathbb{R}^{k_{m_{p}}}$
restricts only the $j^{th}$ component to $[-q_{p}(1-\alpha/2), q_{p}(1-\alpha/2)]$. Then the $1-\alpha$ confidence interval of $\theta_{j}$ in the selected model based on PAIC method is constructed as
$$\left[\widehat{\theta}_{j}-n^{-1/2}q_{p}(1-\alpha/2)
,\widehat{\theta}_{j}+n^{-1/2}q_{p}(1-\alpha/2)\right].$$
The calculation of this confidence interval in the simulation is based on the R code provided by \cite{{Charkhi2018Asymptotic}}.
\subsection{Confidence intervals based on the full model}
In this subsection, we provide one method to construct confidence intervals called FULL. The estimator is the maximum likelihood estimator for the largest model, say the $M\th$ model. The $1-\alpha$ confidence interval of $\theta_{j}$ is given by
\be
CI_{n}=\left[\widehat\theta_{j,M}-z_{1-\alpha/2}s(\widehat\theta_{j,M})
,\widehat\theta_{j,M}+z_{1-\alpha/2}s(\widehat\theta_{j,M})\right],\ee
where $z_{1-\alpha/2}$ is $1-\alpha/2$ quantile of the standard normal distribution and $s(\widehat\theta_{j,M})$ is the
standard error computed based on the $M\th$ model.
The $1-\alpha$ confidence interval of $\mu(\btheta_{0})$ is constructed as
\be
CI_{n}=\left[\widehat{\mu}_{M}-z_{1-\alpha/2}s(\widehat{\mu}_{M})
,\widehat{\mu}_{M}+z_{1-\alpha/2}s(\widehat{\mu}_{M})\right],\ee
where $s(\widehat{\mu}_{M})$ is the standard error computed based on the $M\th$ model.
\section{Simulation Study} \label{sec:sim}
In this section, based on Theorem \ref{th1} and \ref{th2} in Section \ref{sec:Inference}, we first study the coverage probabilities of confidence intervals for $\btheta_{0}$ and $\mu(\btheta_{0})$ based on scaled S-AIC and scaled S-BIC model averaging estimators in comparison with some other commonly used model averaging approaches under general fixed parameter setup. In this section and next section, Let us take $\mu$ as the mean function as an example. Then, we do Q-Q plot to verify that the $\Lambda$ in previous section approximately follows standard normal distribution. In the following, we present some simulation settings and the results of these.

Consider the following estimators for comparison:\\
1. Scaled S-AIC estimator with simulation-based confidence intervals (SAIC$_{F}$);\\
2. Scaled S-BIC estimator with simulation-based confidence intervals (SBIC$_{F}$);\\
3. Scaled S-AIC and Scaled S-BIC estimators with the approach in \cite{S1997Model} (SAIC$_{97}$ and SBIC$_{97}$ );\\
4. Smoothed model-averaged estimators using AIC and BIC with the inference method in \cite{hjort2003averaging} (SAIC$_{L}$ and SBIC$_{L}$);\\
5. Post-AIC estimators with the asymptotic distribution conditional on the selected model in \cite{Charkhi2018Asymptotic} (PAIC);\\
6. Maximum likelihood estimator based on the largest model (FULL).\\
The confidence intervals for each estimator have been constructed in Section \ref{sec:CI}.
\subsection{Simulation Setup}
In this subsection, we present two kinds of simulation settings, one is a linear regression model, the other is a non-linear regression model, Poisson regression model.
\subsubsection{Linear regression model}
We consider a linear regression model with a finite number of
regressors
$$y_{i}=\boldsymbol{x}_{i}^{T}\bb+\boldsymbol{z}_{i}^{T}\bg+e_{i},~~~~i=1,\ldots,n,$$
where $\bb=(1,1)^{T}$,
$\bg=(0,1.5,0)^{T}$, $\boldsymbol{x}_{i}=(1,x_{i,1})^{T},$ $\boldsymbol{z}_{i}=(z_{i,1},z_{i,2},z_{i,3})^{T},$
$(x_{i,1},z_{i,1},z_{i,2},z_{i,3})^{T}\sim \mathcal{N}(\boldsymbol{0},\boldsymbol{Q})$.
The error term $e_{i}$ is generated from a normal distribution $\mathcal{N}(0,\sigma^{2})$ with $\sigma=0.5$, $1$ and $1.5$. Assume the diagonal elements of $\boldsymbol{Q}$ are $1$, and off-diagonal elements are $\rho^{|i-j|}$, where $\rho=0.5$ and $0.8$. Set $n=10$, $50$ and $100$ and study the confidence intervals for parameters $\beta_{1}$, $\gamma_{2}$ and mean $\mu$.

We tried other different cases of distribution for $(x_{i,1},z_{i,1},z_{i,2},z_{i,3})^{T}$, including uniform, exponential and mixture distributions. We found that changing distributions of covariates has little effect on the conclusions. So we only present the result of the normal case.
\subsubsection{Poisson regression model}
We generate $y_{i}$ from Possion distribution with $\lambda_{i}=p_{i}$,
$$\log p_{i}=\boldsymbol{x}_{i}^{T}\bb+\boldsymbol{z}_{i}^{T}\bg,~~~~i=1,\ldots,n,$$
where we set $\bb=(1,0.5)^{T}$,
$\bg=(0,1.2,0)^{T}$, and
the settings others are the same as those in linear regression model.
Let $n=30$, $50$ and $100$ and then study the confidence intervals for the same parameters as in Setting 1.

Set $R_{0}=10,000$ and replicate $R=1,000$ times. For the methods SAIC$_{F}$, SBIC$_{F}$, SAIC$_{97}$, SBIC$_{97}$, SAIC$_{L}$ ,SBIC$_{L}$ and PAIC, we take $\left(\bb^{T},\bg^{T}\right)^{T}$ as our fixed parameter setup. To evaluate the finite sample behavior of each estimator, we report the coverage probability of a nominal 95\% confidence interval (CP(95)), and its average length (Len(95)).
Specially, for the methods SAIC$_{F}$ and SBIC$_{F}$, we report the results with the true model $m_{o}$ being known. In practice, the model selected by BIC can be as an estimator of the real model.

\subsection{Simulation Results}
The performance of methods SAIC$_{F}$, SBIC$_{F}$, SAIC$_{97}$, SBIC$_{97}$, SAIC$_{L}$ ,SBIC$_{L}$, PAIC and FULL under Setting 1 and Setting 2 are provided in Tables \ref{tab:lin1}-\ref{tab:pois}, respectively.
Because
inference post-selection deals with the distributions of the estimators in the selected model,
we merely study the confidence intervals of the parameters included in the selected model.
For simplicity, we show only the results of these parameters for other methods.

First, let us look at the performance of confidence intervals based on the asymptotic theory of model averaging using AIC and BIC. From Tables \ref{tab:lin1}-\ref{tab:pois}, we can see that
the coverage probabilities of SAIC$_{F}$ achieve the nominal level for most parameters when the sample size is small, such as
$n=10$ for linear regression and $n=30$ for Poisson regression. As $n$ increases, the coverage probabilities of SAIC$_{F}$ achieve the nominal level for all parameters and the length of intervals are also decreasing gradually.
The changes of $\rho$ and $\sigma^{2}$ have little effect on its performance.
For the performance of SBIC$_{F}$, we can find almost the same pattern as SAIC$_{F}$ in Tables \ref{tab:lin1}-\ref{tab:pois}. As $n$ increases, the coverage probabilities of SBIC$_{F}$ achieve the nominal level for almost all parameters. However, SBIC$_{F}$ requires more samples than SAIC$_{F}$ to achieve the nominal level (see for example the results in Table \ref{tab:pois}).

Second, when $n$ is large, the coverage probabilities confidence intervals based on SAIC$_{97}$ cannot achieve the nominal level but are near the nominal level, see for example the results in Table \ref{tab:pois}.
While SBIC$_{97}$ achieve the nominal level for all parameters. This verifies our theoretical results in Subsection \ref{D:S1997}.
For Setting 1, with the sample size considered here, the coverage probabilities of SAIC$_{L}$ and SBIC$_{L}$ achieve the nominal level for all parameters, while the coverage probabilities are very low for Setting 2. The simulation results are also consistent with the theory results in Subsection \ref{D:H2003}.
For PAIC, just as the results in \cite{Charkhi2018Asymptotic}, when the sample size is small, such as $n=10$ in Tables \ref{tab:lin1} and \ref{tab:lin2}, the coverage probabilities of confidence intervals are smaller than the nominal level.
For the parameters which are truly non-zero, such as $\gamma_{2}$, the confidence intervals based on the PAIC method are conservative.

In the following, we compare the performance of the above methods.
When $n$ is large, though the coverage probabilities confidence intervals based on SAIC$_{97}$ are near the nominal level and those of SBIC$_{97}$ achieve the nominal level for all parameters, the lengths of those intervals
are wider than those of SAIC$_{F}$ and SBIC$_{F}$.
In linear regression, the performance of SAIC$_{97}$ and SBIC$_{97}$ is superior to the PAIC method and inferior to other methods when the sample size $n$ is small, such as $n=10$ in Tables \ref{tab:lin1} and \ref{tab:lin2}.
The coverage probabilities of SAIC$_{F}$ and SBIC$_{F}$ are closer to those of SAIC$_{L}$ and SBIC$_{L}$, and both achieve the nominal level. However, in Poisson regression, As explained in Subsection \ref{D:H2003}, SAIC$_{L}$ and SBIC$_{L}$ do not work. As shown in Tables \ref{tab:pois}, the coverage probabilities of SAIC$_{L}$ and SBIC$_{L}$ for all parameters are very low and the lengths of their intervals are very short. It means the methods SAIC$_{L}$ and SBIC$_{L}$ are invalid in Poisson regression when the true parameters are fixed. Fortunately, other methods perform well except the coverage probabilities of PAIC for $\gamma_{2}$ are very high.
All in all, our approaches performs best in most cases.
\section{Real-world Data Examples}\label{sec:real}

\subsection{A linear regression example}
We apply the model averaging methods to cross-country growth regressions considered in \cite{Liu2015Distribution}, which has nine regressors and a sample size of $74$ , The challenge of empirical research
on economic growth is that one does not know exactly what explanatory variables should be included in the true model. Many
studies attempt to identify the variables explaining the differences
in growth rates across countries by regressing the average growth
rate of GDP per capita on a large set of potentially relevant variables; see \cite{Durlauf2005Growth} for a literature review. We
estimate the cross-country growth regression $$y_{i}=\boldsymbol{x}_{i}^{T}\bb+\boldsymbol{z}_{i}^{T}\bg+e_{i},$$
where $y_{i}$ is average growth rate of GDP per capita between 1960
and 1996, $\boldsymbol{x}_{i}$ are the Solow variables from the neoclassical growth
theory, and $\boldsymbol{z}_{i}$ are fundamental growth determinants such as geography, institutions, religion, and ethnic fractionalization from the new fundamental growth theory. Here, $\boldsymbol{x}_{i}$ are core regressors,
which appear in every submodel, while $\boldsymbol{z}_{i}$ are the auxiliary regressors, which serve as controls of the neoclassical growth theory and may or may not be included in the submodels. We follow \cite{Liu2015Distribution} and the setting includes six core regressors and four auxiliary regressors. The six core regressors
are the constant term (CONSTANT), the log of GDP per capita in 1960 (GDP60), the 1960¨C1985 equipment investment share of GDP (EQUIPINV), the primary school enrollment rate in 1960 (SCHOOL60), the life expectancy at age zero in 1960 (LIFE60), and the population growth rate between 1960 and 1990 (DPOP). The
four auxiliary regressors are a rule of law index (LAW), a country's fraction of tropical area (TROPICS), an average index of ethnolinguistic fragmentation in a country (AVELF), and the fraction of Confucian population (CONFUC); see \cite{Magnus2010A} for a detailed description of the data. The parameter of interest is the convergence term of the Solow growth model, that is, the
coefficient of the $\log$ GDP per capita in 1960.

We consider all possible submodels, that is, we
have 16 submodels.
We calculate the confidence intervals of all the methods provided in Section \ref{sec:CI}. As shown in Table \ref{T:GDP}, the $95\%$ confidence intervals of the model averaging estimators for GDP60 are very close to that calculated by full model. But the
confidence intervals based on the PAIC method are wider than those of other methods.
Though all methods are work in this example, it is worth pointing out that
the lengths of confidence intervals using SAIC$_{F}$ and SBIC$_{F}$ for GDP60 are shorter than those of using other methods, implying that our methods are the best.
\setcounter{table}{4}
\renewcommand{\thetable}{\arabic{table}}
\begin{table}[h]
\centering  
{
\begin{tabular}{cccccccccc}  
\hline
    Methods&lower&upper&length \\
    \hline
    SAIC$_{F}$ & -0.0202  & -0.0093  & 0.0109  \\
    SAIC$_{97}$ & -0.0204  & -0.0093  & 0.0111  \\
    SAIC$_{L}$ & -0.0211  & -0.0101  & 0.0110  \\
    PAIC  & -0.0226  & -0.0086  & \textbf{0.0140}  \\
    SBIC$_{F}$ & -0.0195  & -0.0088  & 0.0107  \\
    SBIC$_{97}$ & -0.0198  & -0.0084  & 0.0114  \\
    SBIC$_{L}$ & -0.0211  & -0.0101  & 0.0110  \\
    FULL  & -0.0221  & -0.0091  & 0.0129  \\
 \hline
\end{tabular}}
\caption{95\% confidence intervals of GDP60.}
\label{T:GDP}
\end{table}
\subsection{A Poisson regression example}
In this section, we study the dataset taken from \cite{S1997Model} (Table \ref{data:Bird}) and consider a simple multiple regression problem with two correlated explanatory variables, an assumed Poisson error distribution and a log link function. As in Table \ref{data:Bird}, transect counts of singing males of the songbird Troglodytes invisibilis were made on consecutive days. Like in \cite{S1997Model}, we wish to predict future counts, given temperature and wind speed. The problem is to predict the count of on day $19$ and we also estimate the count of on day $11$.
\begin{table}[h]
  \centering
    \begin{tabular}{rrrrrrrr}
    \hline
    \multicolumn{1}{l}{Day} & \multicolumn{1}{l}{Count} & \multicolumn{1}{l}{Temperature} & \multicolumn{1}{l}{Wind speed} & \multicolumn{1}{l}{Day} & \multicolumn{1}{l}{Count} & \multicolumn{1}{l}{Temperature} & \multicolumn{1}{l}{Wind speed} \\
   \hline
    1     & 17    & 22    & 1.1   & 11    & 15    & 15    & 3.7 \\
    2     & 45    & 23    & 0.5   & 12    & 39    & 22    & 0.8 \\
    3     & 9     & 17    & 2.9   & 13    & 18    & 17    & 1.7 \\
    4     & 40    & 22    & 0.4   & 14    & 29    & 24    & 0.8 \\
    5     & 18    & 14    & 4.8   & 15    & 22    & 13    & 3.8 \\
    6     & 15    & 13    & 3.9   & 16    & 10    & 15    & 3.1 \\
    7     & 8     & 14    & 5.7   & 17    & 15    & 16    & 2.3 \\
    8     & 21    & 18    & 2.6   & 18    & 27    & 22    & 0.4 \\
    9     & 42    & 24    & 0.5   & 19    & ??    & 22    & 1.5 \\
    10    & 38    & 26    & 0.3   &       &       &       &  \\
    \hline
    \end{tabular}%
      \caption{Transect counts of a species of songbird in a study area on consecutive days. Covariates temperature ($^\circ$C ) and wind speed (m/s) were also recorded.}
 \label{data:Bird}
\end{table}%

We consider all possible submodels, that is, the models with temperature alone, with wind speed alone, and with both.
We provide the approximate 95\% confidence intervals for the expected count of day $19$ using all the methods in Table \ref{T:19d}.
As shown in Table \ref{T:19d}, the $95\%$ confidence intervals of the model averaging estimators for the expected count of day $19$, calculated by the methods in Section \ref{sec:CI}, are very close to that calculated by full model except the methods SAIC$_{L}$ and SBIC$_{L}$. The length of confidence intervals using SAIC$_{L}$ and SBIC$_{L}$ for all parameters are much shorter than that of using full model.
We do not know the true count of bird on day $19$. If we estimate the count of day 11, from Table \ref{T:11d}, we can see that
the length of confidence intervals using SAIC$_{L}$ and SBIC$_{L}$ are too short to cover the true value.
The results of this real example again show that our methods are the best.
\begin{table}[h]
\centering  
{
\begin{tabular}{cccccccccc}  
\hline
    Methods&lower&upper&length \\
\hline
    SAIC$_{F}$ & 25.2481  & 29.7033  & 4.4553  \\
    SAIC$_{97}$ & 23.4391  & 31.3837  & 7.9446  \\
    SAIC$_{L}$ & 27.0613  & 28.5527  & \textbf{1.4915}  \\
    PAIC  & 23.8067  & 28.6063  & 4.7995  \\
    SBIC$_{F}$ & 25.3888  & 29.2980  & 3.9093  \\
    SBIC$_{97}$ & 23.3354  & 31.3514  & 8.0161  \\
    SBIC$_{L}$ & 27.0613  & 28.5527  & \textbf{1.4915}  \\
    FULL  & 24.2262  & 31.3878  & 7.1616  \\
 \hline
\end{tabular}}
\caption{95\% confidence intervals of the count on day $19$.}
\label{T:19d}
\end{table}
\begin{table}[h]
\centering  
{
\begin{tabular}{cccccccccc}  
\hline
Methods&lower&upper&length \\
\hline
    SAIC$_{F}$ & 12.1263  & 16.8850  & 4.7588  \\
    SAIC$_{97}$ & 11.8865  & 17.3457  & 5.4592  \\
    SAIC$_{L}$ & 14.1297  & 14.9313  & \textbf{0.8016}  \\
    PAIC & 11.9721  & 17.2960  & 5.3240  \\
    SBIC$_{F}$ & 12.3243  & 16.9355  & 4.6112  \\
    SBIC$_{97}$ & 11.8875  & 17.3723  & 5.4848  \\
    SBIC$_{L}$ & 14.1297  & 14.9313  & \textbf{0.8016}  \\
    FULL  & 11.8843  & 17.1767  & 5.2923  \\
 \hline
\end{tabular}}
\caption{95\% confidence intervals of the count on day $11$.}
\label{T:11d}
\end{table}

\section{Concluding remarks}\label{sec:conc}
Smoothed AIC (S-AIC) and Smoothed BIC (S-BIC) are very widely used in model averaging and are very easily to implement.
\cite{S1997Model} provided an intuitive interval construction method. The resulting coverage probability of their confidence interval is not studied accurately, but it is claimed that it will be close to the intended value.
Our derivation in Section \ref{D:S1997} confirm the validity of their method.
In this paper, we study the asymptotic distributions of two commonly used model averaging
estimators, the S-AIC and S-BIC estimators, under the general fixed parameter setup.
The asymptotic distribution of the scaled S-AIC estimator is a nonlinear function of a normal random vector and
that of the scaled S-BIC estimator is a normal distribution.
We construct the confidence intervals based on these asymptotic distributions.
Besides, we also prove that the confidence interval construction method for the linear function of the parameters in \cite{hjort2003averaging} still works in linear regression with the normally distributed error. Both the simulation study and real data analysis support our methods.

It is really meaningful to derive the asymptotic distributions of model averaging estimators under general fixed parameter setup. While our results are derived based on SAIC and SBIC weights, the asymptotic theory based on other weight choice criteria may be studied under such parameter setup.
These remain for future research.

\appendix
\renewcommand{\thesection}{Appendix~\Alph{section}}
\renewcommand{\theequation}{\arabic{equation}}
\section{Proof of Lemmas and Theorem}\label{sec:A}
\subsection{Proof of Lemma \ref{lem1}.} \label{sec:lem1}
When $m\in \cal O,$ the last $q-k_{m}$ components of the true value $\btheta_{0,m}$ are $0$, and given Conditions \ref{c1}, \ref{c3} and \ref{c6}, the assumptions of Theorem 18 and Theorem 22 of \cite{Ferguson1996A} are satisfied.
Expand $l_{n}\left(\widehat{\btheta}_{m}\right)$ about $\widehat{\btheta}_{0}$:
$$l_{n}\left(\widehat{\btheta}_{m}\right)=l_{n}\left(\widehat{\btheta}_{0}\right)+\left\{\dot{l}_{n}\left(\widehat{\btheta}_{0}\right)\right\}^{T}
\left(\widehat{\btheta}_{m}-\widehat{\btheta}_{0}\right)-n\left(\widehat{\btheta}_{m}-\widehat{\btheta}_{0}\right)^{T}
I_{n}\left(\widehat{\btheta}_{m}\right)\left(\widehat{\btheta}_{m}-\widehat{\btheta}_{0}\right),$$
where $I_{n}(\widehat{\btheta}_{m})=-\frac{1}{n}\int_{0}^{1}\int_{0}^{1}v\ddot{l}_{n}
\left(\widehat{\btheta}_{0}+uv\left(\widehat{\btheta}_{m}-\widehat{\btheta}_{0}\right)\right)\mathrm{d}u\mathrm{d}v \overas\frac{1}{2}\mathcal{F}\left(\btheta_{0}\right)$, as in the proof of Theorem 18 of \cite{Ferguson1996A}.
In fact, note that $E\left\{\dot{\Psi}(y,\btheta_{m})\right\}$ is continuous in $\btheta_{m}$ from Condition \ref{c3}(a). Let $\varepsilon>0$, so there is exists $\rho>0$ such that
$|\btheta_{m}-\btheta_{0}|<3\rho$ implies
$$\left|E\left\{\dot{\Psi}(y,\btheta_{m})\right\}+\mathcal{F}\left(\btheta_{0}\right)\right|<\varepsilon.$$
Next note from the Uniform Strong Law of Large Numbers that with probability $1$ there is an integer $N$ such that $n>N$ implies
$$\sup_{\btheta_{m}\in\{\btheta_{m}:|\btheta_{m}-\btheta_{0}|<3\rho\}}
\left|\frac{1}{n}\sum_{t=1}^{n}\dot{\Psi}(y_{t},\btheta_{m})-\mathrm{E}\left\{\dot{\Psi}(y,\btheta_{0})\right\}\right|<\varepsilon.$$
Assuming $N$ is so large that $n>N$ implies $|\widehat{\btheta}_{m}-\btheta_{0}|<\rho$
and $|\widehat{\btheta}_{0}-\btheta_{0}|<\rho$, and then
$\left|\widehat{\btheta}_{0}
+uv\left(\widehat{\btheta}_{m}-\widehat{\btheta}_{0}\right)-\btheta_{0}\right|<3\rho$.
Therefore
\be
&&\left|I_{n}(\widehat{\btheta}_{m})
-\frac{1}{2}\mathcal{F}\left(\btheta_{0}\right)\right|\non\\
&\leq&\int_{0}^{1}\int_{0}^{1}v\left|\frac{1}{n}\sum_{t=1}^{n}\dot{\Psi}\left\{y_{t},\widehat{\btheta}_{0}
+uv\left(\widehat{\btheta}_{m}-\widehat{\btheta}_{0}\right)\right\}+\mathcal{F}\left(\btheta_{0}\right)\right|\mathrm{d}u\mathrm{d}v \non\\
&\leq&\int_{0}^{1}\int_{0}^{1}v\sup_{\btheta_{m}\in\{\btheta_{m}:|\btheta_{m}-\btheta_{0}|<3\rho\}}
\left\{\left|\frac{1}{n}\sum_{t=1}^{n}\dot{\Psi}\left(y_{t},\btheta_{m}\right)-\mathrm{E}\left\{\dot{\Psi}\left(y,\btheta_{0}\right)\right\}\right|\right.\nonumber\\
&&\left.+\left|\mathrm{E}\left\{\dot{\Psi}\left(y,\btheta_{m}\right)+\mathcal{F}\left(\btheta_{0}\right)\right\}\right|\right\}\mathrm{d}u\mathrm{d}v \nonumber\\
 &\leq&\varepsilon.
\ee
Let $o_{a.s.}(1)$ represent a random variable matrix with each element converging almost surely to $0$ as $n\rightarrow\infty$. By $\dot{l}_{n}\left(\widehat{\btheta}_{0}\right)=0,$ we have
\be
-2\log \lambda_{m,n}
&=&-2\left\{l_{n}\left(\widehat{\btheta}_{m}\right)-l_{n}\left(\widehat{\btheta}_{0}\right)\right\}\non\\
&=&2n\left(\widehat{\btheta}_{m}-\widehat{\btheta}_{0}\right)^{T}
I_{n}\left(\widehat{\btheta}_{m}\right)\left(\widehat{\btheta}_{m}-\widehat{\btheta}_{0}\right)\non\\
&=&2n\left\{\Pi_{m}\left(\widehat{\btheta}_{m}-\widehat{\btheta}_{0}\right)\right\}^{T}\Pi_{m}
I_{n}\left(\widehat{\btheta}_{m}\right)\Pi_{m}^{T}\Pi_{m}\left(\widehat{\btheta}_{m}-\widehat{\btheta}_{0}\right)\non\\
&=&2n\left(\widehat{\btheta}_{m}^{\prime}-\widehat{\btheta}_{0,m}\right)^{T}
I_{n}\left(\widehat{\btheta}_{m}^{\prime}\right)\left(\widehat{\btheta}_{m}^{\prime}-\widehat{\btheta}_{0,m}\right)\non\\
&=& n\left(\widehat{\btheta}_{m}^{\prime}-\widehat{\btheta}_{0,m}\right)^{T}\left\{\mathcal{F}_{m}\left(\btheta_{0,m}\right)+o_{a.s.}(1)\right\}
\left(\widehat{\btheta}_{m}^{\prime}-\widehat{\btheta}_{0,m}\right)\non.
\ee
To find the asymptotic distribution of $\sqrt{n}\left(\widehat{\btheta}_{m}^{\prime}-\widehat{\btheta}_{0,m}\right),$ expand $\dot{l}_{n}\left(\widehat{\btheta}_{m}\right)$ about $\widehat{\btheta}_{0}$:
\be
\frac{1}{\sqrt{n}}\Pi_{m}\dot{l}_{n}\left(\widehat{\btheta}_{m}\right)
&=&\frac{1}{\sqrt{n}}\Pi_{m}\dot{l}_{n}\left(\widehat{\btheta}_{0}\right)
+\frac{1}{n}\Pi_{m}\int_{0}^{1}\ddot{l}_{n}\left\{\widehat{\btheta}_{0}+v\left(\widehat{\btheta}_{m}-\widehat{\btheta}_{0}\right)\right\}\mathrm{d}v
\sqrt{n}\left(\widehat{\btheta}_{m}-\widehat{\btheta}_{0}\right)\non\\
&=& \Pi_{m}\left\{-\mathcal{F}\left(\btheta_{0}\right)+o_{a.s.}(1)\right\}\sqrt{n}\left(\widehat{\btheta}_{m}-\widehat{\btheta}_{0}\right)\non\\
&=& \Pi_{m}\left\{-\mathcal{F}\left(\btheta_{0}\right)+o_{a.s.}(1)\right\}\Pi_{m}^{T}\sqrt{n}\Pi_{m}\left(\widehat{\btheta}_{m}-\widehat{\btheta}_{0}\right)\non\\
&=& \left\{-\mathcal{F}_{m}\left(\btheta_{0,m}\right)+o_{a.s.}(1)\right\}\sqrt{n}\left(\widehat{\btheta}_{m}^{\prime}-\widehat{\btheta}_{0,m}\right).\non
\ee
Thus $\sqrt{n}\left(\widehat{\btheta}_{m}^{\prime}-\widehat{\btheta}_{0,m}\right)= -\left\{\mathcal{F}_{m}\left(\btheta_{0,m}\right)^{-1}+o_{a.s.}(1)\right\}\frac{1}{\sqrt{n}}\Pi_{m}\dot{l}_{n}\left(\widehat{\btheta}_{m}\right)$ and \be \label{1}
-2\log \lambda_{m,n}= \frac{1}{\sqrt{n}}\dot{l}_{n}\left(\widehat{\btheta}_{m}\right)^{T}\Pi_{m}^{T}\left\{\mathcal{F}_{m}\left(\btheta_{0,m}\right)^{-1}+o_{a.s.}(1)\right\}
\frac{1}{\sqrt{n}}\Pi_{m}\dot{l}_{n}\left(\widehat{\btheta}_{m}\right).
\ee
To find the asymptotic distribution of $\Pi_{m}\dot{l}_{n}\left(\widehat{\btheta}_{m}\right),$ expand about $\btheta_{0}$ and then
\be \label{2}
\frac{1}{\sqrt{n}}\Pi_{m}\dot{l}_{n}\left(\widehat{\btheta}_{m}\right)=\frac{1}{\sqrt{n}}\Pi_{m}\dot{l}_{n}\left(\btheta_{0}\right)
+\frac{1}{n}\Pi_{m}\int_{0}^{1}\ddot{l}_{n}\left(\btheta_{0}+v\left(\widehat{\btheta}_{m}-\btheta_{0}\right)\right)\mathrm{d}v
\sqrt{n}\left(\widehat{\btheta}_{m}-\btheta_{0}\right).
\ee
For the $m\th$ sub-model, partition $\mathcal{F}_{m}\left(\btheta_{0,m}\right)$ into four matrices,
\begin{center}
\begin{blockarray}{ccc}
&$k_{m}\times k_{m}$     & $k_{m}\times (q-k_{m})$\\
\begin{block}{c(cc)}
&$Q_{m,1}$               & $Q_{m,2}$ \\
$\mathcal{F}_{m}\left(\btheta_{0,m}\right)=$&~&\\
 &$Q_{m,3}$              & $Q_{m,4}$ \\
\end{block}
&$(q-k_{m})\times k_{m}$ & $(q-k_{m})\times (q-k_{m})$
\end{blockarray}~,
\end{center}
and let \begin{equation}
H_{m}=
\left(
\begin{array}{cc}
Q_{m,1}^{-1}&0\\
           0&0\non
\end{array}
\right).
\end{equation}
Note that the first $k_{m}$ components of $\Pi_{m}\dot{l}_{n}\left(\widehat{\btheta}_{m}\right)$ are zero, so that $H_{m}\Pi_{m}\dot{l}_{n}\left(\widehat{\btheta}_{m}\right)=0$ and
$$H_{m}\frac{1}{\sqrt{n}}\Pi_{m}\dot{l}_{n}\left(\btheta_{0}\right)= H_{m}\left\{\mathcal{F}_{m}\left(\btheta_{0,m}\right)+o_{a.s.}(1)\right\}\sqrt{n}\left(\widehat{\btheta}_{m}^{\prime}-\btheta_{0,m}\right)=\sqrt{n}\left(\widehat{\btheta}_{m}^{\prime}-\btheta_{0,m}\right)+o_{a.s.}(1)$$ since the last $q-k_{m}$ components of $\widehat{\btheta}_{m}^{\prime}$ and $\btheta_{0,m}$ are equal.
Substituting into (\ref{2}), we find
\be\label{11}
\frac{1}{\sqrt{n}}\Pi_{m}\dot{l}_{n}\left(\widehat{\btheta}_{m}\right)&=& \left\{I-\mathcal{F}_{m}(\btheta_{0,m})H_{m}\right\}\frac{1}{\sqrt{n}}\Pi_{m}\dot{l}_{n}(\btheta_{0})+o_{a.s.}(1)\non\\
&=&\left\{I-\Pi_{m}\mathcal{F}(\btheta_{0})\Pi_{m}^{T}H_{m}\right\}\frac{1}{\sqrt{n}}\Pi_{m}\dot{l}_{n}\left(\btheta_{0}\right)+o_{a.s.}(1)
.\ee
From the Central Limit Theorem,
$$\frac{1}{\sqrt{n}}\dot{l}_{n}\left(\btheta_{0}\right)=\sqrt{n}\left\{\frac{1}{n}\dot{l}_{n}\left(\btheta_{0}\right)\right\}=\bxi_{n}
\overd \bxi,$$ where $\bxi\sim\mathcal{N}\left(0,\mathcal{F}\left(\btheta_{0}\right)\right)$.
Hence,$$\frac{1}{\sqrt{n}}\Pi_{m}\dot{l}_{n}(\widehat{\btheta}_{m})\overd
[I-\Pi_{m}\mathcal{F}(\btheta_{0})\Pi_{m}^{T}H_{m}]\Pi_{m}\bxi, $$
so that by (\ref{1}), (\ref{11}) and $H_{m}\Pi_{m}\mathcal{F}\left(\btheta_{0}\right)\Pi_{m}^{T}H_{m}=H_{m}$,
\be
-2\log \lambda_{m,n}&=& \frac{1}{\sqrt{n}}\dot{l}_{n}\left(\btheta_{0}\right)^{T}\Pi_{m}^{T}\left\{I-\mathcal{F}_{m}(\btheta_{0,m})H_{m}\right\}^{T}
\mathcal{F}_{m}\left(\btheta_{0,m}\right)^{-1}\non\\
&&\left\{I-\mathcal{F}_{m}\left(\btheta_{0,m}\right)H_{m}\right\}\frac{1}{\sqrt{n}}\Pi_{m}\dot{l}_{n}\left(\btheta_{0}\right)+o_{a.s.}(1)\non\\
&=&\frac{1}{\sqrt{n}}\dot{l}_{n}\left(\btheta_{0}\right)^{T}\Pi_{m}^{T}\left\{\mathcal{F}_{m}\left(\btheta_{0,m}\right)^{-1}-H_{m}\right\}\frac{1}{\sqrt{n}}\Pi_{m}\dot{l}_{n}\left(\btheta_{0}\right)+o_{a.s.}(1)\non\\
&=&\frac{1}{\sqrt{n}}\left\{\Pi_{m}\dot{l}_{n}\left(\btheta_{0}\right)\right\}^{T}
\left[\left\{\Pi_{m}\mathcal{F}\left(\btheta_{0}\right)\Pi_{m}^{T}\right\}^{-1}-H_{m}\right]\frac{1}{\sqrt{n}}\Pi_{m}\dot{l}_{n}\left(\btheta_{0}\right)+o_{a.s.}(1)\non\\
&=&\left(\Pi_{m}\bxi_{n}\right)^{T}\left[\left\{\Pi_{m}\mathcal{F}\left(\btheta_{0}\right)\Pi_{m}^{T}\right\}^{-1}-H_{m}\right]\Pi_{m}\bxi_{n}+o_{a.s.}(1)\non\\
&\overd&(\Pi_{m}\bxi)^{T}\left[\left\{\Pi_{m}\mathcal{F}\left(\btheta_{0}\right)\Pi_{m}^{T}\right\}^{-1}-H_{m}\right]\Pi_{m}\bxi\non\\
&=&\bkappa^{T}\left\{\mathcal{F}\left(\btheta_{0}\right)\right\}^{\frac{1}{2}}\Pi_{m}^{T}
\left[\left\{\Pi_{m}\mathcal{F}\left(\btheta_{0}\right)\Pi_{m}^{T}\right\}^{-1}-H_{m}\right]
\Pi_{m}\left\{\mathcal{F}\left(\btheta_{0}\right)\right\}^{\frac{1}{2}}\bkappa\nonumber\\
&\equiv&\bkappa^{T}P_{m}\bkappa,
\ee
where $\bkappa=\mathcal{F}\left(\btheta_{0}\right)^{-\frac{1}{2}}\bxi\sim\mathcal{N}\left(0,I_{q\times q}\right).$ Note that
$$P_{m}=\left\{\mathcal{F}\left(\btheta_{0}\right)\right\}^{\frac{1}{2}}\Pi_{m}^{T}
\left[\left\{\Pi_{m}\mathcal{F}\left(\btheta_{0}\right)\Pi_{m}^{T}\right\}^{-1}-H_{m}\right]
\Pi_{m}\left\{\mathcal{F}\left(\btheta_{0}\right)\right\}^{\frac{1}{2}}$$
is a projection and that $\mathrm{rank}(P_{m})=\mathrm{trace}(P_{m})=q-k_{m}.$ Therefore
\be
-2\log \lambda_{m,n}\overd \bkappa^{T}P_{m}\bkappa\sim \chi^{2}\left(q-k_{m}\right)\non.
\ee
\subsection{Proof of Lemma \ref{lem2}.}\label{sec:lem2}
For any underfitted model $m\in\cal U,$ under Conditions \ref{c1},\ref{c2} and Conditions \ref{c3}(b)-\ref{c5}, the assumptions of Theorem 2.2 and Theorem 3.2
of \cite{white1982maximum} are satisfied. Then, we have
\be \label{3}
\widehat{\btheta}_{S_{m}}\overas\btheta_{S_{m}}^{\ast}
\ee and
\be
\sqrt{n}\left(\widehat{\btheta}_{S_{m}}-\btheta_{S_{m}}^{\ast}\right)\overd \mathcal{N}\left(0,C_{m}\left(\btheta_{S_{m}}^{\ast}\right)\right).
\ee
Moreover, $C_{m,n}\left(\widehat{\btheta}_{S_{m}}\right)\overas C_{m}\left(\btheta_{S_{m}}^{\ast}\right)$.
By the Taylor's Theorem, we have
\be
l_{n}\left(\widehat{\btheta}_{m}\right)&=&\sum_{t=1}^{n}\log f\left(y_{t},\widehat{\btheta}_{m}\right)\non\\
&=&l_{n}\left(\btheta_{m}^{\ast}\right)+\sum_{t=1}^{n}\Psi\left(y_{t},\btheta_{m}^{\ast}\right)
\left(\widehat{\btheta}_{m}-\btheta_{m}^{\ast}\right)\nonumber\\
&&+\frac{n}{2}\left(\widehat{\btheta}_{m}-\btheta_{m}^{\ast}\right)^{T}
A_{n}\left(\widehat{\btheta}_{m}+\alpha\left(\btheta_{m}^{\ast}-\widehat{\btheta}_{m}\right)\right)
\left(\widehat{\btheta}_{m}-\btheta_{m}^{\ast}\right)\non\\
&=&l_{n}\left(\btheta_{m}^{\ast}\right)+\sum_{t=1}^{n}\Psi_{m}\left(y_{t},\btheta_{S_{m}}^{\ast}\right)
\left(\widehat{\btheta}_{S_{m}}-\btheta_{S_{m}}^{\ast}\right)\nonumber\\
&&+\frac{n}{2}\left(\widehat{\btheta}_{S_{m}}-\btheta_{S_{m}}^{\ast}\right)^{T}
A_{m,n}\left(\widehat{\btheta}_{S_{m}}+\alpha\left(\btheta_{S_{m}}^{\ast}-\widehat{\btheta}_{S_{m}}\right)\right)
\left(\widehat{\btheta}_{S_{m}}-\btheta_{S_{m}}^{\ast}\right),
\ee
where $\alpha\in(0,1).$
In fact, given Conditions \ref{c1}-\ref{c2} and \ref{c3}(b)-\ref{c5}, from the proof of Theorem 3.2 of \cite{white1982maximum}, we have $\mathrm{E}\left\{\Psi_{m}\left(y_{t},\btheta_{S_{m}}^{\ast}\right)\right\}=0$, and by Laws of Large Numbers, we have \be
n^{-1}\sum_{t=1}^{n}\Psi_{m}\left(y_{t},\btheta_{S_{m}}^{\ast}\right)\overp\mathrm{E}\left\{\Psi_{m}\left(y_{t},\btheta_{S_{m}}^{\ast}\right)\right\},
\ee
and
\be
n^{-1}l_{n}\left(\btheta_{m}^{\ast}\right)=n^{-1}\sum_{t=1}^{n}\log f\left(y_{t},\btheta_{m}^{\ast}\right)\overp \mathrm{E}\left\{\log f\left(y_{t},\btheta_{m}^{\ast}\right)\right\}.
\ee
We also have \be \label{4}
A_{m,n}\left(\widehat{\btheta}_{S_{m}}+\alpha\left(\btheta_{S_{m}}^{\ast}-\widehat{\btheta}_{S_{m}}\right)\right)\overas A_{m}\left(\btheta_{S_{m}}^{\ast}\right)
\ee
by Theorem 2.2 of \cite{white1982maximum}. Then, using results \eqref{3}-\eqref{4}, we can conclude that
$$
n^{-1}l_{n}\left(\widehat{\btheta}_{m}\right)
=\mathrm{E}\left\{\log f\left(y_{t},\btheta_{m}^{\ast}\right)\right\}+o_{p}(1).
$$

\subsection{Proof of Lemma \ref{lem3}.}\label{sec:lem3}
From Lemma \ref{lem1}, when $m\in \cal O$, we have
\be
-2\log \lambda_{m,n}
&=&-2\left\{l_{n}\left(\widehat{\btheta}_{m}\right)-l_{n}\left(\widehat{\btheta}_{0}\right)\right\}\non\\
&=&\left(\Pi_{m}\bxi_{n}\right)^{T}\left[\left\{\Pi_{m}\mathcal{F}\left(\btheta_{0}\right)\Pi_{m}^{T}\right\}^{-1}-H_{m}\right]\Pi_{m}\bxi_{n}+o_{a.s.}(1).
\ee
Then
\be \label{8}
&&\widehat w_{\aic,m}\widehat w^{-1}_{\aic,m_{o}}=\exp\left(\AIC_{m_{o}}/2-\AIC_m/2\right)\non\\
&=&\exp\left\{-\sum_{t=1}^{n}\log f\left(y_{t},\widehat{\btheta}_{m_{o}}\right)+\sum_{t=1}^{n}\log f\left(y_{t},\widehat{\btheta}_{m}\right)+k_{m_{o}}-k_m\right\}\non\\
&=&\exp\left[-\log\left\{\prod_{t=1}^{n} f\left(y_{t},\widehat{\btheta}_{m_{o}}\right)\bigg/\prod_{t=1}^{n} f\left(y_{t},\widehat{\btheta}_{o}\right)\right\}\right.\non\\
&&\left.~~~~~~~~~+\log\left\{\prod_{t=1}^{n} f\left(y_{t},\widehat{\btheta}_{m}\right)\bigg/\prod_{t=1}^{n} f\left(y_{t},\widehat{\btheta}_{o}\right)\right\}+k_{m_{o}}-k_m\right]\non\\
&=&\exp\left\{\left(-2\log\lambda_{m_{o},n}+2\log\lambda_{m,n}\right)\big/2+k_{m_{o}}-k_{m}\right\}\non\\
&=&\exp\left\{\left(\left(\Pi_{m_{o}}\bxi_{n}\right)^{T}\left[\left\{\Pi_{m_{o}}
\mathcal{F}\left(\btheta_{0}\right)\Pi_{m_{o}}^{T}\right\}^{-1}-H_{m_{o}}\right]
\Pi_{m_{o}}\bxi_{n}\right.\right.\non\\
&&\left.\left.~~-\left(\Pi_{m}\bxi_{n}\right)^{T}\left[\left\{\Pi_{m}\mathcal{F}\left(\btheta_{0}\right)
\Pi_{m}^{T}\right\}^{-1}-H_{m}\right]\Pi_{m}\bxi_{n}
+o_{a.s.}(1)\right)\big/2+k_{m_{o}}-k_{m}\right\}.
\ee
Therefore, we can conclude that
\begin{eqnarray}\label{eq:waicsum}
\widehat w_{\aic,m}&=&\widehat w_{\aic,m}\widehat w^{-1}_{\aic,m_{o}}/\sum_{m=1}^{M}\widehat w_{\aic,m}\widehat w^{-1}_{\aic,m_{o}}\nonumber\\
&=&\exp\left\{\left(\left(\Pi_{m_{o}}\bxi_{n}\right)^{T}\left[\left\{\Pi_{m_{o}}
\mathcal{F}\left(\btheta_{0}\right)\Pi_{m_{o}}^{T}\right\}^{-1}-H_{m_{o}}\right]
\Pi_{m_{o}}\bxi_{n}-\left(\Pi_{m}\bxi_{n}\right)^{T}\right.\right.\non\\
&&~~\left.\left.
\left[\left\{\Pi_{m}\mathcal{F}\left(\btheta_{0}\right)
\Pi_{m}^{T}\right\}^{-1}-H_{m}\right]\Pi_{m}\bxi_{n}+o_{a.s.}(1)\right)/2+k_{m_{o}}-k_{m}\right\}\non\\
&&~~\bigg/
\left[\sum_{m\in\calO}\exp\left\{\left(\left(\Pi_{m_{o}}\bxi_{n}\right)^{T}\left[\left\{\Pi_{m_{o}}
\mathcal{F}\left(\btheta_{0}\right)\Pi_{m_{o}}^{T}\right\}^{-1}-H_{m_{o}}\right]
\Pi_{m_{o}}\bxi_{n}\right.\right.\right.\nonumber\\
&&~~\left.\left.\left.~~-\left(\Pi_{m}\bxi_{n}\right)^{T}\left[\left\{\Pi_{m}\mathcal{F}\left(\btheta_{0}\right)
\Pi_{m}^{T}\right\}^{-1}-H_{m}\right]\Pi_{m}\bxi_{n}
+o_{a.s.}(1)\right)/2+k_{m_{o}}-k_{m}\right\}\right.\nonumber\\
&&~~\left.+\sum_{m\in\calU}\widehat w_{\aic,m}\widehat w^{-1}_{\aic,m_{o}}\right].
\end{eqnarray}
However, when $m\in \cal U$, from Lemma \ref{lem2}, we have
\be \label{5}
n^{-1}l_{n}\left(\widehat{\btheta}_{m}\right)
=\mathrm{E}\left\{\log f\left(y_{t},\btheta_{m}^{\ast}\right)\right\}+o_{p}(1).
\ee
Similarly, from the proof of Lemma \ref{lem2}, we can prove that
\be
n^{-1}l_{n}\left(\widehat{\btheta}_{m_{o}}\right)
=\mathrm{E}\left\{\log f\left(y_{t},\btheta_{0}\right)\right\}+o_{p}(1).
\ee
By $\btheta_{0}\neq\btheta_{m}^{\ast}$ and the definition of KLIC,
we conclude that there is $\delta_{m}>0$ so that
\be \label{6}
\mathrm{E}\left\{\log f\left(y_{t},\btheta_{0}\right)\right\}-\mathrm{E}\left\{\log f\left(y_{t},\btheta_{m}^{\ast}\right)\right\}=\delta_{m}.
\ee
Thus, when $m\in\cal U,$ by (\ref{5})-(\ref{6}), we have
\be \label{7}
&&\widehat w_{\aic,m}\widehat w^{-1}_{\aic,m_{o}}=\exp\left(\AIC_{m_{o}}/2-\AIC_m/2\right)\non\\
&=&\exp\left\{-\sum_{t=1}^{n}\log f\left(y_{t},\widehat{\btheta}_{m_{o}}\right)+\sum_{t=1}^{n}\log f\left(y_{t},\widehat{\btheta}_{m}\right)+k_{m_{o}}-k_m\right\}\non\\
&=&\exp\left[-n\left\{n^{-1}\sum_{t=1}^{n}\log f\left(y_{t},\widehat{\btheta}_{m_{o}}\right)-n^{-1}\sum_{t=1}^{n}\log f\left(y_{t},\widehat{\btheta}_{m}\right)\right\}+k_{m_{o}}-k_m\right]\non\\
&=&\exp\left[-n\left\{n^{-1}l_{n}\left(\widehat{\btheta}_{m_{o}}\right)
-n^{-1}l_{n}\left(\widehat{\btheta}_{m}\right)\right\}+k_{m_{o}}-k_m\right]\non\\
&=&\left\{\exp(-n)\right\}^{\left\{\delta_{m}+o_{p}(1)+(k_{m_{o}}-k_m)/n\right\}}\non\\
&=&O_{p}\left(e^{-n\delta_{m}}\right)\overp 0\ \ as\ n\rightarrow \infty.
\ee
Since $\boldsymbol{\widehat{w}}_{\aic}\in \mathcal{W},$ (\ref{7}) implies that
\be\label{ord:waics}
\widehat {w}_{\aic,m}=O_{p}\left(e^{-n\delta_{m}}\right)\ee and then when $m\in \cal U$,
\begin{eqnarray}\label{eq:waics}
\widehat {w}_{\aic,m}\overp 0.
\end{eqnarray}
Define $\delta=\min_{m\in \cal U}\delta_{m}$, thus we can conclude that
$$\sum_{m\in\calU}\widehat w_{\aic,m}\widehat w^{-1}_{\aic,m_{o}}=O_{p}\left(e^{-n\delta}\right)=o_{p}(1).$$
By \eqref{eq:waicsum}, when $m\in \cal O$,
\begin{eqnarray}\label{eq:waicm}
\widehat w_{\aic,m}&=&\widehat w_{\aic,m}\widehat w^{-1}_{\aic,m_{o}}\big/\sum_{m=1}^{M}\widehat w_{\aic,m}\widehat w^{-1}_{\aic,m_{o}}\nonumber\\
&=&\exp\left\{\left(\left(\Pi_{m_{o}}\bxi_{n}\right)^{T}\left[\left\{\Pi_{m_{o}}
\mathcal{F}\left(\btheta_{0}\right)\Pi_{m_{o}}^{T}\right\}^{-1}-H_{m_{o}}\right]
\Pi_{m_{o}}\bxi_{n}-\left(\Pi_{m}\bxi_{n}\right)^{T}\right.\right.\non\\
&&~~\left.\left.
\left[\left\{\Pi_{m}\mathcal{F}\left(\btheta_{0}\right)
\Pi_{m}^{T}\right\}^{-1}-H_{m}\right]\Pi_{m}\bxi_{n}+o_{a.s.}(1)\right)/2+k_{m_{o}}-k_{m}\right\}\non\\
&&~~\bigg/
\left[\sum_{m\in\calO}\exp\left\{\left(\left(\Pi_{m_{o}}\bxi_{n}\right)^{T}\left[\left\{\Pi_{m_{o}}
\mathcal{F}\left(\btheta_{0}\right)\Pi_{m_{o}}^{T}\right\}^{-1}-H_{m_{o}}\right]
\Pi_{m_{o}}\bxi_{n}\right.\right.\right.\nonumber\\
&&~~\left.\left.\left.~~-\left(\Pi_{m}\bxi_{n}\right)^{T}\left[\left\{\Pi_{m}\mathcal{F}\left(\btheta_{0}\right)
\Pi_{m}^{T}\right\}^{-1}-H_{m}\right]\Pi_{m}\bxi_{n}
+o_{a.s.}(1)\right)/2+k_{m_{o}}-k_{m}\right\}+o_{p}(1)\right]\nonumber\\
&=&\exp\left\{(\Pi_{m}\bxi_{n})^{T}\left[H_{m}-\left\{\Pi_{m}\mathcal{F}\left(\btheta_{0}\right)
\Pi_{m}^{T}\right\}^{-1}\right]\Pi_{m}\bxi_{n}/2-k_{m}+o_{a.s.}(1)\right\}\non\\
&&\bigg/\left\{\sum_{m\in\calO}\exp\left((\Pi_{m}\bxi_{n})^{T}\left[H_{m}-\left\{\Pi_{m}\mathcal{F}\left(\btheta_{0}\right)
\Pi_{m}^{T}\right\}^{-1}\right]\Pi_{m}\bxi_{n}/2-k_{m}+o_{p}(1)\right)\right\}\nonumber\\
&\overd&\exp\left\{\left(\Pi_{m}\bxi\right)^{T}\left[H_{m}-\left\{\Pi_{m}\mathcal{F}\left(\btheta_{0}\right)
\Pi_{m}^{T}\right\}^{-1}\right]\Pi_{m}\bxi/2-k_{m}\right\}\non\\
&&\bigg/\left\{\sum_{m\in\calO}\exp\left\{(\Pi_{m}\bxi)^{T}\left[H_{m}-\left\{\Pi_{m}\mathcal{F}\left(\btheta_{0}\right)
\Pi_{m}^{T}\right\}^{-1}\right]\Pi_{m}\bxi/2-k_{m}\right\}\right\}\nonumber\\
&=&\exp\left\{\left\{\Pi_{m}\mathcal{F}(\btheta_{0})\bbeta\right\}^{T}\left[H_{m}-\left\{\Pi_{m}\mathcal{F}\left(\btheta_{0}\right)
\Pi_{m}^{T}\right\}^{-1}\right]\Pi_{m}\mathcal{F}(\btheta_{0})\bbeta/2-k_{m}\right\}\non\\
&&~~\bigg/\left\{\sum_{m\in\calO}\exp\left\{\left\{\Pi_{m}\mathcal{F}(\btheta_{0})\bbeta\right\}^{T}\left[H_{m}-\left\{\Pi_{m}\mathcal{F}\left(\btheta_{0}\right)
\Pi_{m}^{T}\right\}^{-1}\right]\Pi_{m}\mathcal{F}(\btheta_{0})\bbeta/2-k_{m}\right\}\right\}\non\\
&=&G_{m}/\sum_{m\in\calO}G_{m}=G_{m}/G,
\end{eqnarray}
where   $$G_{m}=\exp\left\{\left\{\Pi_{m}\mathcal{F}(\btheta_{0})\bbeta\right\}^{T}\left[H_{m}-\left\{\Pi_{m}\mathcal{F}\left(\btheta_{0}\right)
\Pi_{m}^{T}\right\}^{-1}\right]\Pi_{m}\mathcal{F}(\btheta_{0})\bbeta/2-k_{m}\right\},$$
$\bbeta\sim \mathcal{N}\left(0,\mathcal{F}^{-1}\left(\btheta_{0}\right)\right)$ and $G=\sum_{m\in\calO}G_{m}$.

Combining \eqref{eq:waics} and \eqref{eq:waicm}, by Theorem 6 in \cite{Ferguson1996A},
we have $$\widehat{\boldsymbol{w}}_\aic\overd\boldsymbol{w}_\aic,$$
where $\boldsymbol{w}_\aic$ is a $M\times1$ vector with the $m^{th}$ element $w_{\aic,m}=\boldsymbol{1}\{m\in \calO\}G_m/\sum_{m\in\calO}G_m$.

Similarly, we can obtain that for any overfitted model $m$ and $m\neq m_{o}$,
\be\label{eq:wbicm}
&&\widehat w_{\bic,m}\widehat w^{-1}_{\bic,m_{o}}
=O_{p}(1)\cdot n^{\left(k_{m_{o}}-k_m\right)/2}\overp 0, \non
\ee
and that for any underfitted model $m\in\calU$,
\be\label{eq:wbics}
\widehat{w}_{\bic,m}\widehat w^{-1}_{\bic,m_{o}}
&=&\left\{\exp(-n)\right\}^{n^{-1}l_{n}\left(\widehat{\btheta}_{m_{o}}\right)-n^{-1}l_{n}\left(\widehat{\btheta}_{m}\right)}n^{(k_{m_{o}}-k_m)/2}\nonumber\\ &=&\left\{\exp(-n)\right\}^{\left\{\delta_{m}+o_{p}(1)\right\}}n^{(k_{m_{o}}-k_m)/2}\non\\
&=&\left\{\exp(-n)\right\}^{\left\{\delta_{m}+o_{p}(1)\right\}}n^{(k_{m_{o}}-k_m)/2}\non\\
&=&O_{p}\left(e^{-n\delta_{m}}\right)=o_{p}(1).
\ee
From the above two formulas, we have $\boldsymbol{w}_\bic \overp \boldsymbol{w}_{m_o}^o$.
\vskip 0.6cm
\subsection{Proof of Lemma \ref{lem4}.}\label{sec:lem4}
Since $\bTheta$ is a compact subset of $\mathbb{R}^{q}$ and by Theorem 2.2 and Theorem 3.2 of \cite{white1982maximum}, we can conclude that for any sub-model $m$,
\be \label{9}
\btheta_{m}^{\ast}=O(1),\ \btheta_{0}=O(1),\
 \widehat{\btheta}_{m}\overas\btheta_{m}^{\ast},
\ee
\be \label{10}
and \ \sqrt{n}\left(\widehat{\btheta}_{S_{m}}-\btheta_{S_{m}}^{\ast}\right)
=-A_{m}^{-1}\left(\btheta_{S_{m}}^{\ast}\right)\frac{1}{\sqrt{n}}\sum_{t=1}^{n}\frac{\partial\log f\left(y_{t},\btheta_{m}^{\ast}\right)}{\partial\btheta_{S_{m}}}+o_{p}(1).
\ee
When $m\in\calO$, we have $\btheta_{S_{m}}^{\ast}=\btheta_{0,S_{m}}$ and $\btheta_{m}^{\ast}=\btheta_{0}.$
Then, by Theorem 3.3 of \cite{white1982maximum}, we have $-A_{m}^{-1}\left(\btheta_{S_{m}}^{\ast}\right)=Q^{-1}_{m,1}$. Further,
\be
\sqrt{n}\left(\widehat{\btheta}_{S_{m}}-\btheta_{S_{m}}^{\ast}\right)
=Q^{-1}_{m,1}\sqrt{n}\left\{\frac{1}{n}\sum_{t=1}^{n}\frac{\partial\log f\left(y_{t},\btheta_{m}^{\ast}\right)}{\partial\btheta_{S_{m}}}\right\}+o_{p}(1).
\ee
From the proof of Lemma \ref{lem1}, we conclude that $$\sqrt{n}\left\{\frac{1}{n}\sum_{t=1}^{n}\frac{\partial\log f\left(y_{t},\btheta_{m}^{\ast}\right)}{\partial\btheta_{S_{m}}}\right\}=\bxi_{m,n}\overd\boldsymbol{\zeta}_{m},$$ where $\boldsymbol{\zeta}_{m}\sim \mathcal{N}\left(\boldsymbol0,Q_{m,1}\right)$. And by \eqref{ord:waics} in Lemma \ref{lem3},
Then we have \be
&&\sqrt{n}\left\{\widehat{\btheta}\left(\boldsymbol{\widehat{w}}_{\aic}\right)-\btheta_{0}\right\}
  =\sum_{m=1}^{M}\widehat{w}_{\aic,m}\sqrt{n}\left(\widehat{\btheta}_{m}-\btheta_{0}\right)\non\\
&=&\sum_{m\in\calU}\widehat{w}_{\aic,m}\sqrt{n}\left(\widehat{\btheta}_{m}-\btheta_{m}^{\ast}+\btheta_{m}^{\ast}-\btheta_{0}\right)
+\sum_{m\in\calO}\widehat{w}_{\aic,m}\sqrt{n}\left(\widehat{\btheta}_{m}-\btheta_{0}\right)\non\\
&=&\sum_{m\in\calU}O_{p}\left(e^{-n\delta_{m}}\right)\sqrt{n}O_{p}(1)+\sum_{m\in\calO}\widehat{w}_{\aic,m}\sqrt{n}\left(\widehat{\btheta}_{m}
-\btheta_{0}\right)\non\\
&=&\sum_{m\in\calO}\widehat{w}_{\aic,m}\sqrt{n}\left(\widehat{\btheta}_{m}-\btheta_{0}\right)+O_{p}\left(e^{-n\delta}\right)\non\\
&=&\sum_{m\in\calO}\widehat{w}_{\aic,m}\sqrt{n}\left(\widehat{\btheta}_{m}-\btheta_{m}^{\ast}\right)+o_{p}(1)\non\\
&=&\sum_{m\in\calO}\widehat{w}_{\aic,m}\sqrt{n}\Pi_{m}^{-1}\left(\widehat{\btheta}_{m}^{\prime}
-\btheta_{m}^{\prime}\right)+o_{p}(1)\non\\
&=&\sum_{m\in\calO}\widehat{w}_{\aic,m}\sqrt{n}\Pi_{m}^{T}\left\{\left(\widehat{\btheta}_{S_{m}}^{T},\boldsymbol{0}^{T}\right)^{T}
-\left(\btheta_{S_{m}}^{\ast T},\boldsymbol{0}^{T}\right)^{T}\right\}+o_{p}(1)\non\\
&=&\sum_{m\in\calO}\widehat{w}_{\aic,m}\sqrt{n}\Pi_{m}^{T}\left(\left(\widehat{\btheta}_{S_{m}}
-\btheta_{S_{m}}^{\ast}\right)^{T},\boldsymbol{0}_{q-k_{m}}^{T}\right)^{T}+o_{p}(1)\non\\
&=&\sum_{m\in\calO}\widehat{w}_{\aic,m}\Pi_{m}^{T}\left(
\begin{array}{cc}
G^{-1}_{S_{m},1}&0\\
           0&0\non
\end{array}
\right)
\left(
\begin{array}{c}
\frac{1}{\sqrt{n}}\sum_{t=1}^{n}\frac{\partial\log f\left(y_{t},\btheta_{m}^{\ast}\right)}{\partial\btheta_{S_{m}}}\\
           0\non
\end{array}
\right)+o_{p}(1)\non\\
&=&\sum_{m\in\calO}\widehat{w}_{\aic,m}\Pi_{m}^{T}H_{S_{m}}\Pi_{m}\Delta_{m}\bxi_{n}+o_{p}(1).
\ee
By \eqref{eq:waicm}, note that $\widehat{w}_{\aic,m}$ is also the function of random vector $\bxi_{n}$.
Then we can obtain that
\be
&&\sqrt{n}\left\{\widehat{\btheta}\left(\boldsymbol{\widehat{w}}_{\aic}\right)-\btheta_{0}\right\}\non\\
&=&\sum_{m\in\calO}\widehat{w}_{\aic,m}\Pi_{m}^{T}H_{S_{m}}\Pi_{m}\Delta_{m}\bxi_{n}+o_{p}(1)\non\\
&\overd&\sum_{m\in\calO}\left(G_{m}/G\right)\Pi_{m}^{T}H_{S_{m}}\Pi_{m}\Delta_{m}\mathcal{F}\left(\btheta_{0}\right)\bbeta.
\ee
Similarly, we can prove that
\be
&&\sqrt{n}\left\{\widehat{\btheta}\left(\boldsymbol{\widehat{w}}_{\bic}\right)-\btheta_{0}\right\}\non\\ &=&\sum_{m=1}^{M}\widehat{w}_{\bic,m}\sqrt{n}\left(\widehat{\btheta}_{m}-\btheta_{0}\right)\non\\
&=&\sum_{m\in\calU}\widehat{w}_{\bic,m}\sqrt{n}\left(\widehat{\btheta}_{m}-\btheta_{0}\right)+\sum_{m\in\calO}\widehat{w}_{\bic,m}\sqrt{n}\left(\widehat{\btheta}_{m}-\btheta_{0}\right)\non\\
&=&\sum_{m\in\calU}O_{p}\left(e^{-n\delta_{m}}\right)\sqrt{n}O_{p}(1)+\sum_{m\in\calO}\widehat{w}_{\bic,m}\sqrt{n}\left(\widehat{\btheta}_{m}-\btheta_{0}\right)\non\\
&=&O_{p}\left(e^{-n\delta}\right)+\sum_{m\in\calO}\widehat{w}_{\bic,m}\sqrt{n}\left(\widehat{\btheta}_{m}-\btheta_{0}\right)\non\\
&=&\sum_{m\in\calO}\widehat{w}_{\bic,m}\sqrt{n}\left(\widehat{\btheta}_{m}-\btheta_{0}\right)+o_{p}(1)\non\\
&\overd&
\Pi_{m_{o}}^{T}H_{S_{m_{o}}}\Pi_{m_{o}}\Delta_{m_{o}}\mathcal{F}(\btheta_{0})\bbeta,
\ee
where $$\Pi_{m_{o}}^{T}H_{S_{m_{o}}}\Pi_{m_{o}}\Delta_{m_{o}}\mathcal{F}(\btheta_{0})\bbeta\sim \mathcal{N}\left(0,\Sigma_{m_{o}}\right)$$ with $$\Sigma_{m_{o}}=\Pi_{m_{o}}^{T}H_{S_{m_{o}}}\Pi_{m_{o}}\Delta_{m_{o}}\mathcal{F}\left(\btheta_{0}\right)\left(\Pi_{m_{o}}^{T}H_{S_{m_{o}}}\Pi_{m_{o}}\Delta_{m_{o}}\right)^{T}.$$
By the definition of $\Delta_{m_{o}}$
and because the $m_{o}^{th}$ candidate model is the true model, when $j\not\in S_{m_{o}}$, it means $\theta_{j,0}=0$. Suppose $\sigma_{j,m_{o}}^{2}$ is the $j^{th}$ element on the diagonal of matrix $\Sigma_{m_{o}}$, then $\sigma_{j,m_{o}}^{2}=0$. We can conclude that $\widehat{\theta}_{j}\left(\boldsymbol{\widehat{w}}_{\bic}\right)\overp \theta_{j,0}=0$.
On the other hand, for $j\in S_{m_{o}}$, $$\sqrt{n}\left\{\widehat{\theta}_{j}\left(\boldsymbol{\widehat{w}}_{\bic}\right)-\theta_{j,0}\right\}\overd Z_{j},$$
where $Z_{j}\sim \mathcal{N}\left(0,\sigma_{j,m_{o}}^{2}\right)$.
\subsection{Proof of Theorem \ref{th2}.}\label{sec:th2}
From the proof of Lemma \ref{lem4}, for any sub-model $m$,
\be
\btheta_{m}^{\ast}=O(1),\ \btheta_{0}=O(1),\
 \widehat{\btheta}_{m}\overas\btheta_{m}^{\ast}.
\ee
$\mu\left(\btheta\right)$ is continuous so that we have
\be
 \mu\left(\widehat{\btheta}_{m}\right)\overas\mu\left(\btheta_{m}^{\ast}\right).
\ee

When $m\in\calU$, from the proof of Lemma \ref{lem3}, we conclude that
\be
\widehat{w}_{\aic,m}=O_{p}\left(e^{-n\delta_{m}}\right)\ \ \text{and}\ \ \widehat{w}_{\bic,m}=O_{p}\left(e^{-n\delta_{m}}\right).
\ee
Then we have
\be\label{eq:waicmuU}
&&\sum_{m\in\calU}\widehat{w}_{\aic,m}\sn\left\{\mu\left(\widehat{\btheta}_{m}\right)-\mu\left(\btheta_{0}\right)\right\}\nonumber\\
&=&\sum_{m\in\calU}\widehat{w}_{\aic,m}\sn\left\{\mu\left(\widehat{\btheta}_{m}\right)-\mu\left(\btheta_{m}^{*}\right)+\mu\left(\btheta_{m}^{*}\right)-\mu\left(\btheta_{0}\right)\right\}\nonumber\\
&=&\sum_{m\in\calU}O_{p}\left(e^{-n\delta_{m}}\right)\sn O_{p}(1)\nonumber\\
&=&O_{p}\left(e^{-n\delta}\right)=o_{p}(1).
\ee

When $m\in\calO$,  from the proof of Lemma \ref{lem4}, we have
\begin{eqnarray}
\widehat w_{\aic,m}
&=&\exp\left\{(\Pi_{m}\bxi_{n})^{T}\left[H_{m}-\left\{\Pi_{m}\mathcal{F}\left(\btheta_{0}\right)
\Pi_{m}^{T}\right\}^{-1}\right]\Pi_{m}\bxi_{n}/2-k_{m}+o_{a.s.}(1)\right\}\non\\
&&
\bigg/\left\{\sum_{m\in\calO}\exp\left\{(\Pi_{m}\bxi_{n})^{T}
\left[H_{m}-\left\{\Pi_{m}\mathcal{F}\left(\btheta_{0}\right)
\Pi_{m}^{T}\right\}^{-1}\right]\Pi_{m}\bxi_{n}/2-k_{m}+o_{a.s.}(1)\right\}+o_{p}(1)\right\}\nonumber\\
\ee and
\be
\sqrt{n}\left(\widehat{\btheta}_{m}-\btheta_{0}\right)=\Pi_{m}^{T}H_{S_{m}}\Pi_{m}\Delta_{m}\bxi_{n}+o_{p}(1).
\ee
By the theorem's proof in Subsection 3.1 in \cite{Vaart2000Asymptotic},
\be
&&\sn\left\{\mu\left(\widehat{\btheta}_{m}\right)-\mu\left(\btheta_{0}\right)\right\}\nonumber\\
&=&\dot{\mu}\left(\btheta_{0}\right)^{T}\sqrt{n}\left(\widehat{\btheta}_{m}-\btheta_{0}\right)+o_{p}(1)\nonumber\\
&=&\dot{\mu}\left(\btheta_{0}\right)^{T}\Pi_{m}^{T}H_{S_{m}}\Pi_{m}\Delta_{m}\bxi_{n}+o_{p}(1).
\ee
Therefore,
\be\label{eq:waicmuO}
&&\sum_{m\in\calO}\widehat{w}_{\aic,m}\sn\left\{\mu\left(\widehat{\btheta}_{m}\right)-\mu\left(\btheta_{0}\right)\right\}\nonumber\\
&\overd&\sum_{m\in\calO}\left(G_{m}/G\right)\dot{\mu}(\btheta_{0})^{T}\Pi_{m}^{T}H_{S_{m}}\Pi_{m}\Delta_{m}\mathcal{F}(\btheta_{0})\bbeta.
\ee

Combing \eqref{eq:waicmuU} and \eqref{eq:waicmuO}, we can draw the following conclusion,
\be
&&\sn\left\{\widehat{\mu}\left(\widehat{\boldsymbol{w}}_{\aic}\right)-\mu\left(\btheta_{0}\right)\right\}\nonumber\\
&=&\sum_{m=1}^{M}\widehat{w}_{\aic,m}\sn\left\{\mu\left(\widehat{\btheta}_{m}\right)-\mu\left(\btheta_{0}\right)\right\}\nonumber\\
&\overd& \sum_{m\in\calO}\left(G_{m}/G\right)\dot{\mu}(\btheta_{0})^{T}\Pi_{m}^{T}H_{S_{m}}\Pi_{m}\Delta_{m}\mathcal{F}(\btheta_{0})\bbeta.\ee
Naturally, by the similar proof technique, we have
\be \sn\left\{\widehat{\mu}\left(\widehat{\boldsymbol{w}}_{\bic}\right)-\mu(\btheta_{0})\right\}
&=&\sum_{m\in\calO}\widehat{w}_{\bic,m}\dot{\mu}(\btheta_{0})^{T}\sqrt{n}\left(\widehat{\btheta}_{m}-\btheta_{0}\right)+o_{p}(1)\non\\
&\overd&
\dot{\mu}(\btheta_{0})^{T}\Pi_{m_{o}}^{T}H_{S_{m_{o}}}\Pi_{m_{o}}\Delta_{m_{o}}\mathcal{F}(\btheta_{0})\bbeta.
\ee

\bibliographystyle{plainnat}
\bibliography{SAIC-BIC}
\setcounter{figure}{0}
\setcounter{table}{0}
\renewcommand{\thetable}{\arabic{table}}
\begin{table}[b]
\centering  
\scalebox{0.9}{
    \begin{tabular}{ccccccccccc}
    \toprule
          &       & \multicolumn{3}{c}{$\sigma=0.5$} & \multicolumn{3}{c}{$\sigma=1$} & \multicolumn{3}{c}{$\sigma=1.5$} \\
          & \multicolumn{1}{c}{Methods} & \multicolumn{1}{c}{$\beta_{0}$} & \multicolumn{1}{c}{$\beta_{1}$} & \multicolumn{1}{c}{$\mu$} & \multicolumn{1}{c}{$\beta_{0}$} & \multicolumn{1}{c}{$\beta_{1}$} & \multicolumn{1}{c}{$\mu$} & \multicolumn{1}{c}{$\beta_{0}$} & \multicolumn{1}{c}{$\beta_{1}$} & \multicolumn{1}{c}{$\mu$} \\
    \midrule
    \multirow{8}[2]{*}{CP(95)} & SAIC$_{F}$ & 0.919  & 0.918  & 0.929  & 0.916  & 0.909  & 0.919  & 0.899  & 0.947  & 0.919  \\
          & SAIC$_{97}$ & 0.884  & 0.888  & 0.918  & 0.872  & 0.872  & 0.913  & 0.870  & 0.909  & 0.916  \\
          & SAIC$_{L}$ & 0.931  & 0.946  & 0.950  & 0.930  & 0.937  & 0.944  & 0.916  & 0.959  & 0.949  \\
          & PAIC  & 0.800  & 0.788  & 0.835  & 0.749  & 0.748  & 0.814  & 0.728  & 0.763  & 0.793  \\
          & SBIC$_{F}$ & 0.902  & 0.892  & 0.856  & 0.897  & 0.863  & 0.839  & 0.876  & 0.918  & 0.841  \\
          & SBIC$_{97}$ & 0.883  & 0.890  & 0.918  & 0.872  & 0.873  & 0.912  & 0.872  & 0.913  & 0.914  \\
          & SBIC$_{L}$ & 0.931  & 0.946  & 0.950  & 0.930  & 0.937  & 0.944  & 0.916  & 0.959  & 0.949  \\
          & FULL  & 0.893  & 0.891  & 0.898  & 0.886  & 0.883  & 0.892  & 0.866  & 0.904  & 0.895  \\
    \midrule
    \multirow{8}[2]{*}{Len(95)} & SAIC$_{F}$ & 0.7962  & 0.9606  & 1.6436  & 1.5786  & 1.8568  & 3.1849  & 2.3578  & 2.7960  & 4.7358  \\
          & SAIC$_{97}$ & 0.7277  & 0.8909  & 1.6754  & 1.4500  & 1.7406  & 3.2627  & 2.1794  & 2.6141  & 4.8475  \\
          & SAIC$_{L}$ & 0.8913  & 1.1425  & 2.0703  & 1.7584  & 2.2072  & 4.0190  & 2.6415  & 3.2808  & 5.9535  \\
          & PAIC  & 0.5382  & 0.6445  & 1.2529  & 1.0581  & 1.2409  & 2.3867  & 1.5755  & 1.8592  & 3.5171  \\
          & SBIC$_{F}$ & 0.7412  & 0.8405  & 1.3017  & 1.4730  & 1.6226  & 2.5035  & 2.1927  & 2.4674  & 3.7725  \\
          & SBIC$_{97}$ & 0.7232  & 0.8824  & 1.6506  & 1.4452  & 1.7293  & 3.2236  & 2.1731  & 2.5994  & 4.7914  \\
          & SBIC$_{L}$ & 0.8913  & 1.1425  & 2.0703  & 1.7584  & 2.2072  & 4.0190  & 2.6415  & 3.2808  & 5.9535  \\
          & FULL  & 0.8190  & 1.0513  & 1.8965  & 1.5976  & 2.0065  & 3.6554  & 2.3914  & 2.9755  & 5.3714  \\
    \bottomrule
    \end{tabular}}%
\caption{Coverage Probability and Length of 95\% confidence intervals (CP(95) and Len(95)): Linear regression, $n=10$, $\rho=0.5$.}
\label{tab:lin1}
\end{table}
\begin{table}[b]
\centering  
\scalebox{0.9}{
    \begin{tabular}{ccccccccccc}
    \toprule
          &       & \multicolumn{3}{c}{$\sigma=0.5$} & \multicolumn{3}{c}{$\sigma=1$} & \multicolumn{3}{c}{$\sigma=1.5$} \\
          & \multicolumn{1}{c}{Methods} & \multicolumn{1}{c}{$\beta_{0}$} & \multicolumn{1}{c}{$\beta_{1}$} & \multicolumn{1}{c}{$\mu$} & \multicolumn{1}{c}{$\beta_{0}$} & \multicolumn{1}{c}{$\beta_{1}$} & \multicolumn{1}{c}{$\mu$} & \multicolumn{1}{c}{$\beta_{0}$} & \multicolumn{1}{c}{$\beta_{1}$} & \multicolumn{1}{c}{$\mu$} \\
    \midrule
    \multirow{8}[2]{*}{CP(95)} & SAIC$_{F}$ & 0.927  & 0.929  & 0.903  & 0.924  & 0.924  & 0.919  & 0.916  & 0.917  & 0.902  \\
          & SAIC$_{97}$ & 0.888  & 0.903  & 0.912  & 0.879  & 0.897  & 0.906  & 0.871  & 0.886  & 0.887  \\
          & SAIC$_{L}$ & 0.940  & 0.952  & 0.950  & 0.935  & 0.951  & 0.955  & 0.934  & 0.955  & 0.944  \\
          & PAIC  & 0.775  & 0.792  & 0.795  & 0.778  & 0.764  & 0.777  & 0.739  & 0.742  & 0.733  \\
          & SBIC$_{F}$ & 0.905  & 0.862  & 0.822  & 0.910  & 0.864  & 0.843  & 0.889  & 0.852  & 0.825  \\
          & SBIC$_{97}$ & 0.887  & 0.900  & 0.912  & 0.878  & 0.894  & 0.905  & 0.872  & 0.884  & 0.884  \\
          & SBIC$_{L}$ & 0.940  & 0.952  & 0.950  & 0.935  & 0.951  & 0.955  & 0.934  & 0.955  & 0.944  \\
          & FULL  & 0.897  & 0.901  & 0.882  & 0.886  & 0.903  & 0.901  & 0.878  & 0.884  & 0.871  \\
    \midrule
    \multirow{8}[2]{*}{Len(95)} & SAIC$_{F}$ & 0.8022  & 1.2875  & 1.6250  & 1.5776  & 2.5101  & 3.1716  & 2.3524  & 3.7287  & 4.7741  \\
          & SAIC$_{97}$ & 0.7363  & 1.2316  & 1.6913  & 1.4461  & 2.3762  & 3.2138  & 2.1309  & 3.4837  & 4.7235  \\
          & SAIC$_{L}$ & 0.8966  & 1.6026  & 2.0575  & 1.7602  & 3.1275  & 3.9895  & 2.6425  & 4.6764  & 6.0432  \\
          & PAIC  & 0.5432  & 0.9131  & 1.2463  & 1.0473  & 1.7203  & 2.2909  & 1.5466  & 2.5100  & 3.3704  \\
          & SBIC$_{F}$ & 0.7455  & 1.0578  & 1.2719  & 1.4700  & 2.0625  & 2.5170  & 2.1728  & 3.0253  & 3.7600  \\
          & SBIC$_{97}$ & 0.7329  & 1.2194  & 1.6697  & 1.4406  & 2.3540  & 3.1747  & 2.1229  & 3.4515  & 4.6586  \\
          & SBIC$_{L}$ & 0.8966  & 1.6026  & 2.0575  & 1.7602  & 3.1275  & 3.9895  & 2.6425  & 4.6764  & 6.0432  \\
          & FULL  & 0.8176  & 1.4679  & 1.8765  & 1.6022  & 2.8476  & 3.6119  & 2.3816  & 4.2121  & 5.4696  \\
    \bottomrule
    \end{tabular}}%
\caption{Coverage Probability and Length of 95\% confidence intervals (CP(95) and Len(95)): Linear regression, $n=10$, $\rho=0.8$.}
\label{tab:lin2}
\end{table}
\begin{table}[b]
\centering  
\scalebox{0.9}{
    \begin{tabular}{cccccccccccc}
    \toprule
          &       &       & \multicolumn{3}{c}{$\sigma=0.5$} & \multicolumn{3}{c}{$\sigma=1$} & \multicolumn{3}{c}{$\sigma=1.5$} \\
    n     &       & \multicolumn{1}{l}{Methods} & \multicolumn{1}{c}{$\beta_{1}$} & \multicolumn{1}{c}{$\gamma_{2}$} & \multicolumn{1}{c}{$\mu$} & \multicolumn{1}{c}{$\beta_{1}$} & \multicolumn{1}{c}{$\gamma_{2}$} & \multicolumn{1}{c}{$\mu$} & \multicolumn{1}{c}{$\beta_{1}$} & \multicolumn{1}{c}{$\gamma_{2}$} & \multicolumn{1}{c}{$\mu$} \\
    \midrule
    \multirow{16}[4]{*}{50} & \multirow{8}[2]{*}{CP(95)} & SAIC$_{F}$ & 0.944  & 0.942  & 0.954  & 0.946  & 0.938  & 0.951  & 0.933  & 0.941  & 0.945  \\
          &       & SAIC$_{97}$ & 0.950  & 0.954  & 0.965  & 0.952  & 0.947  & 0.964  & 0.938  & 0.954  & 0.953  \\
          &       & SAIC$_{L}$ & 0.949  & 0.944  & 0.961  & 0.954  & 0.939  & 0.944  & 0.938  & 0.945  & 0.955  \\
          &       & PAIC  & 0.944  & 0.979  & 0.975  & 0.948  & 0.984  & 0.973  & 0.935  & 0.989  & 0.967  \\
          &       & SBIC$_{F}$ & 0.937  & 0.931  & 0.943  & 0.942  & 0.929  & 0.933  & 0.928  & 0.931  & 0.929  \\
          &       & SBIC$_{97}$ & 0.951  & 0.953  & 0.965  & 0.952  & 0.947  & 0.967  & 0.940  & 0.955  & 0.957  \\
          &       & SBIC$_{L}$ & 0.949  & 0.944  & 0.961  & 0.954  & 0.939  & 0.944  & 0.938  & 0.945  & 0.955  \\
          &       & FULL  & 0.939  & 0.938  & 0.956  & 0.947  & 0.933  & 0.942  & 0.931  & 0.942  & 0.947  \\
\cmidrule{2-12}          & \multirow{8}[2]{*}{Len(95)} & SAIC$_{F}$ & 0.3107  & 0.3302  & 0.5346  & 0.6268  & 0.6654  & 1.0979  & 0.9473  & 0.9930  & 1.6533  \\
          &       & SAIC$_{97}$ & 0.3143  & 0.3387  & 0.5537  & 0.6331  & 0.6853  & 1.1349  & 0.9598  & 1.0221  & 1.7118  \\
          &       & SAIC$_{L}$ & 0.3370  & 0.3791  & 0.6212  & 0.6806  & 0.7640  & 1.2779  & 1.0281  & 1.1415  & 1.9283  \\
          &       & PAIC  & 0.3111  & 0.4271  & 0.6142  & 0.6257  & 0.8674  & 1.2629  & 0.9488  & 1.2935  & 1.8814  \\
          &       & SBIC$_{F}$ & 0.2948  & 0.2963  & 0.4668  & 0.5948  & 0.6000  & 0.9562  & 0.9001  & 0.8914  & 1.4282  \\
          &       & SBIC$_{97}$ & 0.3067  & 0.3237  & 0.5245  & 0.6177  & 0.6551  & 1.0732  & 0.9371  & 0.9774  & 1.6191  \\
          &       & SBIC$_{L}$ & 0.3370  & 0.3791  & 0.6212  & 0.6806  & 0.7640  & 1.2779  & 1.0281  & 1.1415  & 1.9283  \\
          &       & FULL  & 0.3330  & 0.3746  & 0.6137  & 0.6730  & 0.7556  & 1.2640  & 1.0158  & 1.1279  & 1.9053  \\
    \midrule
    \multirow{16}[4]{*}{100} & \multirow{8}[2]{*}{CP(95)} & SAIC$_{F}$ & 0.944  & 0.936  & 0.939  & 0.955  & 0.939  & 0.948  & 0.945  & 0.942  & 0.952  \\
          &       & SAIC$_{97}$ & 0.947  & 0.957  & 0.953  & 0.964  & 0.957  & 0.971  & 0.951  & 0.957  & 0.975  \\
          &       & SAIC$_{L}$ & 0.948  & 0.947  & 0.942  & 0.956  & 0.947  & 0.953  & 0.946  & 0.950  & 0.952  \\
          &       & PAIC  & 0.938  & 0.990  & 0.968  & 0.957  & 0.989  & 0.976  & 0.947  & 0.990  & 0.978  \\
          &       & SBIC$_{F}$ & 0.938  & 0.926  & 0.925  & 0.955  & 0.932  & 0.930  & 0.946  & 0.936  & 0.942  \\
          &       & SBIC$_{97}$ & 0.942  & 0.954  & 0.960  & 0.959  & 0.953  & 0.962  & 0.952  & 0.955  & 0.972  \\
          &       & SBIC$_{L}$ & 0.948  & 0.947  & 0.942  & 0.956  & 0.947  & 0.953  & 0.946  & 0.950  & 0.952  \\
          &       & FULL  & 0.946  & 0.945  & 0.941  & 0.955  & 0.946  & 0.951  & 0.946  & 0.947  & 0.949  \\
\cmidrule{2-12}          & \multirow{8}[2]{*}{Len(95)} & SAIC$_{F}$ & 0.2165  & 0.2284  & 0.3762  & 0.4339  & 0.4562  & 0.7491  & 0.6507  & 0.6908  & 1.1201  \\
          &       & SAIC$_{97}$ & 0.2203  & 0.2370  & 0.3924  & 0.4415  & 0.4718  & 0.7732  & 0.6631  & 0.7137  & 1.1579  \\
          &       & SAIC$_{L}$ & 0.2333  & 0.2610  & 0.4337  & 0.4683  & 0.5226  & 0.8689  & 0.7028  & 0.7876  & 1.2983  \\
          &       & PAIC  & 0.2173  & 0.2990  & 0.4410  & 0.4341  & 0.5935  & 0.8579  & 0.6538  & 0.9019  & 1.2906  \\
          &       & SBIC$_{F}$ & 0.2063  & 0.2068  & 0.3310  & 0.4139  & 0.4120  & 0.6524  & 0.6210  & 0.6245  & 0.9773  \\
          &       & SBIC$_{97}$ & 0.2136  & 0.2236  & 0.3658  & 0.4278  & 0.4442  & 0.7195  & 0.6433  & 0.6733  & 1.0750  \\
          &       & SBIC$_{L}$ & 0.2333  & 0.2610  & 0.4337  & 0.4683  & 0.5226  & 0.8689  & 0.7028  & 0.7876  & 1.2983  \\
          &       & FULL  & 0.2320  & 0.2595  & 0.4311  & 0.4660  & 0.5201  & 0.8647  & 0.6991  & 0.7834  & 1.2915  \\
    \bottomrule
    \end{tabular}}%
\caption{Coverage Probability and Length of 95\% confidence intervals (CP(95) and Len(95)): Linear regression, $\rho=0.5$.}
\label{tab:lin3}
\end{table}
\begin{table}[b]
\centering  
\scalebox{0.9}{
    \begin{tabular}{cccccccccccc}
    \toprule
          &       &       & \multicolumn{3}{c}{$\sigma=0.5$} & \multicolumn{3}{c}{$\sigma=1$} & \multicolumn{3}{c}{$\sigma=1.5$} \\
    n     &       & \multicolumn{1}{l}{Methods} & \multicolumn{1}{c}{$\beta_{1}$} & \multicolumn{1}{c}{$\gamma_{2}$} & \multicolumn{1}{c}{$\mu$} & \multicolumn{1}{c}{$\beta_{1}$} & \multicolumn{1}{c}{$\gamma_{2}$} & \multicolumn{1}{c}{$\mu$} & \multicolumn{1}{c}{$\beta_{1}$} & \multicolumn{1}{c}{$\gamma_{2}$} & \multicolumn{1}{c}{$\mu$} \\
\midrule
    \multirow{16}[4]{*}{50} & \multirow{8}[2]{*}{CP(95)} & SAIC$_{F}$ & 0.943  & 0.945  & 0.936  & 0.956  & 0.936  & 0.927  & 0.952  & 0.947  & 0.948  \\
          &       & SAIC$_{97}$ & 0.949  & 0.962  & 0.956  & 0.961  & 0.960  & 0.946  & 0.962  & 0.971  & 0.959  \\
          &       & SAIC$_{L}$ & 0.953  & 0.957  & 0.943  & 0.956  & 0.944  & 0.935  & 0.950  & 0.954  & 0.960  \\
          &       & PAIC  & 0.950  & 0.983  & 0.972  & 0.965  & 0.983  & 0.966  & 0.959  & 0.994  & 0.966  \\
          &       & SBIC$_{F}$ & 0.924  & 0.895  & 0.923  & 0.934  & 0.892  & 0.914  & 0.924  & 0.911  & 0.926  \\
          &       & SBIC$_{97}$ & 0.950  & 0.965  & 0.954  & 0.960  & 0.968  & 0.950  & 0.956  & 0.968  & 0.958  \\
          &       & SBIC$_{L}$ & 0.953  & 0.957  & 0.943  & 0.956  & 0.944  & 0.935  & 0.950  & 0.954  & 0.960  \\
          &       & FULL  & 0.947  & 0.953  & 0.938  & 0.952  & 0.938  & 0.932  & 0.944  & 0.945  & 0.957  \\
\cmidrule{2-12}          & \multirow{8}[2]{*}{Len(95)} & SAIC$_{F}$ & 0.4236  & 0.4912  & 0.5408  & 0.8448  & 0.9822  & 1.0798  & 1.2686  & 1.4739  & 1.6413  \\
          &       & SAIC$_{97}$ & 0.4334  & 0.5152  & 0.5559  & 0.8673  & 1.0367  & 1.1148  & 1.3091  & 1.5533  & 1.7212  \\
          &       & SAIC$_{L}$ & 0.4911  & 0.6313  & 0.6319  & 0.9830  & 1.2643  & 1.2611  & 1.4720  & 1.8881  & 1.9103  \\
          &       & PAIC  & 0.4430  & 0.5898  & 0.6126  & 0.8848  & 1.1935  & 1.2311  & 1.3318  & 1.7733  & 1.8737  \\
          &       & SBIC$_{F}$ & 0.3751  & 0.3747  & 0.4674  & 0.7466  & 0.7464  & 0.9339  & 1.1293  & 1.1284  & 1.4266  \\
          &       & SBIC$_{97}$ & 0.4115  & 0.4665  & 0.5235  & 0.8240  & 0.9397  & 1.0537  & 1.2517  & 1.4080  & 1.6437  \\
          &       & SBIC$_{L}$ & 0.4911  & 0.6313  & 0.6319  & 0.9830  & 1.2643  & 1.2611  & 1.4720  & 1.8881  & 1.9103  \\
          &       & FULL  & 0.4863  & 0.6250  & 0.6256  & 0.9722  & 1.2502  & 1.2474  & 1.4564  & 1.8682  & 1.8897  \\
    \midrule
    \multirow{16}[4]{*}{100} & \multirow{8}[2]{*}{CP(95)} & SAIC$_{F}$ & 0.947  & 0.944  & 0.944  & 0.951  & 0.950  & 0.950  & 0.957  & 0.948  & 0.940  \\
          &       & SAIC$_{97}$ & 0.963  & 0.974  & 0.954  & 0.970  & 0.968  & 0.961  & 0.962  & 0.976  & 0.953  \\
          &       & SAIC$_{L}$ & 0.951  & 0.951  & 0.948  & 0.946  & 0.952  & 0.953  & 0.959  & 0.950  & 0.949  \\
          &       & PAIC  & 0.963  & 0.990  & 0.969  & 0.968  & 0.983  & 0.967  & 0.958  & 0.992  & 0.970  \\
          &       & SBIC$_{F}$ & 0.937  & 0.916  & 0.923  & 0.944  & 0.925  & 0.927  & 0.939  & 0.919  & 0.927  \\
          &       & SBIC$_{97}$ & 0.958  & 0.978  & 0.957  & 0.964  & 0.965  & 0.959  & 0.960  & 0.969  & 0.950  \\
          &       & SBIC$_{L}$ & 0.951  & 0.951  & 0.948  & 0.946  & 0.952  & 0.953  & 0.959  & 0.950  & 0.949  \\
          &       & FULL  & 0.950  & 0.944  & 0.945  & 0.941  & 0.948  & 0.951  & 0.952  & 0.943  & 0.947  \\
\cmidrule{2-12}          & \multirow{8}[2]{*}{Len(95)} & SAIC$_{F}$ & 0.2936  & 0.3381  & 0.3851  & 0.5808  & 0.6747  & 0.7526  & 0.8797  & 1.0155  & 1.1435  \\
          &       & SAIC$_{97}$ & 0.3037  & 0.3575  & 0.3971  & 0.6010  & 0.7190  & 0.7787  & 0.9091  & 1.0829  & 1.1810  \\
          &       & SAIC$_{L}$ & 0.3387  & 0.4326  & 0.4463  & 0.6712  & 0.8628  & 0.8684  & 1.0131  & 1.2991  & 1.3296  \\
          &       & PAIC  & 0.3102  & 0.4067  & 0.4389  & 0.6157  & 0.8286  & 0.8712  & 0.9303  & 1.2464  & 1.3101  \\
          &       & SBIC$_{F}$ & 0.2615  & 0.2610  & 0.3361  & 0.5193  & 0.5181  & 0.6614  & 0.7848  & 0.7830  & 0.9920  \\
          &       & SBIC$_{97}$ & 0.2841  & 0.3137  & 0.3695  & 0.5631  & 0.6315  & 0.7267  & 0.8512  & 0.9497  & 1.0961  \\
          &       & SBIC$_{L}$ & 0.3387  & 0.4326  & 0.4463  & 0.6712  & 0.8628  & 0.8684  & 1.0131  & 1.2991  & 1.3296  \\
          &       & FULL  & 0.3370  & 0.4304  & 0.4440  & 0.6674  & 0.8577  & 0.8635  & 1.0077  & 1.2922  & 1.3225  \\
    \bottomrule
    \end{tabular}}%
\caption{Coverage Probability and Length of 95\% confidence intervals (CP(95) and Len(95)): Linear regression, $\rho=0.8$.}
\label{tab:lin4}
\end{table}
\begin{table}[htbp]
  \centering
  \scalebox{0.73}{
    \begin{tabular}{ccccccccc}
    \toprule
          &       &       & \multicolumn{3}{c}{$\rho=0.5$} & \multicolumn{3}{c}{$\rho=0.8$} \\
    \multicolumn{1}{l}{n} &       & Methods & \multicolumn{1}{l}{$\beta_{1}$} & \multicolumn{1}{l}{$\gamma_{2}$} & \multicolumn{1}{l}{$\mu$} & \multicolumn{1}{l}{$\beta_{1}$} & \multicolumn{1}{l}{$\gamma_{2}$} & \multicolumn{1}{l}{$\mu$} \\
    \midrule
    \multirow{16}[4]{*}{30} & \multirow{8}[2]{*}{CP(95)} & SAIC$_{F}$ & 0.943  & 0.942  & 0.933  & 0.944  & 0.957  & 0.927  \\
          &       & SAIC$_{97}$ & 0.960  & 0.959  & 0.948  & 0.963  & 0.979  & 0.948  \\
          &       & SAIC$_{L}$ & 0.282  & 0.228  & 0.205  & 0.240  & 0.251  & 0.233  \\
          &       & PAIC  & 0.957  & 0.990  & 0.973  & 0.956  & 0.992  & 0.974  \\
          &       & SBIC$_{F}$ & 0.933  & 0.925  & 0.919  & 0.919  & 0.904  & 0.910  \\
          &       & SBIC$_{97}$ & 0.962  & 0.957  & 0.949  & 0.959  & 0.974  & 0.947  \\
          &       & SBIC$_{L}$ & 0.285  & 0.217  & 0.210  & 0.241  & 0.249  & 0.240  \\
          &       & FULL  & 0.950  & 0.944  & 0.932  & 0.949  & 0.952  & 0.935  \\
\cmidrule{2-9}          & \multirow{8}[2]{*}{Len(95)} & SAIC$_{F}$ & 0.3946  & 0.4488  & 4.8367  & 0.5094  & 0.6178  & 5.7910  \\
          &       & SAIC$_{97}$ & 0.4083  & 0.4689  & 5.0453  & 0.5315  & 0.6528  & 5.9100  \\
          &       & SAIC$_{L}$ & 0.0733  & 0.0777  & 1.0429  & 0.1085  & 0.1309  & 1.2440  \\
          &       & PAIC  & 0.4010  & 0.5753  & 5.8368  & 0.5268  & 0.7212  & 6.8963  \\
          &       & SBIC$_{F}$ & 0.3610  & 0.3943  & 3.9774  & 0.4288  & 0.4584  & 4.7257  \\
          &       & SBIC$_{97}$ & 0.3957  & 0.4496  & 4.7583  & 0.5034  & 0.6007  & 5.5382  \\
          &       & SBIC$_{L}$ & 0.0733  & 0.0777  & 1.0429  & 0.1085  & 0.1309  & 1.2440  \\
          &       & FULL  & 0.4468  & 0.5281  & 5.9641  & 0.6203  & 0.8077  & 7.0149  \\
    \midrule
    \multirow{16}[4]{*}{50} & \multirow{8}[2]{*}{CP(95)} & SAIC$_{F}$ & 0.935  & 0.940  & 0.935  & 0.960  & 0.950  & 0.946  \\
          &       & SAIC$_{97}$ & 0.952  & 0.958  & 0.946  & 0.971  & 0.981  & 0.959  \\
          &       & SAIC$_{L}$ & 0.214  & 0.165  & 0.173  & 0.231  & 0.196  & 0.207  \\
          &       & PAIC  & 0.949  & 0.984  & 0.965  & 0.973  & 0.992  & 0.974  \\
          &       & SBIC$_{F}$ & 0.931  & 0.927  & 0.920  & 0.939  & 0.915  & 0.929  \\
          &       & SBIC$_{97}$ & 0.953  & 0.950  & 0.947  & 0.974  & 0.980  & 0.958  \\
          &       & SBIC$_{L}$ & 0.213  & 0.164  & 0.178  & 0.228  & 0.191  & 0.205  \\
          &       & FULL  & 0.937  & 0.941  & 0.935  & 0.960  & 0.956  & 0.948  \\
\cmidrule{3-9}          & \multirow{8}[2]{*}{Len(95)} & SAIC$_{F}$ & 0.2743  & 0.3110  & 4.1097  & 0.3474  & 0.4141  & 5.4156  \\
          &       & SAIC$_{97}$ & 0.2828  & 0.3235  & 4.2536  & 0.3621  & 0.4352  & 5.5940  \\
          &       & SAIC$_{L}$ & 0.0400  & 0.0429  & 0.6725  & 0.0568  & 0.0686  & 0.8760  \\
          &       & PAIC  & 0.2800  & 0.4037  & 4.9884  & 0.3644  & 0.4877  & 6.4575  \\
          &       & SBIC$_{F}$ & 0.2561  & 0.2781  & 3.5754  & 0.2983  & 0.3130  & 4.6923  \\
          &       & SBIC$_{97}$ & 0.2734  & 0.3081  & 4.0233  & 0.3390  & 0.3921  & 5.2512  \\
          &       & SBIC$_{L}$ & 0.0400  & 0.0429  & 0.6725  & 0.0568  & 0.0686  & 0.8760  \\
          &       & FULL  & 0.3036  & 0.3586  & 4.7645  & 0.4144  & 0.5329  & 6.4744  \\
    \midrule
    \multirow{16}[4]{*}{100} & \multirow{8}[2]{*}{CP(95)} & SAIC$_{F}$ & 0.951  & 0.958  & 0.939  & 0.944  & 0.947  & 0.942  \\
          &       & SAIC$_{97}$ & 0.964  & 0.972  & 0.957  & 0.963  & 0.969  & 0.964  \\
          &       & SAIC$_{L}$ & 0.148  & 0.129  & 0.140  & 0.163  & 0.163  & 0.135  \\
          &       & PAIC  & 0.959  & 0.998  & 0.970  & 0.966  & 0.986  & 0.978  \\
          &       & SBIC$_{F}$ & 0.953  & 0.947  & 0.926  & 0.935  & 0.915  & 0.945  \\
          &       & SBIC$_{97}$ & 0.962  & 0.963  & 0.959  & 0.959  & 0.969  & 0.961  \\
          &       & SBIC$_{L}$ & 0.150  & 0.132  & 0.141  & 0.157  & 0.153  & 0.138  \\
          &       & FULL  & 0.953  & 0.967  & 0.941  & 0.954  & 0.951  & 0.950  \\
\cmidrule{3-9}          & \multirow{8}[2]{*}{Len(95)} & SAIC$_{F}$ & 0.1780  & 0.1960  & 2.3035  & 0.2199  & 0.2616  & 2.4519  \\
          &       & SAIC$_{97}$ & 0.1836  & 0.2044  & 2.4338  & 0.2287  & 0.2779  & 2.5435  \\
          &       & SAIC$_{L}$ & 0.0185  & 0.0201  & 0.2789  & 0.0253  & 0.0310  & 0.2974  \\
          &       & PAIC  & 0.1822  & 0.2556  & 2.7329  & 0.2358  & 0.3129  & 2.7757  \\
          &       & SBIC$_{F}$ & 0.1682  & 0.1760  & 1.9549  & 0.1934  & 0.1990  & 2.0850  \\
          &       & SBIC$_{97}$ & 0.1766  & 0.1918  & 2.2195  & 0.2124  & 0.2417  & 2.3300  \\
          &       & SBIC$_{L}$ & 0.0185  & 0.0201  & 0.2789  & 0.0253  & 0.0310  & 0.2974  \\
          &       & FULL  & 0.1950  & 0.2253  & 2.7357  & 0.2573  & 0.3349  & 2.8827  \\
    \bottomrule
    \end{tabular}}%
    \caption{Coverage Probability and Length of 95\% confidence intervals (CP(95) and Len(95)): Poisson regression.}%
  \label{tab:pois}%
\end{table}%

\end{document}